\documentclass[english]{aa}
\usepackage{mathptmx}
\usepackage[T1]{fontenc}
\usepackage[latin9]{inputenc}
\usepackage{url}
\usepackage{graphicx}
\usepackage[authoryear]{natbib}

\makeatletter

\newcommand{\noun}[1]{\textsc{#1}}
\providecommand{\tabularnewline}{\\}

\makeatother

\usepackage{babel}

\begin{document}
\abstract{}{Up to now, the study of the star formation rate in galaxies
has been mainly based on the H$\alpha$ emission-line luminosity.
However, this standard calibration cannot be applied at all redshifts
given one instrumental setup. Surveys based on optical spectroscopy
do not observe the H$\alpha$ emission line at redshifts higher than
$z\sim0.5$. Our goal is to study existing star formation rate calibrations
and to provide new ones, still based on emission-line luminosities,
which can be applied for various instrumental setups.}{We use the
SDSS public data release DR4, which gives star formation rates and
emission-line luminosities of more than $100\,000$ star-forming galaxies
observed at low redshifts. We take advantage of this statistically
significant sample in order to study the relations, based on these
data, between the star formation rate and the luminosities of some
well-chosen emission lines. We correct the emission-line measurements
for dust attenuation using the same attenuation curve as the one used
to derive the star formation rates.}{We confirm that the best results
are obtained when relating star formation rates to the H$\alpha$
emission line luminosity, itself corrected for dust attenuation. This
calibration has an uncertainty of $0.17$ dex. We show that one has
to check carefully the method used to derive the dust attenuation
and to use the adequate calibration: in some cases, the standard scaling
law has to be replaced by a more general power law. When data is corrected
for dust attenuation but the H$\alpha$ emission line not observed,
the use of the H$\beta$ emission line, if possible, has to be preferred
to the {[}O\noun{ii}]$\lambda$3727 emission line. In the case of
uncorrected data, the correction for dust attenuation can be assumed
as a constant mean value but we show that such method leads to poor
results, in terms of dispersion and residual slope. Self-consistent
corrections such as previous studies based on the absolute magnitude
give better results in terms of dispersion but still suffer from systematic
shifts, and/or residual slopes. The best results with data not corrected
for dust attenuation are obtained when using the observed {[}O\noun{ii}]$\lambda$3727
and H$\beta$ emission lines together. This calibration has an uncertainty
of $0.23$ dex.}{}


\title{Emission-lines calibrations of the Star Formation Rate from the Sloan
Digital Sky Survey}

\author{B. Argence\inst{1,2}\and F. Lamareille\inst{1,3}}

\institute{Laboratoire d\textquoteright{}Astrophysique de Toulouse-Tarbes, Université
de Toulouse, CNRS, 14 avenue Edouard Belin, F-31400 Toulouse, France\and APC,
UMR 7164, CNRS, Université Paris 7 Denis Diderot, 10 rue Alice Domon
et Léonie Duquet, F-75025 Paris Cedex 13, France\and Osservatorio
Astronomico di Bologna, via Ranzani 1, I-40127 Bologna, Italy\\
\email{argence@apc.univ-paris7.fr, flamare@ast.obs-mip.fr}}

\date{Received ; Accepted}

\maketitle

\section{Introduction}

Massive ongoing spectroscopic surveys of the high-redshift Universe
allow us to draw a new picture of how the galaxies evolve through
cosmic times. Many physical properties are used as tracers of the
evolutionary stage and the past history of one galaxy (e.g. stellar
mass, metallicity, amount of dust, star formation rate). Thanks to
extensive works on stellar population models \citep{Bruzual:2003MNRAS.344.1000B,Fioc:1997A&A...326..950F}
and photo-ionization models \citep{Charlot:2001MNRAS.323..887C},
most of these parameters can now be well recovered from rest-frame
optical spectroscopic observations, using all information available
from the stellar continuum and absorption- or emission-line measurements.

Unfortunately, these models often require a better signal-to-noise
ratio, wavelength coverage, and/or spectral resolution than the ones
available on recent deep surveys (e.g. VVDS: \citealp{LeFevre:2005A&A...439..845L};
zCOSMOS: Lily et al. 2007). Thus, a simpler approach is to use instead
emission-line diagnostics that would be calibrated on high quality
samples, taken from local Universe surveys (e.g. 2dFGRS: \citealp{Colless:2001MNRAS.328.1039C};
SDSS: \citealp{York:2000AJ....120.1579Y}). This approach has already
been extensively used with older data to recover metallicities with
oxygen-to-hydrogen \citep{McGaugh:1991ApJ...380..140M,Kewley:2002ApJS..142...35K}
or nitrogen-to-hydrogen \citep{Pettini:2004MNRAS.348L..59P,VanZee:1998AJ....116.2805V}
line ratios, dust content with the H$\alpha$/H$\beta$ Balmer decrement
\citep{Seaton:1979MNRAS.187P..73S,Calzetti:2001PASP..113.1449C},
or star formation rates with the H$\alpha$ or {[}O\noun{ii}]$\lambda$3727
emission-line luminosities \citep{Kennicutt:1998ARA&A..36..189K,Kewley:2004AJ....127.2002K}.

Recently, the star formation rates of $>100\,000$ SDSS star-forming
galaxies have been estimated using the CL01 method \citep{Charlot:2001MNRAS.323..887C}
which takes into account all information available in emission-line
measurements \citep{Brinchmann:2004MNRAS.351.1151B}. Then, the emission-line
measurements of the same galaxies have also been made publicly available.
We can thus use this large amount of data to derive new good quality
calibrations of the star formation rates (hereafter SFR) from emission-line
luminosities. 

The standard calibration of the SFR versus the H$\alpha$ luminosity
is based on a very simple scaling relation, driven by the amount of
H$\alpha$ recombination photons being directly proportional to the
intensity of the ionizing source, i.e. the amount of young, hot stars,
which the SFR is an indicator of \citep{Kennicutt:1998ARA&A..36..189K}.
The only drawback to this simple assumption is that the H$\alpha$
luminosity has to be corrected for dust attenuation. 

As shown by \citet{Kewley:2004AJ....127.2002K}, the calibration of
the SFR versus the {[}O\noun{ii}]$\lambda$3727 luminosity is not
only more affected by dust attenuation, but also depends on metallicity,
since interstellar medium with higher gas-phase metallicities produces
less {[}O\noun{ii}]$\lambda$3727 photons because of increased oxygen
cooling. Starting from this result, they have derived new calibrations
of the SFR versus {[}O\noun{ii}]$\lambda$3727 luminosity which take
into account the metallicity. The idea was to calibrate the {[}O\noun{ii}]$\lambda$3727/H$\alpha$
emission-line ratio versus the metallicity, and then to use it to
recover the SFR from the H$\alpha$ luminosity, itself derived from
the {[}O\noun{ii}]$\lambda$3727 luminosity. The dependence on the
metallicity of the {[}O\noun{ii}]$\lambda$3727/H$\alpha$ emission-line
ratio, or more generally the {[}O\noun{ii}]$\lambda$3727 based SFR
calibrations, has been later confirmed by \citet{Mouhcine:2005MNRAS.362.1143M}
and \citet{Bicker:2005A&A...443L..19B}.

Recently \citet{Moustakas:2006ApJ...642..775M} and \citet{Weiner:2006astro.ph.10842W}
have derived a new calibration of the SFR of galaxies versus {[}O\noun{ii}]$\lambda$3727
luminosity. They take care of the effect of the metallicity with a
slightly different approach, by parametrizing their calibration in
terms of the $B$-band luminosity, which can be used as a rough metallicity
indicator because of the metallicity-luminosity relation \citep{Lequeux:1979A&A....80..155L,Skillman:1989ApJ...347..875S,Richer:1995ApJ...445..642R,Lamareille:2004MNRAS.350..396L,Tremonti:2004astro.ph..5537T,Lamareille:2006A&A...448..907L}.

We point out that these previous studies base their new calibrations
on either a direct estimate or a secondary indicator (e.g. rest-frame
luminosity) of the metallicity, which in both cases \emph{suffers
from intrinsic uncertainties and degeneracies} (see \citealp{Kewley:2008arXiv0801.1849K}
for a detailed discussion on the uncertainties of metallicity calibrations,
or \citealt{Lamareille:2004MNRAS.350..396L} for a discussion on the
dispersion of the luminosity-metallicity relation). Moreover, some
of these previous works also \emph{assume that the dust attenuation
is known}, whereas it is difficult to estimate this quantity when
H$\alpha$ emission line is not measured (which is the case when one
wants to used {[}O\noun{ii}]$\lambda$3727 instead to derive SFR).

The paper is organized as follows: we first present the sample we
have used and the selection we applied to it (Sect.~\ref{sec:Description-of-the}),
then we calibrate the SFR as a function of emission-line luminosities
when a correction for dust attenuation is available (Sect.~\ref{sec:SFR-calibration-with}),
or when it is not available (Sect.~\ref{sec:SFR-calibration-without}).
We finally draw some conclusions (Sect.~\ref{sec:Conclusions}).

\section{Description of the sample\label{sec:Description-of-the}}

The physical properties of a total of $567\,486$ galaxy spectra inside
SDSS Data Release 4 (DR4, \citealp{Adelman-McCarthy:2006ApJS..162...38A})
have been made publicy available on the following website: \url{http://www.mpa-garching.mpg.de/SDSS/}.
Taking advantage of this huge amount of data, we made a sub catalog
with the SFR, the emission-line measurements, the spectral classification
and the metallicities of a sub sample of galaxies. The selection will
be described later in this section as we first describe, for the benefit
of the reader, the main ingredients of the methods used to derive
the physical parameters.

\subsection{Physical properties}

The spectral properties of SDSS galaxies have been measured with {}``platefit''
software by \citet{Tremonti:2004astro.ph..5537T}. For each spectrum,
they perform a careful subtraction of the stellar continuum and absorption
lines by fitting a linear combination of single stellar population
models of different ages \citep{Bruzual:2003MNRAS.344.1000B}. Then
they fit all emission lines at the same time as a unique nebular spectrum
in which each line is modeled by a Gaussian. Thanks to the subtraction
of the stellar component, Balmer emission lines are automatically
corrected for underlying absorption. We note that the {[}O\noun{ii}]$\lambda$3727
line that we use in this study is the sum of two Gaussians of the
same width modeling the {[}O\noun{ii}]$\lambda\lambda$3726,3729 doublet.
The AGN classification is based on the {[}O\noun{iii}]$\lambda$5007/H$\beta$
vs. {[}N\noun{ii}]$\lambda$6584/H$\alpha$ emission-line diagnostic
diagram and has been performed by \citet{Kauffmann:2003MNRAS.346.1055K}.

The SFR and gas-phase oxygen abundances have been computed from emission-lines
fluxes using the \citet[hereafter CL01]{Charlot:2001MNRAS.323..887C}
method by \citet{Brinchmann:2004MNRAS.351.1151B} and \citet{Tremonti:2004astro.ph..5537T}
respectively. This method compares all emission-lines fluxes together
to a set of theoretical nebular spectra, which are modeled with five
parameters: the metallicity, the ionization degree, the dust-to-metal
ratio, the dust attenuation, and the SFR efficiency factor. For each
galaxy, one estimate of each observed parameter is computed through
a Bayesian approach (see \citealp{Brinchmann:2004MNRAS.351.1151B}
for more details). We note that the SFRs are computed assuming the
\citet{Kroupa:2001MNRAS.322..231K} initial mass function (IMF). One
may scale the SFR estimates discussed in this paper to the \citet{Salpeter:1955ApJ...121..161S}
IMF by multiplying them by a factor $1.5$ (or by adding $0.176$
dex to their logarithm), and to the \citet{Chabrier:2003PASP..115..763C}
IMF by multiplying them by a factor $0.88$ (or by subtracting $0.056$
dex to their logarithm). The CL01 SFR will be used as the reference
SFR throughout this study.

We compute the luminosity $L$ of each line starting from its measured
flux $F$, and the redshift $z$ of the galaxy, using the following
equation:

\begin{equation}
\begin{array}{c}
L=4\pi\left(\frac{c}{H_{\mathrm{0}}}\cdot(1+z)\cdot\int_{0}^{z}f(z')^{-1/2}dz'\right)^{2}\times F\\
f(z')=(1+z')^{2}(1+\Omega_{\mathrm{m}}z')-\Omega_{\Lambda}z'(2+z')\end{array}\label{eq:lum}\end{equation}

The luminosities are calculated with the same cosmology based on WMAP
results \citep{Spergel:2003ApJS..148..175S} as the one used by \citet{Brinchmann:2004MNRAS.351.1151B}
to estimate the SFR: $H_{\mathrm{0}}=70$ km s$^{-1}$ Mpc$^{-1}$,
$\Omega_{\Lambda}=0.7$ and $\Omega_{\mathrm{m}}=0.3$. We note that
all the calibrations provided in this study are \emph{independent
of the cosmology}. They indeed compare SFR and emission-line luminosities
which are affected in the same way by the distance modulus.

Finally, we have cross-matched this catalog with the VAGC catalog
\citep{Blanton:2005AJ....129.2562B} to get the $k$-corrected absolute
magnitudes which are used in this study. The absolute magnitude in
the $B$-band is derived from SDSS $u'$- and $g'$-bands using Eq.~1
of \citet{Moustakas:2006ApJ...642..775M}.

\subsection{Dust attenuation}

We will use the following notations in all this work: the observed
flux of the three emission lines used in this study are designed by
the name of the line: H$\alpha$, H$\beta$ and {[}O\noun{ii}] (stands
for {[}O\noun{ii}]$\lambda$3727). The intrinsic flux corrected for
dust attenuation is designed by a superscript \emph{i}: H$\alpha^{\mathrm{i}}$,
H$\beta^{\mathrm{i}}$ and {[}O\noun{ii}]$^{\mathrm{i}}$. The attenuation
law is designed by the function $\tau(\lambda)$, and the effective
dust attenuation in $V$-band by the symbol $\tau_{V}$. For simplification
purposes, we also note: $\tau(V)=\tau(5500\,\mbox{\AA})$ ($V$-band),
$\tau(\beta)=\tau(4861\,\mbox{\AA})$, $\tau(\alpha)=\tau(6563\,\mbox{\AA})$,
and $\tau(\mathrm{O}\mathsc{ii})=\tau(3727\,\mbox{\AA})$. The intrinsic
flux $f_{\lambda}^{\mathrm{i}}(\lambda)$ at any wavelength $\lambda$
as a function of the observed flux $f_{\lambda}(\lambda)$ is given
by the following equation:\begin{equation}
f_{\lambda}^{\mathrm{i}}(\lambda)=f_{\lambda}(\lambda)\cdot\exp\left(\frac{\tau_{V}}{\tau(V)}\cdot\tau(\lambda)\right)\label{eq:intrflux}\end{equation}

We compute the dust attenuation by comparing the observed to the intrinsic
H$\alpha$/H$\beta$ emission-line ratios, using the following equation:\begin{equation}
\tau_{V}^{\mathrm{Balmer}}=\frac{\ln(\mathrm{H}\alpha/\mathrm{H}\beta)-\ln(\mathrm{H}\alpha^{\mathrm{i}}/\mathrm{H}\beta^{\mathrm{i}})}{\tau(\beta)-\tau(\alpha)}\cdot\tau(V)\label{eq:tauV}\end{equation}

In the following work, we use the standard case B recombination ratio
(hereafter standard intrinsic Balmer ratio): $\mathrm{H}\alpha^{\mathrm{i}}/\mathrm{H}\beta^{\mathrm{i}}=2.85$
\citep{Osterbrock:1989agna.book.....O} and the \citet{Charlot:2000ApJ...539..718C}
mean attenuation curve: $\tau(\lambda)\propto\lambda^{-0.7}$. The
choice of this curve is made to be consistent with the setup used
to compute the SFRs in the SDSS DR4 data (the SFR is computed self-consistently
with correction for dust attenuation).

The dust attenuation recovered with an assumed constant intrinsic
Balmer ratio ($\tau_{V}^{\mathrm{Balmer}}$) does not take into account
the variations of this ratio with other physical parameters (e.g.
metallicity). We thus also use in this study the true dust attenuation
recovered with CL01 models by \citet{Brinchmann:2004MNRAS.351.1151B}.
We design it by the symbol $\tau_{V}^{\mathrm{CL01}}$. The intrinsic
flux corrected for dust attenuation using this parameter is designed
by a superscript $iC$: H$\alpha^{\mathrm{iC}}$, H$\beta^{\mathrm{iC}}$
and {[}O\noun{ii}]$^{\mathrm{iC}}$. 

Finally, another way to estimate the dust attenuation is to compare
the observed colors of the galaxies to a library of stellar population
synthesis models (hereafter SED fitting). This work has been carried
on with \citet{Bruzual:2003MNRAS.344.1000B} models on SDSS galaxies
by \citet{Kauffmann:2003MNRAS.341...33K}, who provide an estimate
of the continuum attenuation in the $z$-band ($A_{z}$) with \citet{Charlot:2000ApJ...539..718C}
attenuation curve. It is straightforward to convert the $A_{z}$ parameter
to $\tau_{V}$ using the following formula:\begin{equation}
\tau_{V}^{\mathrm{SED}}=\frac{A_{z}}{1.086}\cdot\left(\frac{5500}{8800}\right)^{-0.7}\cdot\frac{1}{A_{\star}/A_{\mathrm{g}}}\label{eq:tauVSED}\end{equation}
Please note that the $1.086$ factor is the conversion from magnitudes
to opacities and $A_{\star}/A_{\mathrm{g}}$ is the mean ratio between
the dust attenuation affecting stars (which dominate the colors of
the galaxies) and the one affecting gas. The dust attenuation estimated
from SED fitting is characterized by a high uncertainty but might
be used in a statistical way.

\subsection{Sample selection}

Thanks to the spectral classification provided by \citet{Kauffmann:2003MNRAS.346.1055K},
we select only star-forming galaxies and remove all AGN from our sample.
In the whole study, we additionally remove all galaxies for which
H$\alpha$ and H$\beta$ emission lines (used for estimating the dust
attenuation) are not measured with a signal-to-noise ratio of at least
$5$. We apply the same selection to the {[}O\noun{ii}] emission line
in order to derive the calibrations in which this line is involved.
We note that the signal-to-noise ratios have been corrected according
to the values provided in the SDSS/MPA website (based on the analysis
of duplicated observations): they have been divided by $2.473$ for
H$\alpha$ line, $1.882$ for H$\beta$ line and $2.199$ for {[}O\noun{ii}]
line. We also kept only galaxies with an available estimate of $\tau_{V}^{\mathrm{CL01}}$.
This selection was applied to remove a number of spectra with unreliable
flux measurements. Please note that a number of objects may show a
negative $\tau_{V}^{\mathrm{Balmer}}$ estimate. This is due to statistical
variations around the wrong assumption of a constant intrinsic Balmer
ratio: these objects have not to be removed.

Finally, some objects in SDSS DR4 catalog have been observed twice
or more. We decided to keep only one observation of each duplicated
galaxy by selecting the one with the best signal-to-noise ratio on
the H$\alpha$ emission line. We end up with $123\,713$ and $84\,733$
star-forming galaxies for H$\alpha$-based and {[}O\noun{ii}]-based
studies respectively.

All the parameters used in this study (SFR, metallicity, dust attenuation)
are the median estimates.

\subsection{Properties of the studied sample}

\begin{figure}
\begin{centering}
\includegraphics[width=0.49\columnwidth]{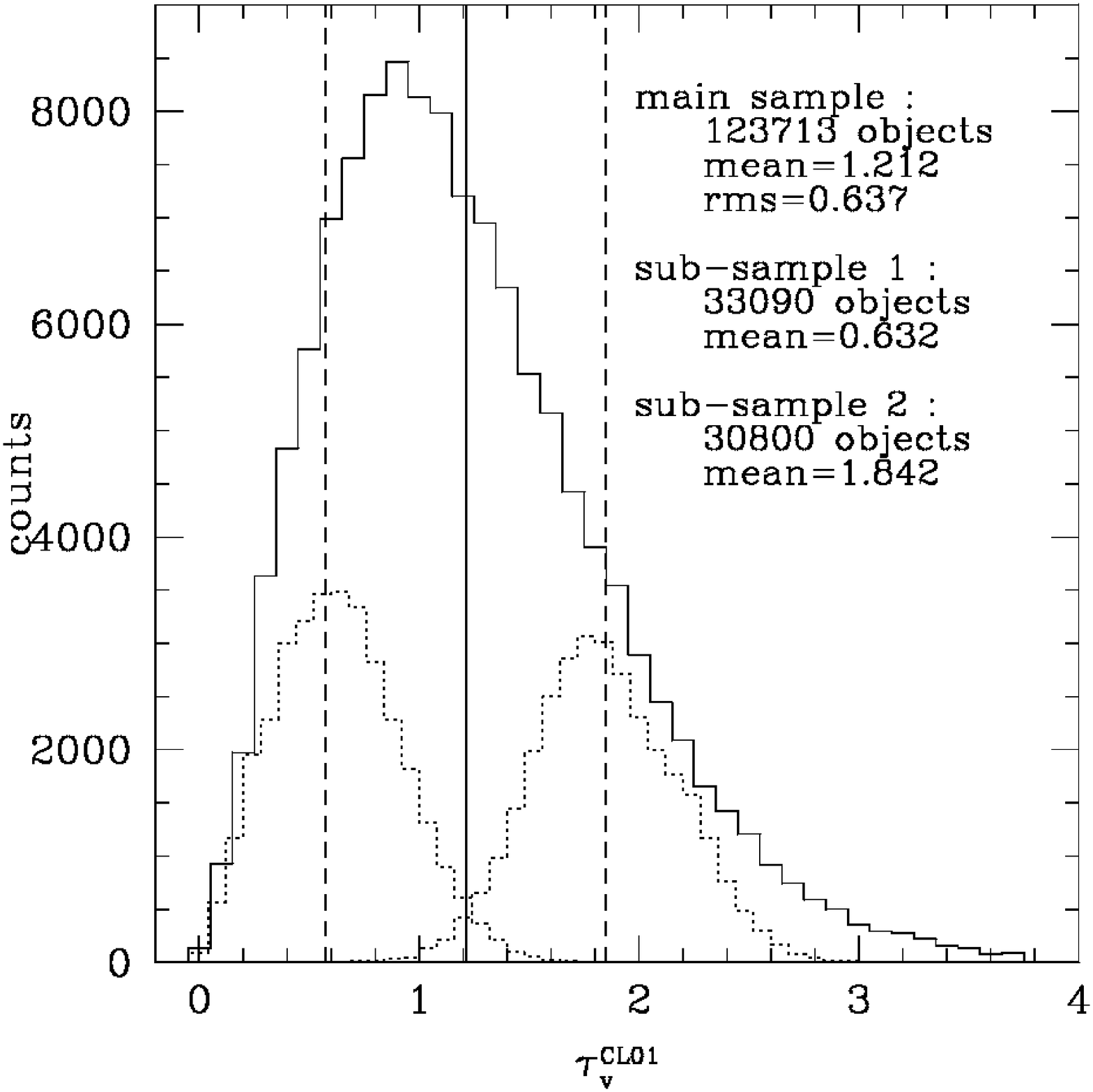}
\includegraphics[width=0.49\columnwidth]{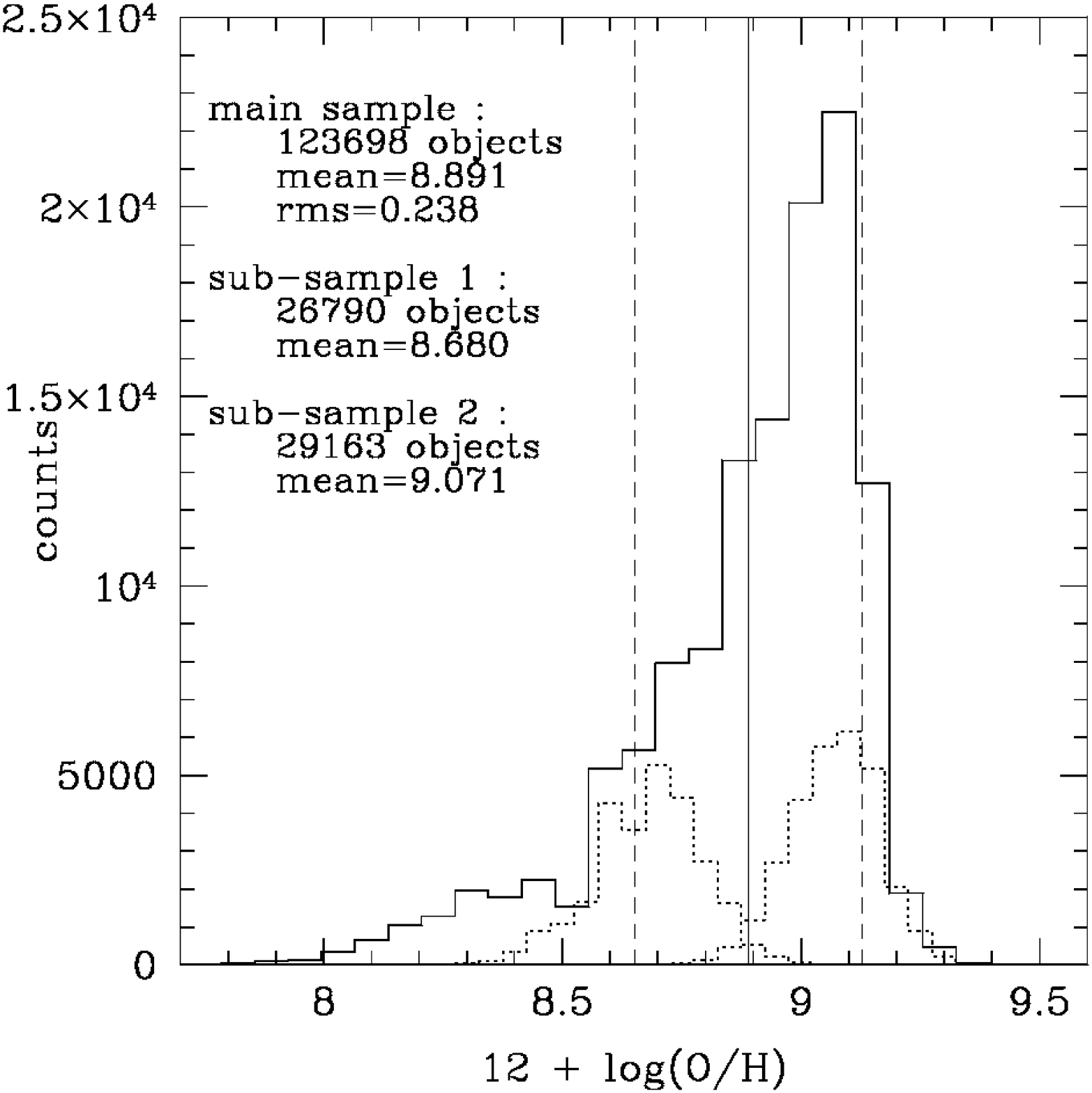}
\par\end{centering}

\caption{This plot shows the histogram of the dust attenuation (left), or of
the gas-phase oxygen abundance (right), both estimated with the CL01
method, for the star-forming galaxies in the SDSS DR4 data. The solid
vertical line shows the mean value of the distribution, and the dashed
vertical lines the 1-$\sigma$ tails. The values are given in the
plot. The two dotted histograms shows the distribution of our two
sub-samples (see text for details).}

\label{fig:histotau}
\end{figure}

Fig.~\ref{fig:histotau} (left) shows the distribution of the effective
dust attenuation in the $V$-band (estimated with the CL01 method)
in the sample we have obtained. The mean value is $1.21$ and the
dispersion is $0.637$. We note that this values are obtained with
the selection on H$\alpha$ and H$\beta$ lines only. The new mean
and dispersion when we add the selection on the {[}O\noun{ii}] line
are $1.10$ and $0.547$ respectively. Fig.~\ref{fig:histotau} (right)
shows the distribution of the gas-phase oxygen abundance in the same
sample (for objects with an available measurement of this parameter).
The mean value is $8.89$ with a dispersion of $0.24$ ($8.88$ mean
value with a dispersion of $0.21$ if the selection on the {[}O\noun{ii}]
line is applied).

Fig.~\ref{fig:histotau} also shows the distributions of four sub-samples
(two in each panel). These sub-samples have been randomly selected
from the main sample, forcing their mean values to be closest as possible
to the mean value of the main sample plus or minus its dispersion,
and their dispersion to be half the dispersion of the main sample.
We end up with four sub-samples: low dust attenuation, high dust attenuation,
low metallicity, and high metallicity. They will be used to test the
calibrations on samples with different dust properties.

\section{SFR calibration with an available dust estimate\label{sec:SFR-calibration-with}}

We present in this section emission-line calibrations of the SFR based
on data already corrected for dust attenuation. We first focus in
Sect.~\ref{sub:dustbalmer} on data with dust attenuation estimated
from the Balmer decrement, which is the common case in the literature.
We then study in Sect.~\ref{sub:otherdust} the different calibrations
we obtain when using the dust attenuation estimated from the CL01
method or SED fitting.

\subsection{Base theory}

The common way to calibrate the SFR against one emission line luminosity
(e.g. H$\alpha$) is a simple scaling relation of the following form:\begin{equation}
\mathrm{SFR}=L(\mathrm{H}\alpha)/\eta_{\mathrm{H}\alpha}\label{eq:defstandard}\end{equation}

This formula, where $\eta_{H\alpha}$ is the efficiency factor, reproduces
the simple relation between the amount of ionizing photons (the SFR)
and the emission-line luminosity. However, it has been already shown
that the H$\alpha$ efficiency factor depends on other physical properties
of the galaxies: \citet{Charlot:2001MNRAS.323..887C} have shown that
$\eta_{H\alpha}$ depends on the metallicity, metal-rich stars being
less luminous and thus producing less H$\alpha$ photons in the interstellar
medium for a same SFR. More recently, \citet{Brinchmann:2004MNRAS.351.1151B}
have shown that $\eta_{H\alpha}$ is also linked to the stellar mass
of SDSS galaxies, massive galaxies producing less H$\alpha$ photons
than dwarfs for a given SFR. We note that these two results are obviously
linked by the mass-metallicity relation \citep{Tremonti:2004astro.ph..5537T,Lamareille:2008AA}.
Fortunately, \citet{Brinchmann:2004MNRAS.351.1151B} have also shown
that this effect is canceled by the commonly used wrong assumption
of a constant intrinsic H$\alpha$/H$\beta$ intrinsic Balmer ratio.
In fact this ratio increases with metallicity, ending up with an overestimate
of the dust attenuation for metal-rich galaxies. This latter effect
almost exactly compensates the decrease of $\eta_{H\alpha}$ when
metallicity increases.

Nevertheless, we decided to adopt a more general approach using a
power-law as described by the following formula:\begin{equation}
\log(\mathrm{SFR})=\epsilon_{\mathrm{H}\alpha}\log L(\mathrm{H}\alpha)-\log(\eta_{\mathrm{H}\alpha})\label{eq:defpower}\end{equation}

The parameter $\epsilon_{\mathrm{H}\alpha}$ is the exponent of the
power law. The simple calibration described by Eq.~\ref{eq:defstandard}
corresponds to the $\epsilon_{\mathrm{H}\alpha}=1$ special case.
We show in the following sections that this more general approach
is not mandatory for the cases with a dust attenuation estimated from
the Balmer decrement ($\epsilon_{\mathrm{H}\alpha}$ is very close
to $1$), but that it is useful in other cases.

We note that the correlation between dust attenuation and metallicity
\citep{Cortese:2006ApJ...637..242C,Mouhcine:2005MNRAS.362.1143M}
also has an effect on the slope. However the goal of our work in not
to study in details such phenomenon.

In the following sections, the efficiency factors are expressed in
erg s$^{-1}$ (M$_{\odot}$ yr$^{-1}$)$^{-1}$ units and the exponents
of the power laws are unit-less. All fits are least square fits with
errors in $x$ and $y$. The errors on the fitted parameters are given
by the rms of 40 bootstrap estimates. The term {}``dispersion''
relates to the rms of the residuals around the fitted calibrations,
the term {}``shift'' to the mean of these residuals. The given residual
slopes are unit-less.

\subsection{Dust estimated from the Balmer decrement\label{sub:dustbalmer}}

We now present possible calibrations of the SFR based on data corrected
with dust estimated from the Balmer decrement (see Eq.~\ref{eq:tauV}).
We remind the reader that the results presented in this subsection
are valid only if we correct the observed flux for dust attenuation
using Eqs.~\ref{eq:intrflux} and~\ref{eq:tauV}, i.e. making the
wrong assumption that the intrinsic Balmer ratio is constant over
all the range of stellar masses.

\subsubsection{Improved standard calibrations\label{sub:improved}}

\begin{figure}
\begin{centering}
\includegraphics[width=0.49\columnwidth]{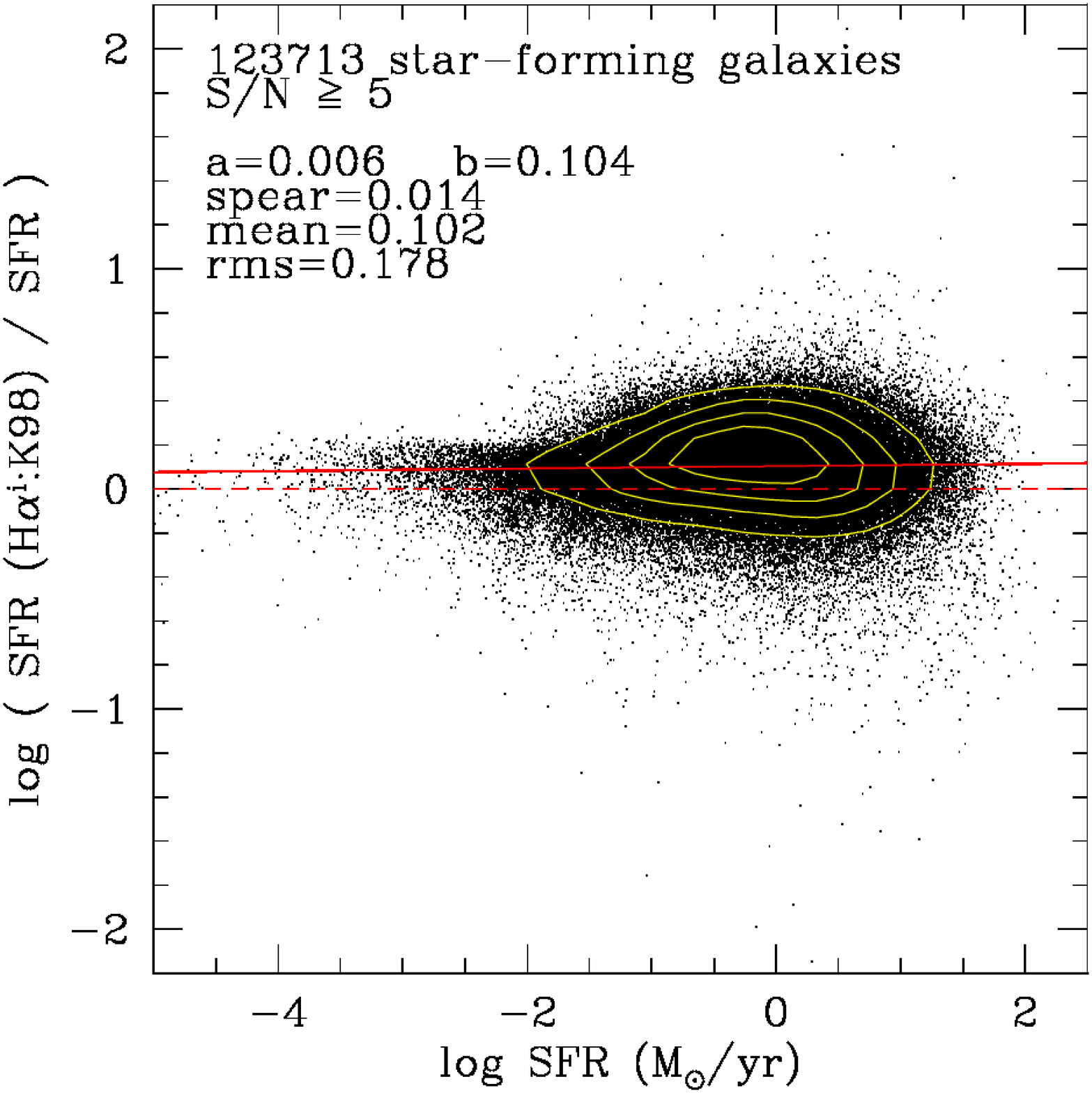}
\includegraphics[width=0.49\columnwidth]{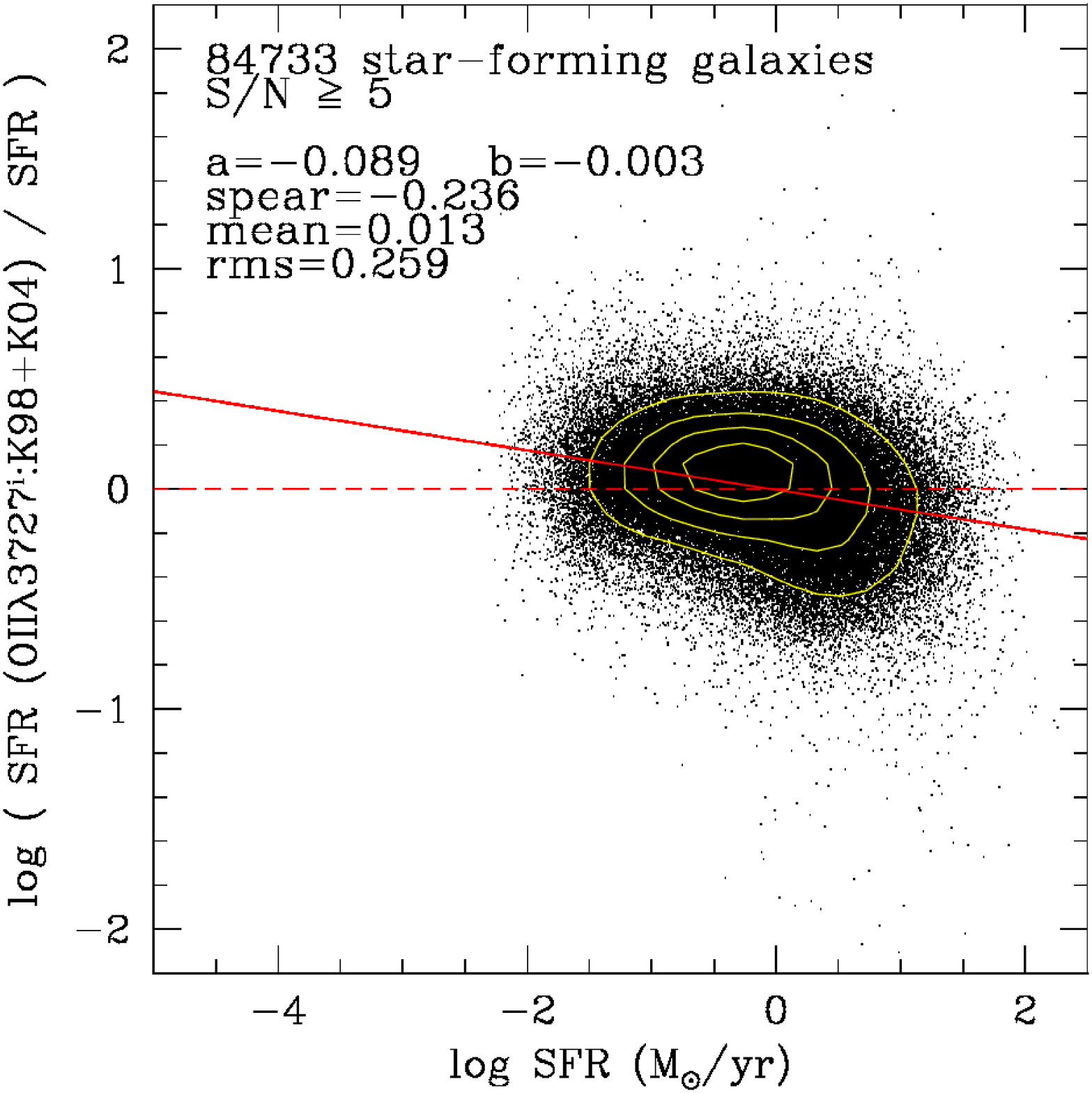}
\includegraphics[width=0.49\columnwidth]{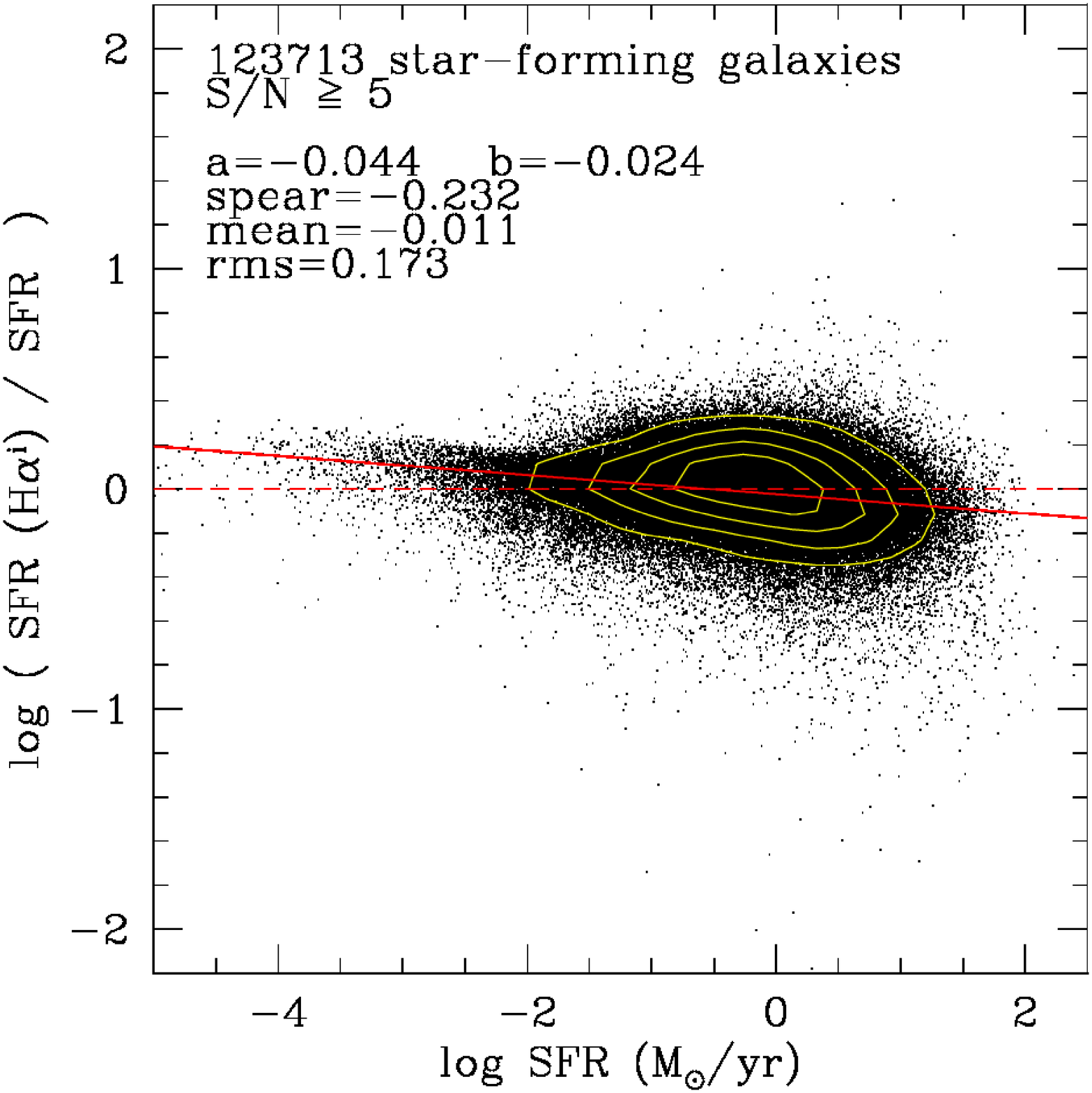}
\includegraphics[width=0.49\columnwidth]{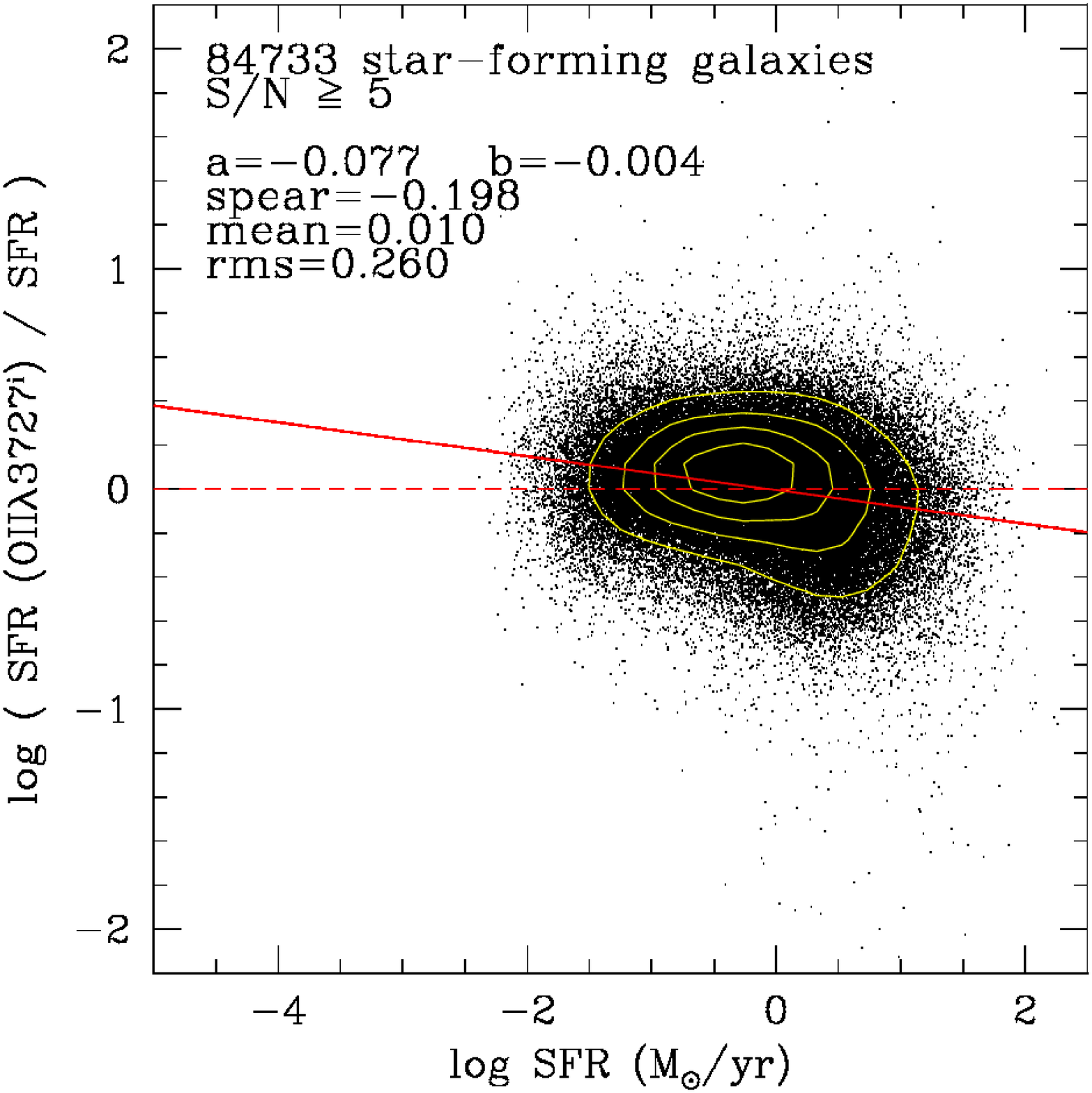}
\par\end{centering}

\caption{These four plots show the residuals of the SFR recovered from an emission-line
calibration as compared to the reference CL01 SFR, for star-forming
galaxies in the SDSS DR4 data. This comparison is shown as the logarithm
of the ratio between the derived and the reference SFR, as a function
of the logarithm of the reference SFR (in M$_{\odot}$/yr). The number
of objects used in the comparison, the shift, the dispersion, and
the Spearman rank correlation coefficient are given in the plot. The
dashed horizontal line is the $y=0$ curve, the solid line is the
$y=a\cdot x+b$ curve where $a$ and $b$ are the parameters of the
linear regression given in the plot. Isodensity contours are overplotted
in white. The studied calibrations are: top-left: \citet{Kennicutt:1998ARA&A..36..189K}
H$\alpha^{\mathrm{i}}$; top-right: \citet{Kennicutt:1998ARA&A..36..189K}
\citep[corrected by][]{Kewley:2004AJ....127.2002K} {[}O\noun{ii}]$^{\mathrm{i}}$;
bottom-left: our improved H$\alpha^{\mathrm{i}}$ calibration; bottom-right:
our improved {[}O\noun{ii}]$^{\mathrm{i}}$ calibration.}

\label{fig:compstandard}
\end{figure}

\begin{figure*}
\begin{centering}
\includegraphics[width=0.4\linewidth]{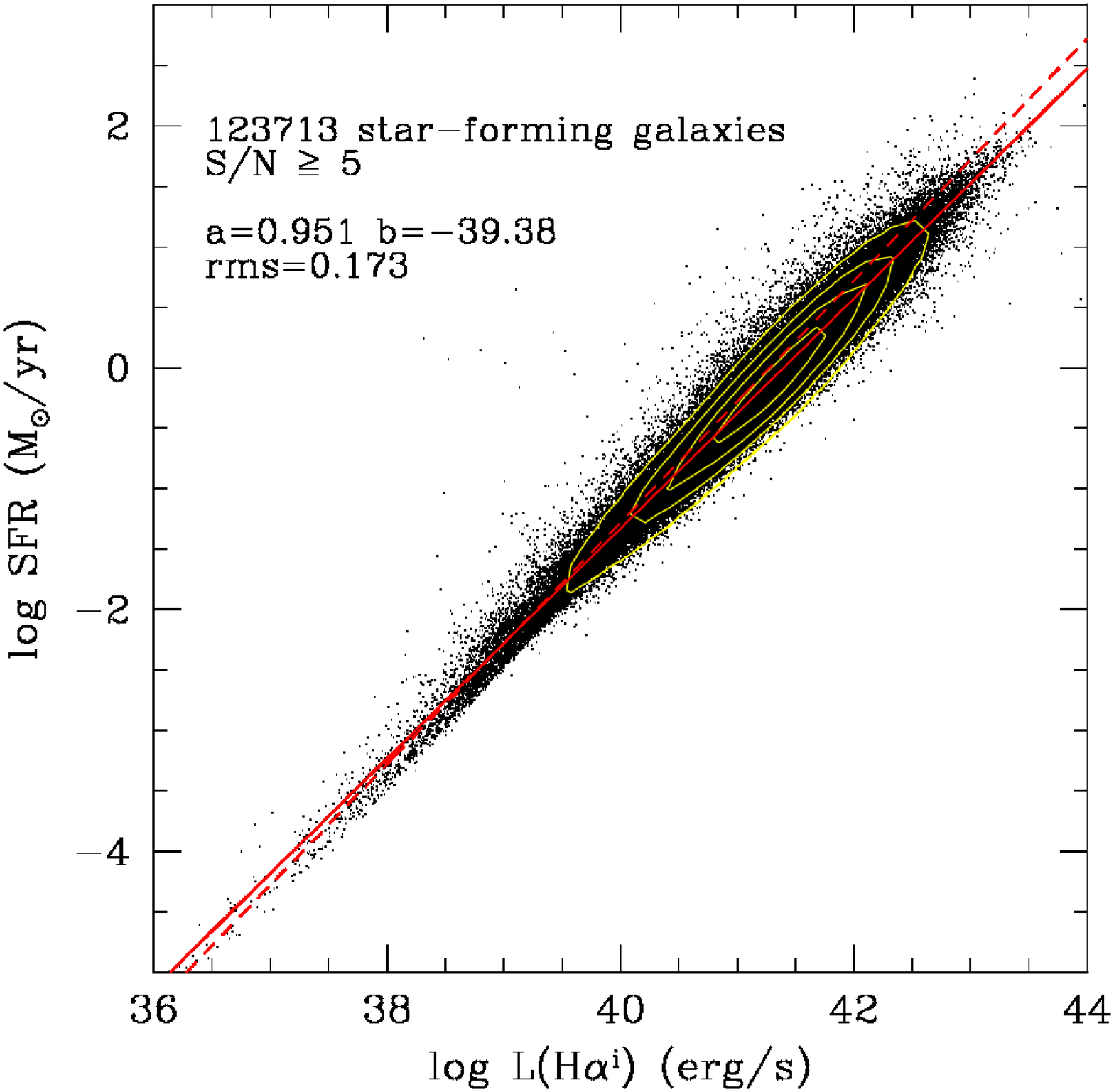}
\includegraphics[width=0.4\linewidth]{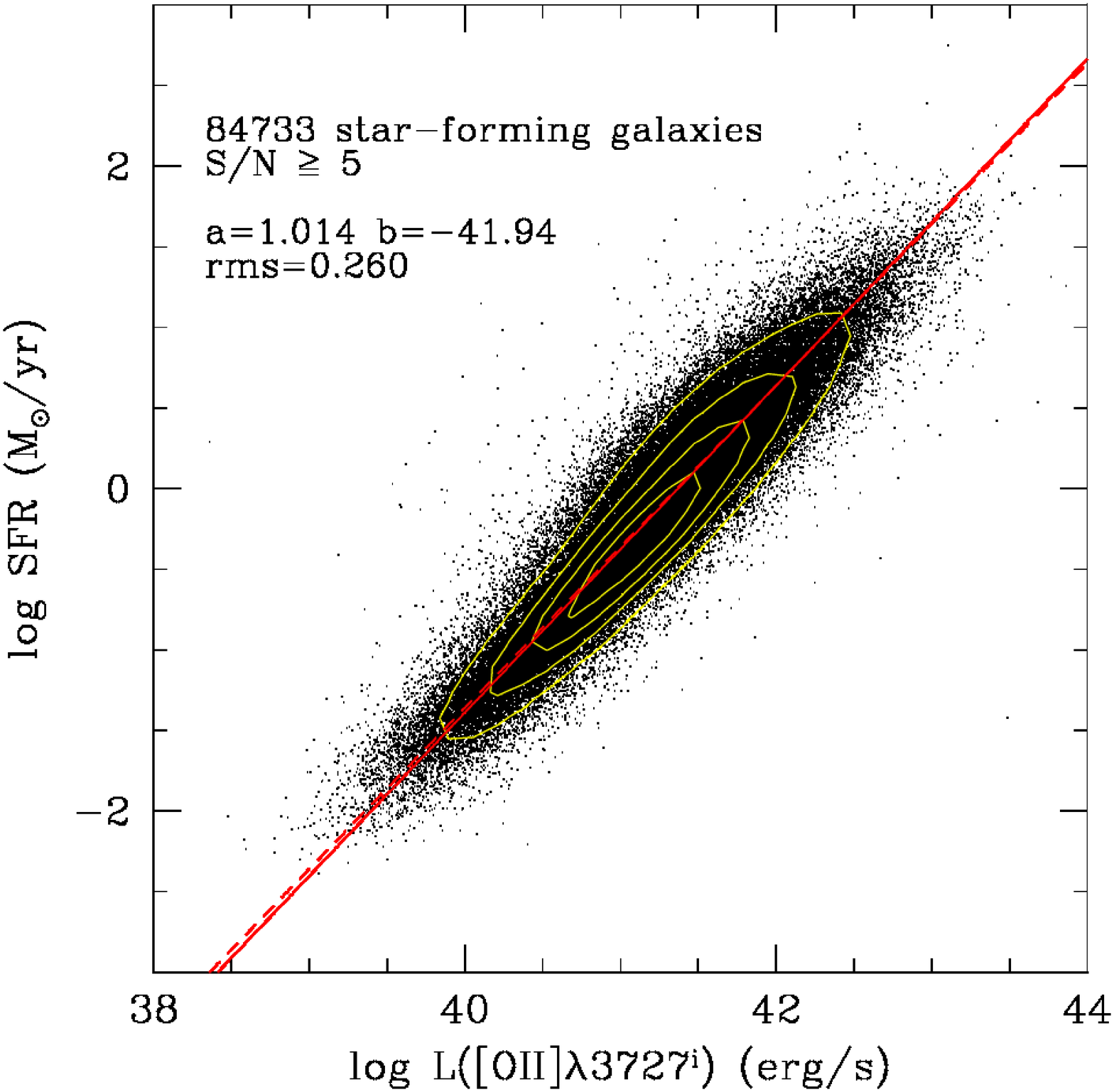}
\par\end{centering}

\caption{Relation between the SFR (logarithm of M$_{\odot}$/yr) and the H$\alpha^{\mathrm{i}}$
(left) or {[}O\noun{ii}]$^{\mathrm{i}}$ (right) emission lines luminosities
(logarithm of erg/s) for star-forming galaxies in the SDSS DR4 data.
The solid line is a least-square fit to the data (errors on $x$ and
$y$). The dashed line is the \citet{Kennicutt:1998ARA&A..36..189K}
calibration corrected to our assumed IMF, and corrected for dust by
\citet{Kewley:2004AJ....127.2002K} on the right panel. Isodensity
contours are overplotted in white.}

\label{fig:newstandard}
\end{figure*}

\citet{Kennicutt:1998ARA&A..36..189K} has developed two calibrations
of the SFR versus H$\alpha^{\mathrm{i}}$ or the {[}O\noun{ii}]$^{\mathrm{i}}$
emission lines which are now widely used in the literature. We take
advantage of the SDSS DR4 data to test these two calibrations as compared
to the high quality CL01 estimate of the SFR. We also derive two new
improved calibrations based on Eq.~\ref{eq:defpower} rather than
Eq.~\ref{eq:defstandard}.

Fig.~\ref{fig:compstandard} (top-left) shows the comparison between
the CL01 SFR and the one recovered with the standard \citet{Kennicutt:1998ARA&A..36..189K}
H$\alpha^{\mathrm{i}}$ calibration, which uses $\log(\eta_{\mathrm{H}\alpha}^{\mathrm{i}})=41.28$
if we scale their results to our adopted IMF. We find a good agreement:
the dispersion is $0.17$ dex, which is very similar to the minimum
uncertainty associated to the CL01 SFR estimates (i.e. $0.16$ dex).
Thus, \emph{we confirm that the H$\alpha$ emission line luminosity
alone is sufficient to recover almost perfectly the SFR}.

However, we see that the reduced efficiency factor and the dust attenuation
overestimate do not exactly cancel each other on the massive end of
the sample, which induces a mean of the residuals of $+0.1$ dex.
One might decide to subtract $0.1$ dex to the \citet{Kennicutt:1998ARA&A..36..189K}
calibration in order to find a null mean of the residuals, or we can
try to derive an improved calibration.

Fig.~\ref{fig:newstandard} (left) shows our improved calibration
for the correlation between the H$\alpha^{\mathrm{i}}$ emission-line
luminosity and the SFR in SDSS DR4 data. We find the following best
fit values:

\begin{equation}
\left\{ \begin{array}{ccl}
\log\eta_{\mathrm{H}\alpha}^{\mathrm{i}} & = & 39.38\pm0.03\\
\epsilon_{\mathrm{H}\alpha}^{\mathrm{i}} & = & 0.951\pm0.001\end{array}\right.\label{eq:ha}\end{equation}

One can notice that the formal errors of these two parameters are
very low thanks to the large number of objects, which makes this calibration
quite reliable. The uncertainty on the exponent of the power law especially
tells us that this parameter is very well constrained, while not equal
to $1$ as compared to all previous studies. As shown in Fig.~\ref{fig:compstandard}
(bottom-left), the mean on the residuals is now almost null ($0.01$
dex) with the improved calibration. A small residual slope ($-0.04$)
is present as the improved calibration is now less accurate for very
low SFR values. This residual slope is not significant given the low
Spearman rank correlation coefficient: $-0.23$.

\medskip{}

We now consider the case where the H$\alpha$ line is not observed,
or not measurable, and that we decide to use the {[}O\noun{ii}]$\lambda$3727
emission line instead. In such case, the dust attenuation might be
derived from another Balmer ratio not involving the H$\alpha$ line
(e.g. H$\gamma$/H$\beta$, H$\delta$/H$\gamma$, ...)

As already stated in the literature, the relation between the SFR
and the {[}O\noun{ii}] line has to be considered with caution because
of the stronger dependence of this line on metallicity (compared to
H$\alpha$), as well as a stronger sensibility to dust attenuation.
However, as shown in Fig.~\ref{fig:compstandard} (top-right), we
find a good agreement with the standard \citet{Kennicutt:1998ARA&A..36..189K}
{[}O\noun{ii}]$^{\mathrm{i}}$ calibration, which uses $\log(\eta_{[\mathrm{O}\mathsc{ii}]}^{\mathrm{i}})=41.36$
if we correct their results for dust attenuation \citep{Kewley:2004AJ....127.2002K}
and scale them to our adopted IMF.

It seems that the effect of metallicity on the {[}O\noun{ii}] line
does not produce any systematic shift and only leads to a higher dispersion:
$0.25$ dex. To be more precise, the three effects of the metallicity
on the relative strength of the {[}O\noun{ii}] line, on the efficiency
factor, and on the estimate of the dust attenuation exactly compensate
in order to produce a exponent of the power law very close to unity. 

Fig.~\ref{fig:newstandard} (right) shows our improved calibration
in the correlation between the {[}O\noun{ii}]$^{\mathrm{i}}$ emission-line
luminosity and the SFR in SDSS DR4 data. We find the following best-fit
values:

\begin{equation}
\left\{ \begin{array}{ccl}
\log\eta_{[\mathrm{O}\mathsc{ii}]}^{\mathrm{i}} & = & 41.94\pm0.7\\
\epsilon_{[\mathrm{O}\mathsc{ii}]}^{\mathrm{i}} & = & 1.014\pm0.02\end{array}\right.\label{eq:o2}\end{equation}

As expected, and as shown in Fig.~\ref{fig:compstandard} (bottom-right),
this calibration does not produce any significant improvement compared
to the \citet{Kennicutt:1998ARA&A..36..189K} \citep[corrected by][]{Kewley:2004AJ....127.2002K}
calibration. In both cases, the observed residual slope of $-0.09$
is not significant (Spearman rank correlation coefficient of $-0.24$).

\subsubsection{New calibrations}

\begin{figure}
\begin{centering}
\includegraphics[width=0.8\columnwidth]{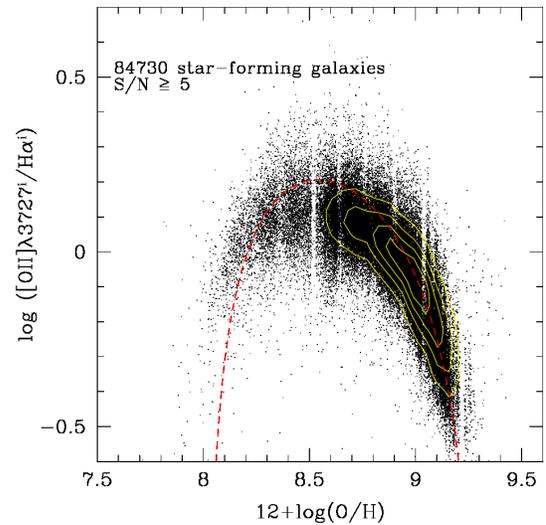}
\par\end{centering}

\caption{Relation between the {[}O\noun{ii}]$\lambda$3727$^{\mathrm{i}}$/H$\alpha^{\mathrm{i}}$
emission line ratio and gas-phase oxygen abundance (estimated with
the CL01 method) for the star-forming galaxies in the SDSS DR4 catalog.
The dashed curve is a semi-empirical estimation of this relation \citep{Kewley:2004AJ....127.2002K}.}

\label{fig:o2haoh}
\end{figure}

\begin{figure}
\begin{centering}
\includegraphics[width=0.49\columnwidth]{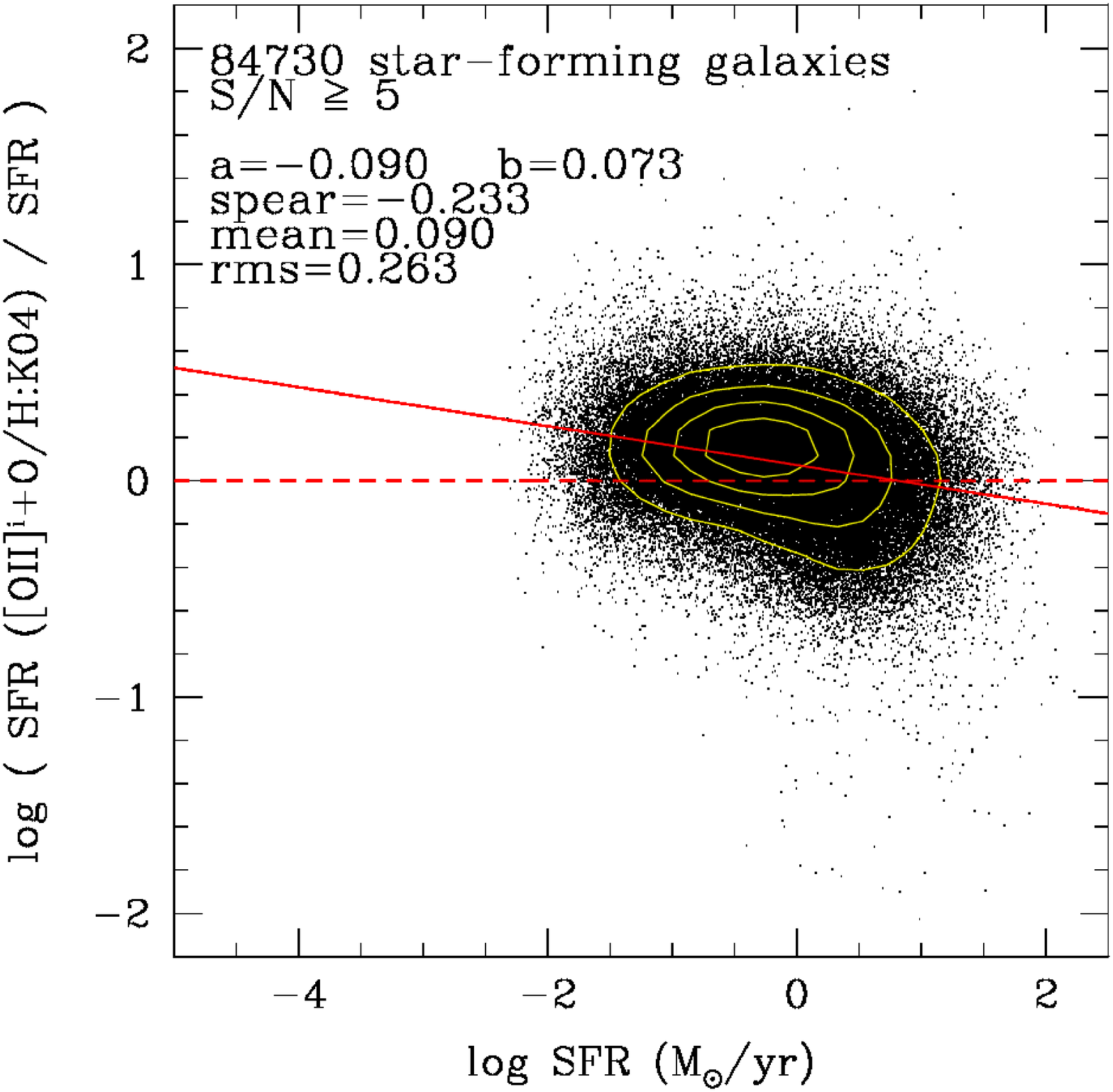}
\includegraphics[width=0.49\columnwidth]{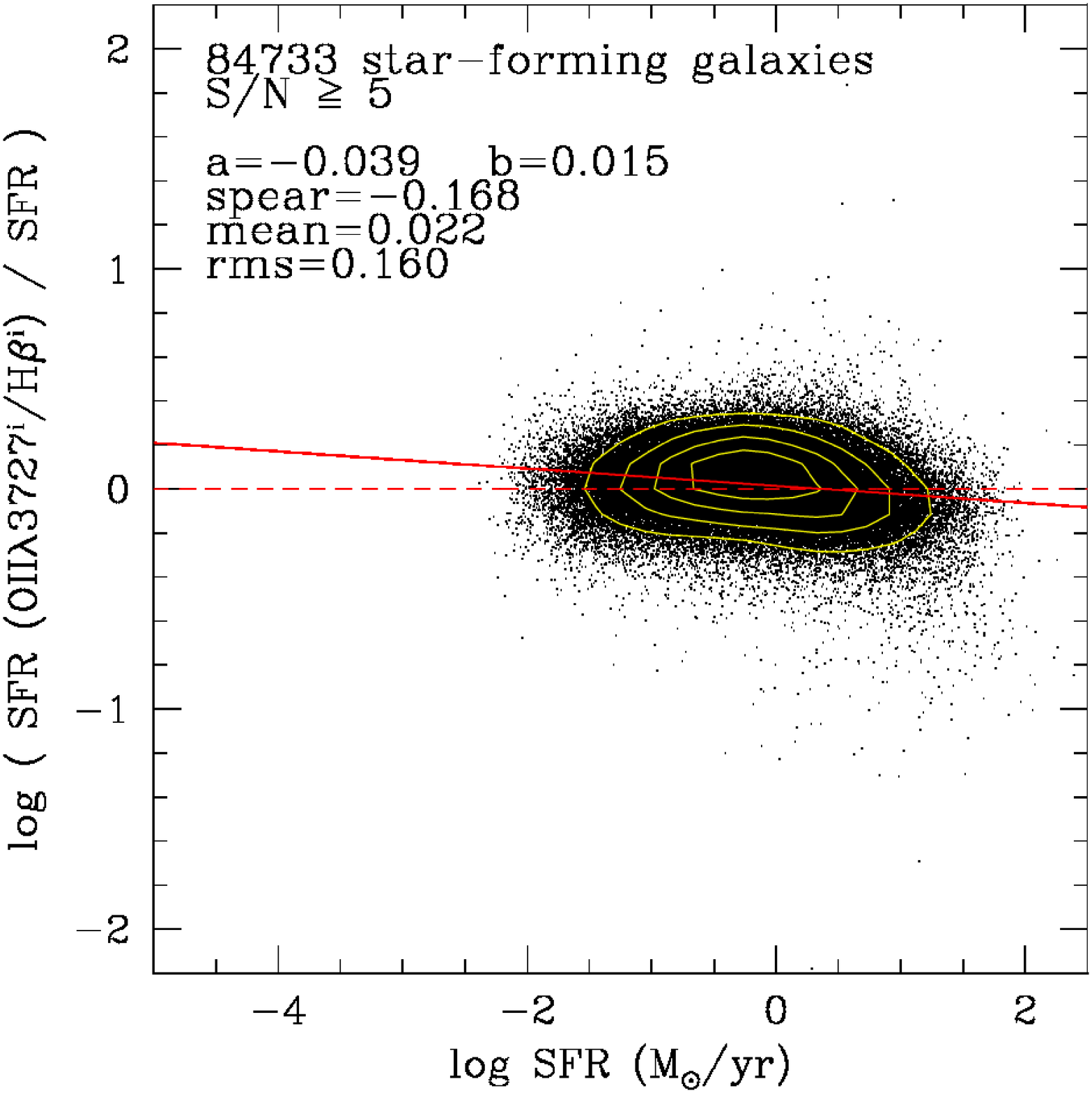}
\par\end{centering}

\caption{Same legend as in Fig.~\ref{fig:compstandard}. The studied calibrations
are: left: \citet{Kewley:2004AJ....127.2002K} {[}O\noun{ii}]$^{\mathrm{i}}$
with a correction for metallicity; right: our new {[}O\noun{ii}]$^{\mathrm{i}}$+H$\beta^{\mathrm{i}}$
calibration.}

\label{fig:compo2hbi}
\end{figure}

As already shown in previous works \citep{Kewley:2004AJ....127.2002K,Moustakas:2006ApJ...642..775M,Weiner:2006astro.ph.10842W},
the best way to provide a reliable SFR vs. {[}O\noun{ii}] calibration
is to actually use the SFR vs. H$\alpha^{\mathrm{i}}$ calibration,
and to calibrate either the {[}O\noun{ii}]/H$\alpha^{\mathrm{i}}$
(observed-to-intrinsic) or the {[}O\noun{ii}]$^{\mathrm{i}}$/H$\alpha^{\mathrm{i}}$
(intrinsic-to-intrinsic) line ratio against a well chosen parameter.
Then the star formation rate is given by the following formula:\begin{equation}
\log\mathrm{SFR}=\epsilon_{\mathrm{H}\alpha}^{\mathrm{i}}\left(\log L(\mathrm{[O}\mathsc{ii}\mathrm{]}^{\mathrm{i?}})-\log(\mathrm{[O}\mathsc{ii}]^{\mathrm{i?}}/\mathrm{H}\alpha^{\mathrm{i}})\right)-\log\eta_{\mathrm{H}\alpha}^{\mathrm{i}}\label{eq:defo2ha}\end{equation}

\citet{Kewley:2004AJ....127.2002K} has calibrated the {[}O\noun{ii}]$^{\mathrm{i}}$/H$\alpha^{\mathrm{i}}$
line ratio against the gas-phase oxygen abundance ($12+\log(\mathrm{O}/\mathrm{H})$),
ending up in an appreciable improvement of the calibration. As also
shown by \citet{Mouhcine:2005MNRAS.362.1143M} with 2dFGRS data and
in Fig.~\ref{fig:o2haoh}, there is indeed a robust correlation between
these two parameters. The dashed line in Fig.~\ref{fig:o2haoh} shows
the relation derived semi-empirically by \citet{Kewley:2004AJ....127.2002K}.
The residuals around this relation are characterized by a mean of
$-0.04$ dex and a rms of $0.1$ dex in SDSS DR4 data.

However, we see three drawbacks against the use of this relation:

\begin{itemize}
\item First the gas-phase oxygen abundance is not an easy parameter to derive.
There are many different calibrations which can be different from
each other by up to $0.5$ dex \citep[see][for a detailed discussion]{Kewley:2008arXiv0801.1849K}.
Moreover, most of these calibrations need many emission lines to be
measured and we emphasize that, in such cases, it might be easier
to use directly the SFR estimated from multiple-lines fitting methods
like CL01.
\item Second, we already stated before that correcting the {[}O\noun{ii}]
emission line for dust attenuation requires in most common cases that
the H$\alpha$ emission line is being observed: \emph{in this case
one would estimate the SFR directly from the H$\alpha$} \emph{line},
rather from the {[}O\noun{ii}] line. Thus, trying to improve the {[}O\noun{ii}]
calibration with data corrected for dust attenuation is actually not
applicable in most common cases.
\item For the other cases (e.g. dust attenuation estimated from another
Balmer ratio), we also remark that using the metallicity estimate
(or any other derived parameter) to correct the calibration is \emph{not
necessary}. Indeed a very simple relation, which do not need a metallicity
estimate, already exists between {[}O\noun{ii}]$^{\mathrm{i}}$, H$\alpha^{\mathrm{i}}$
and H$\beta^{\mathrm{i}}$ emission lines. This simple relation is
expressed by the following equation assuming the standard intrinsic
Balmer ratio:\begin{equation}
\frac{[\mathrm{O}\mathsc{ii}]^{\mathrm{i}}}{\mathrm{H}\alpha^{\mathrm{i}}}=\frac{[\mathrm{O}\mathsc{ii}]^{\mathrm{i}}}{\mathrm{H}\beta^{\mathrm{i}}}/\frac{\mathrm{H}\alpha^{\mathrm{i}}}{\mathrm{H}\beta^{\mathrm{i}}}=\frac{[\mathrm{O}\mathsc{ii}]^{\mathrm{i}}}{\mathrm{H}\beta^{\mathrm{i}}}/2.85\label{eq:o2hadust}\end{equation}

\end{itemize}
We applied Eq.~15 of \citet{Kewley:2004AJ....127.2002K} in order
to compare their metallicity- and dust-corrected calibration to the
reference CL01 SFR. We have done that only for the $84\,730$ galaxies
for which an estimate of the gas-phase oxygen abundance was available.
Before doing that, we have corrected our available CL01 metallicities
to \citet{Kewley:2002ApJS..142...35K} metallicities, using the formula
in Table~B3 of \citet{Kewley:2008arXiv0801.1849K}.

As shown in Fig.~\ref{fig:compo2hbi} (left), the \citet{Kewley:2004AJ....127.2002K}
calibration shows a non-negligible $0.1$ dex shift. Moreover, the
dispersion of the relation is $0.26$ dex, which is \emph{not better}
than the standard {[}O\noun{ii}]$^{\mathrm{i}}$ calibration without
a correction for metallicity. Their is also a small, not significant,
residual slope of $-0.09$ (Spearman rank correlation coefficient
of $-0.23$) .

\medskip{}

Eqs.~\ref{eq:defo2ha} and~\ref{eq:o2hadust} are now combined to
provide an estimate of the SFR, which is compared to the CL01 SFR
on Fig.~\ref{fig:compo2hbi}\textbf{ }(right). The relation is, as
expected, very good as no dispersion is added by Eq.~\ref{eq:o2hadust}.
The small dispersion of the relation ($0.16$ dex) comes only from
the uncertainty of the H$\alpha^{\mathrm{i}}$ calibration that we
derived previously. This calibration do not show a significant residual
slope (Spearman rank correlation coefficient: $-0.17$). In fact this
new {[}O\noun{ii}]$^{\mathrm{i}}$+H$\beta^{\mathrm{i}}$ calibration
is expected to give exactly the same results as the underlying H$\alpha^{\mathrm{i}}$
calibration since the relation between them is the constant $2.85$
factor. Fig.~\ref{fig:compo2hbi} (right) is nothing else than Fig.~\ref{fig:compstandard}
(bottom-left) without some missing points because of the additional
selection on the {[}O\noun{ii}] line. These missing points explain
the smaller scatter.

Following this conclusion, we emphasize that a simple H$\beta^{\mathrm{i}}$
calibration is even easier to derive and to apply to the observations
using this relation:\begin{equation}
\mathrm{H}\alpha^{\mathrm{i}}=\mathrm{H}\beta^{\mathrm{i}}\times2.85\label{eq:hahbdust}\end{equation}

The H$\beta^{\mathrm{i}}$ is expected to give exactly the same results
as the H$\alpha^{\mathrm{i}}$ calibration, provided that a correction
for dust attenuation has been estimated from the assumption of a constant
intrinsic Balmer ratio.

\subsection{Dust estimated from another method\label{sub:otherdust}}

The dust attenuation might not be estimated with the assumption of
a constant intrinsic Balmer decrement. Other methods such as CL01
can provide a better estimate of the dust attenuation, not biased
towards the metallicity dependence of the intrinsic Balmer ratio (thanks
to the use of all available emission-line measurements). The dust
attenuation estimated from SED fitting methods does not either make
the assumption of a constant intrinsic Balmer ratio.

As stated before, the calibrations derived in previous subsection
should show an exponent of the power law greater than $1.0$, because
of the metallicity dependence of the relation between the SFR and
the emission-line luminosities (assuming that the higher is the SFR
the higher are the stellar mass and the metallicity). However, this
effect is not seen since it is almost exactly compensated by the overestimate
of the dust attenuation at high metallicities. 

We thus study in this subsection the results obtained with dust estimated
from other methods.

\subsubsection{Metallicity-unbiased calibrations\label{sub:metalunbiased}}

\begin{figure*}
\begin{centering}
\includegraphics[width=0.27\linewidth]{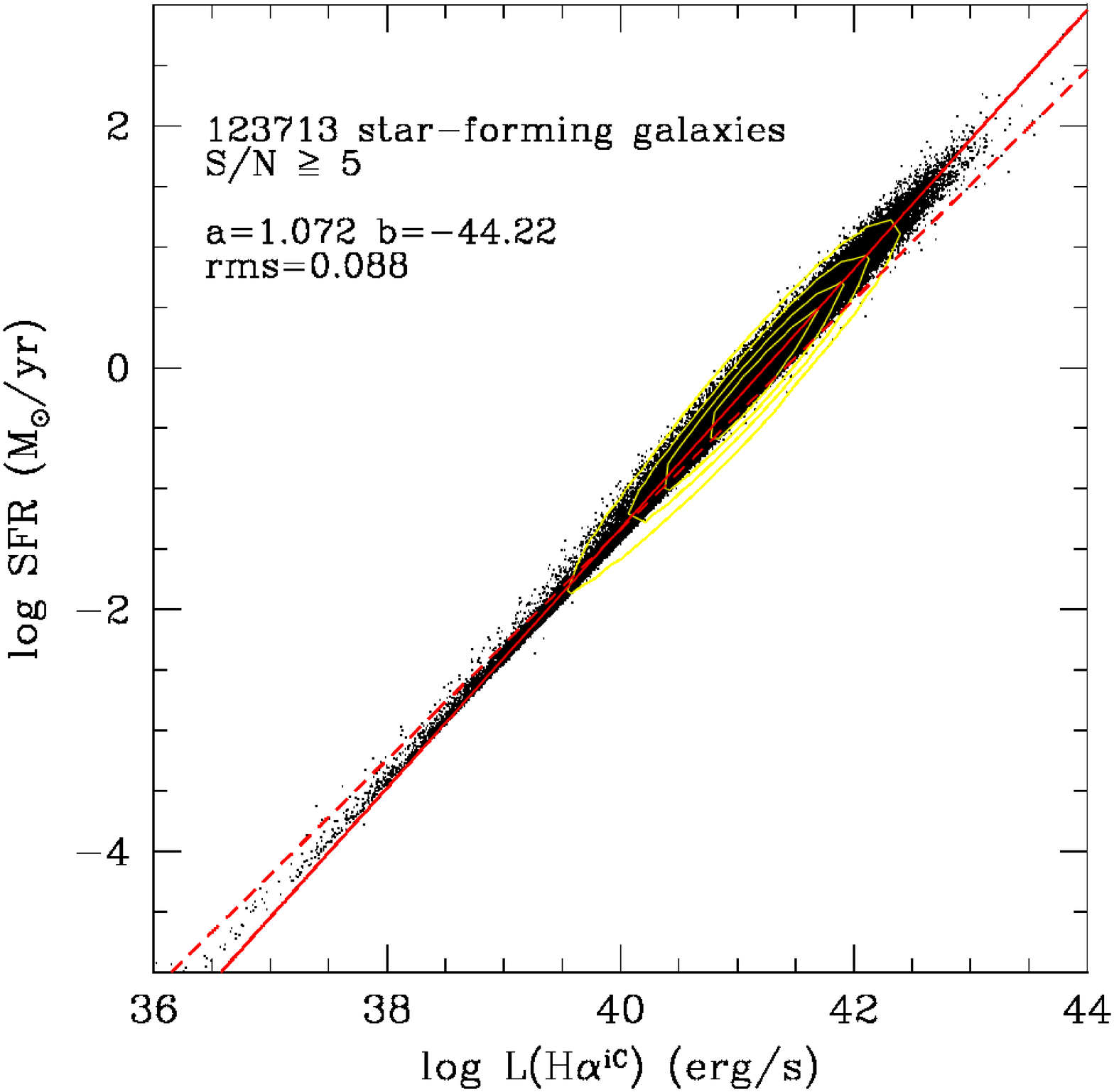}
\includegraphics[width=0.27\linewidth]{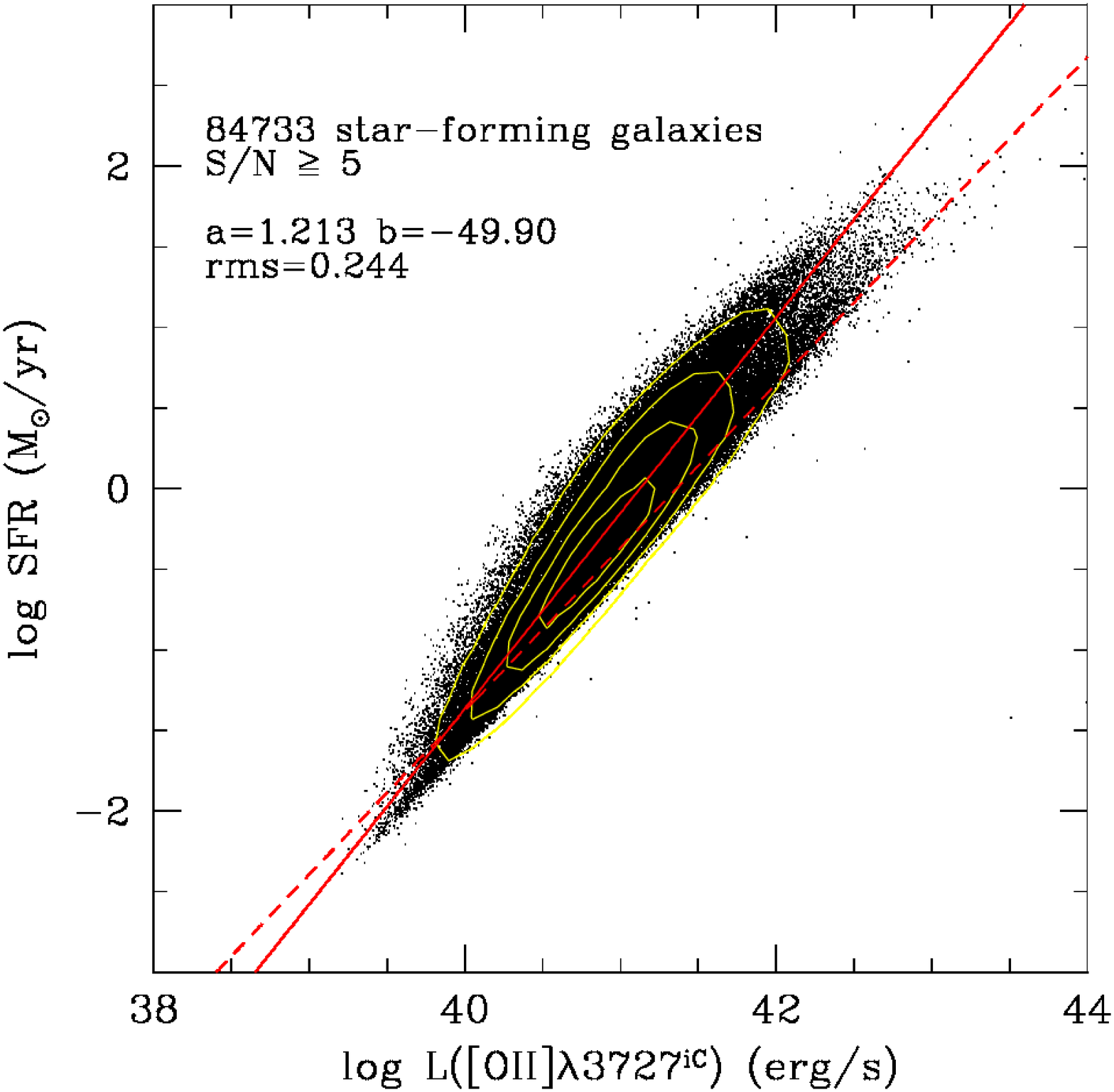}
\includegraphics[width=0.27\linewidth]{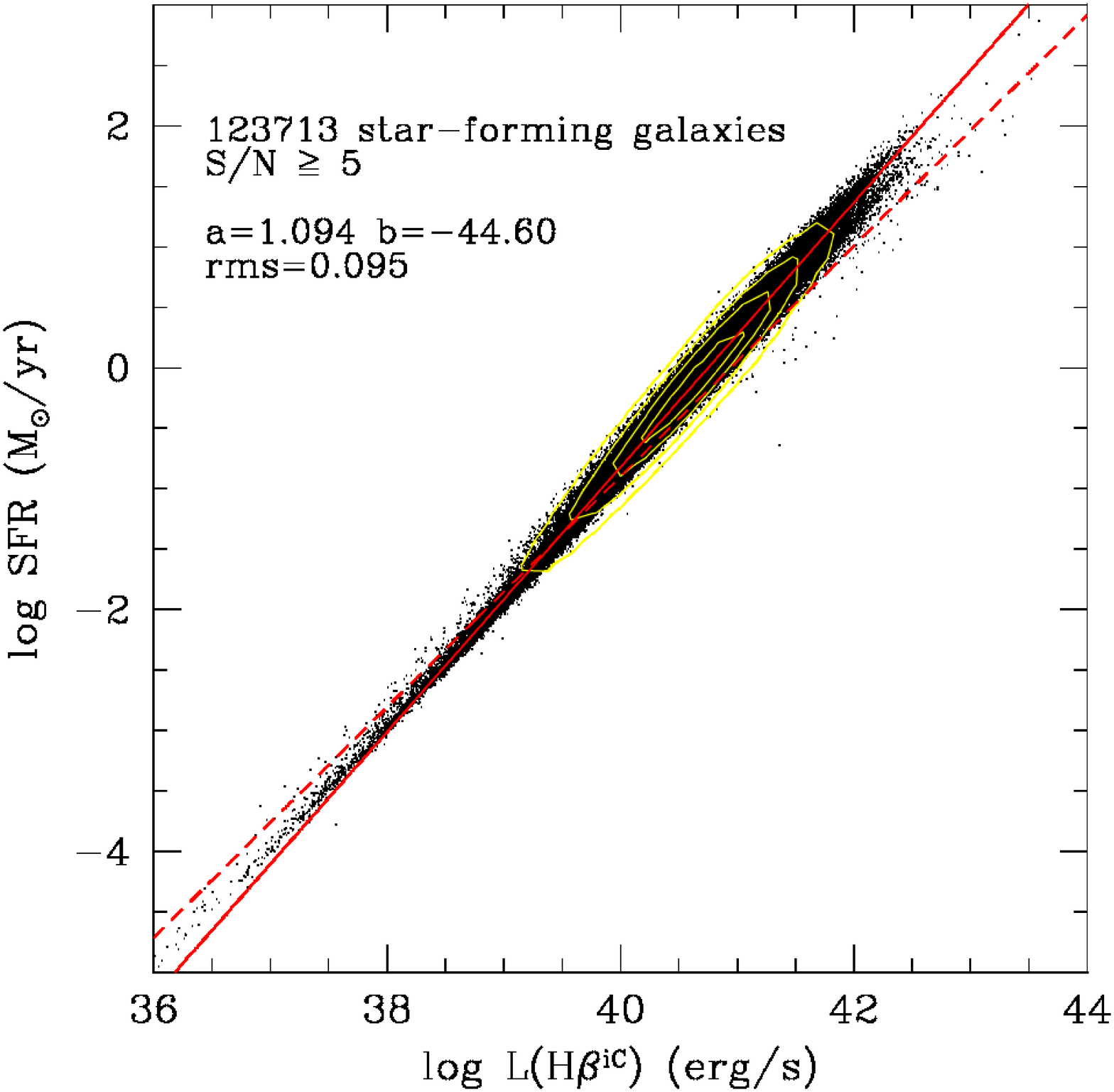}
\par\end{centering}

\caption{Relation between the SFR (logarithm of M$_{\odot}$/yr) and the H$\alpha^{\mathrm{iC}}$
(left), {[}O\noun{ii}]$^{\mathrm{iC}}$ (center) or H$\beta^{\mathrm{iC}}$
emission lines luminosities (logarithm of erg/s) for star-forming
galaxies in the SDSS DR4 data. The solid line is a least-square fit
to the data (errors on $x$ and $y$). These data have been corrected
using the unbiased CL01 estimate of dust attenuation. The dashed lines
show the previous calibrations obtained with dust estimated from the
assumption of a constant intrinsic Balmer ratio (see Fig.~\ref{fig:newstandard}).
Isodensity contours are overplotted in white.}

\label{fig:newCL01}
\end{figure*}

We provide new calibrations to be used on data corrected with a good
estimate of the dust attenuation, i.e. not biased towards metallicity.
To this end, we corrected the emission-line luminosities using the
dust attenuation estimated with the CL01 method, instead of the one
estimated from the Balmer decrement. We emphasize that this work is
mainly done for comparison purposes: using the CL01 method to estimate
the dust attenuation makes our SFR calibration useless since the CL01
SFR is already better constrained.

Fig.~\ref{fig:newCL01} shows the relations between the SFR and H$\alpha^{\mathrm{iC}}$,
{[}O\noun{ii}]$^{\mathrm{iC}}$ or H$\beta^{\mathrm{iC}}$ emission
lines. In the three cases, we find as expected an exponent of the
power law greater than $1$. However we see that the metallicity dependence
of the efficiency factor has not a strong effect on the slope for
the H$\alpha^{\mathrm{iC}}$ and H$\beta^{\mathrm{iC}}$ calibrations.
We remind the reader that, in contrary to what we have said in Sect.~\ref{sub:dustbalmer},
the calibrations based on H$\alpha^{\mathrm{iC}}$ or H$\beta^{\mathrm{iC}}$
lines are now slightly different since the ratio between these two
lines is not any more assumed constant. We obtain the following results:

\begin{equation}
\left\{ \begin{array}{ccl}
\log\eta_{\mathrm{H}\alpha}^{\mathrm{iC}} & = & 44.22\pm0.03\\
\epsilon_{\mathrm{H}\alpha}^{\mathrm{iC}} & = & 1.072\pm0.001\end{array}\right.\label{eq:haiC}\end{equation}
\begin{equation}
\left\{ \begin{array}{ccl}
\log\eta_{[\mathrm{O}\mathsc{ii}]}^{\mathrm{iC}} & = & 49.90\pm0.9\\
\epsilon_{[\mathrm{O}\mathsc{ii}]}^{\mathrm{iC}} & = & 1.213\pm0.03\end{array}\right.\label{eq:o2iC}\end{equation}
\begin{equation}
\left\{ \begin{array}{ccl}
\log\eta_{\mathrm{H}\beta}^{\mathrm{iC}} & = & 44.60\pm0.03\\
\epsilon_{\mathrm{H}\beta}^{\mathrm{iC}} & = & 1.094\pm0.001\end{array}\right.\label{eq:hbiC}\end{equation}

In the three cases, the dispersion is reduced thanks to the better
estimate of the dust attenuation provided by the CL01 method. Is it
now $<0.1$ dex in the H$\alpha^{\mathrm{iC}}$ and H$\beta^{\mathrm{iC}}$
calibrations and $0.24$ dex in the {[}O\noun{ii}]$^{\mathrm{iC}}$
calibration.

\subsubsection{Dust estimated from SED fitting}

\begin{figure}
\begin{centering}
\includegraphics[width=0.45\columnwidth]{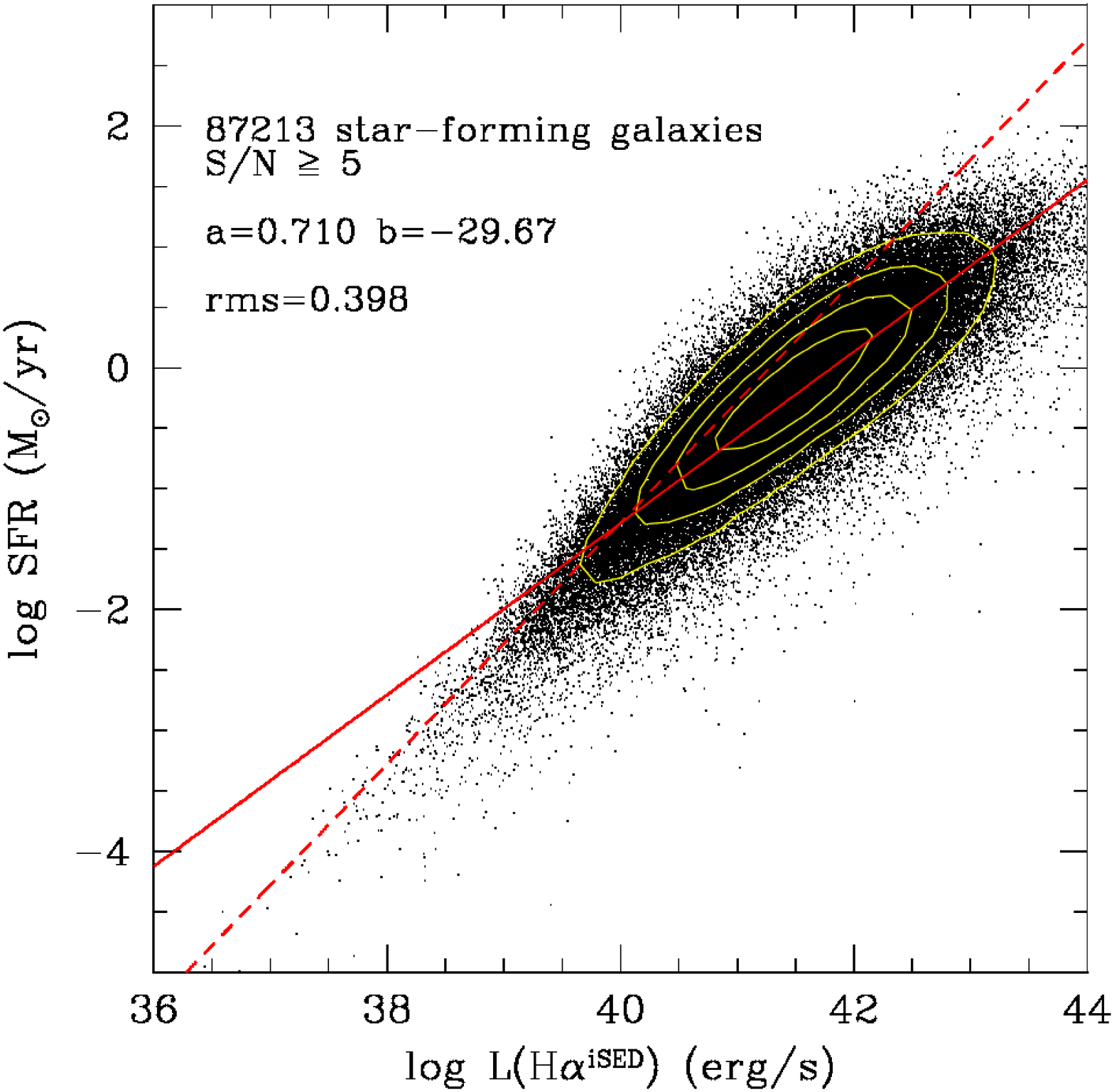}
\includegraphics[width=0.45\columnwidth]{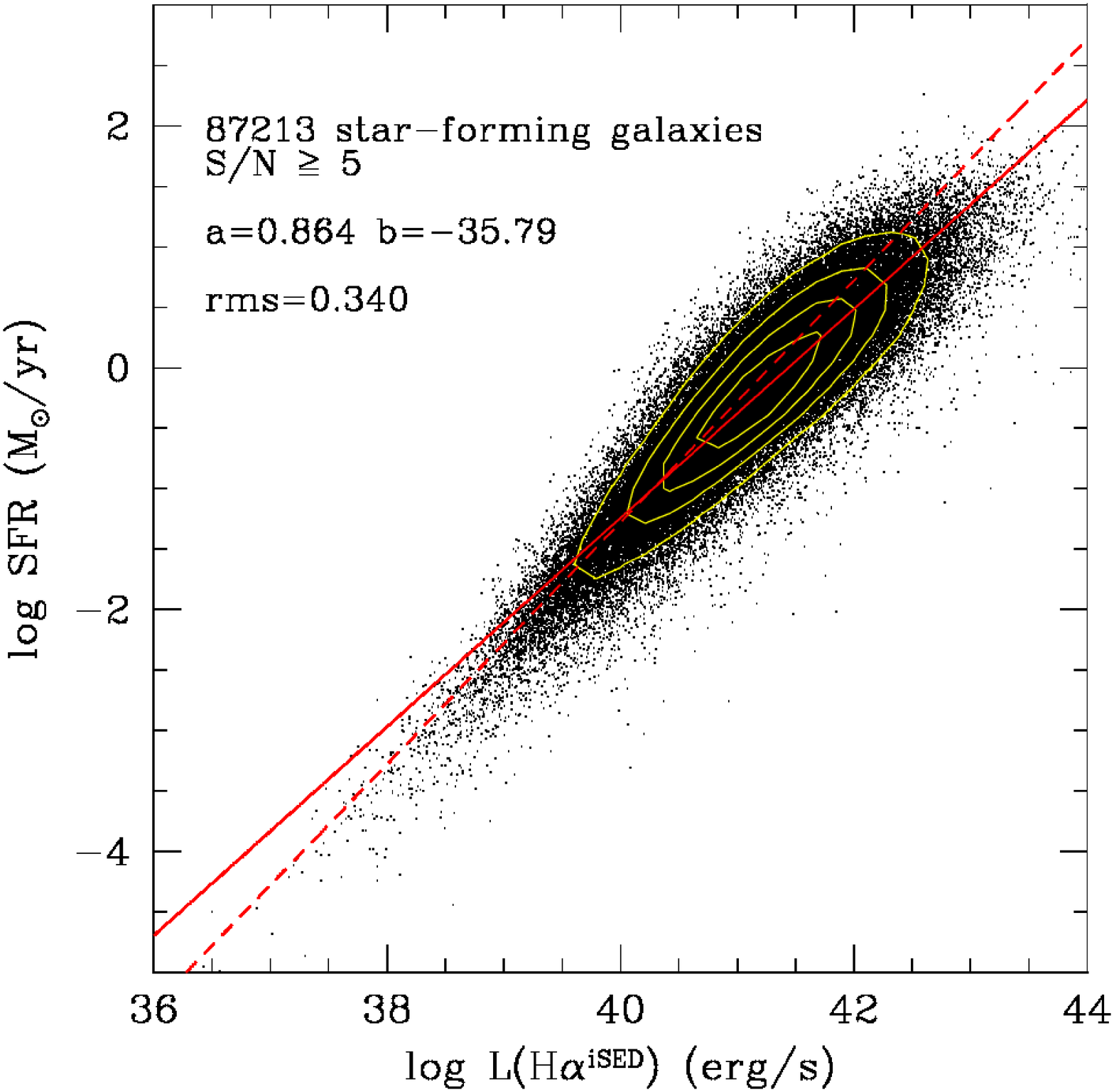}
\par\end{centering}

\caption{Relation between the SFR (logarithm of M$_{\odot}$/yr) and the H$\alpha^{\mathrm{iSED}}$
emission lines luminosity (logarithm of erg/s) for star-forming galaxies
in the SDSS DR4 data. The solid line is a least-square fit to the
data (errors on $x$ and $y$). The data has been corrected for dust
attenuation estimated with the SED fitting method (see Eq.~\ref{eq:tauVSED}).
The stellar-to-gas attenuation ratio has been assumed to $0.3$ (left:
SDSS mean value), or $0.44$ (right: \citealp{Calzetti:2001PASP..113.1449C}).
The dashed line is the \citet{Kennicutt:1998ARA&A..36..189K} calibration
corrected to our assumed IMF. Isodensity contours are overplotted
in white.}

\label{fig:seddust}
\end{figure}

One may ask if emission lines can also be used to derived SFR with
an estimation of dust attenuation coming from SED fitting. Fig.~\ref{fig:seddust}
(left) shows the correlation between the intrinsic H$\alpha$ emission-line
luminosity and the SFR in SDSS DR4 data, if we correct H$\alpha$
luminosities with the dust attenuation computed from SED fitting by
\citet{Kauffmann:2003MNRAS.341...33K} (see Eq.~\ref{eq:tauVSED}).
Note that only the $87\,213$ galaxies with an available estimate
of the dust attenuation are plotted.

We use a stellar-to-gas attenuation ratio of $0.3$ which corresponds
to the mean value observed in the SDSS data (Brinchmann, private communication).
We see that the exponent of the power law is quite smaller ($0.710\pm0.001$),
telling us that the SED fitting dust attenuation is even more overestimated
at higher masses than with the Balmer decrement method.

But no strong conclusion can be drawn: the dispersion that we find
($0.40$ dex) is indeed quite high and moreover, it does not take
into account the likely variations of the stellar-to-gas attenuation
ratio between the SDSS data and any other sample. Using instead a
stellar-to-gas attenuation ratio of $0.44$ \citep{Calzetti:2001PASP..113.1449C},
we found in Fig.~\ref{fig:seddust} (right) an exponent of the power
law of $0.864\pm0.001$. The derived efficiency factors with a mean
stellar-to-gas attenuation ratio of $0.3$ and $0.44$ are respectively
$29.67\pm0.02$ and $35.79\pm0.03$. We emphasize that a stellar-to-gas
attenuation ratio of $0.44$ is \emph{not} the right value representing
these data, even if it gives by chance a less dispersed calibration
($0.34$ dex).

We conclude that the most critical issue in using dust estimated from
SED fitting, in order to correct emission lines, comes from the uncertainty
in the stellar-to-gas attenuation ratio. This problem makes SFR derived
using this method almost completely random on a galaxy per galaxy
basis. We note that this method might be used in a statistical way,
but with great uncertainties coming both from the dispersion observed
in Fig.~\ref{fig:seddust}, and from the fact that one has to know
the right stellar-to-gas attenuation ratio to apply to his data.

\section{SFR calibration without a dust estimate\label{sec:SFR-calibration-without}}

We present in this section emission-line calibrations of the SFR based
on data \emph{not corrected} for dust attenuation. Because of the
small wavelength coverage of many spectroscopic surveys, it is common
that not all emission lines necessary to derive a correction for dust
attenuation, and to compute a SFR, are present in a given spectrum.

There are two ways to handle such data: \emph{(i)} use an assumed
mean extinction and apply a statistical correction for dust attenuation
(see Sect.~\ref{sub:assumedust}); or \emph{(ii)} use a SFR calibration
which provides, still in a statistical approach, a self-consistent
correction for dust-attenuation, i.e. with dust properties recovered
from another parameter (see Sect.~\ref{sub:selfdust}). We remark
that such self-consistent correction depends on the properties of
the sample used to do the calibration. Thus, we present in Sect.~\ref{sub:selfbaddust}
a way to correct for this bias.

\subsection{Use of an assumed mean correction\label{sub:assumedust}}

When dust attenuation cannot be reliably derived from the data, it
is common in the literature to use an assumed mean correction (frequently
$A_{V}=1$).

\subsubsection{Standard calibrations with $A_{V}=1$\label{sub:meanav1}}

\begin{figure}
\begin{centering}
\includegraphics[width=0.49\columnwidth]{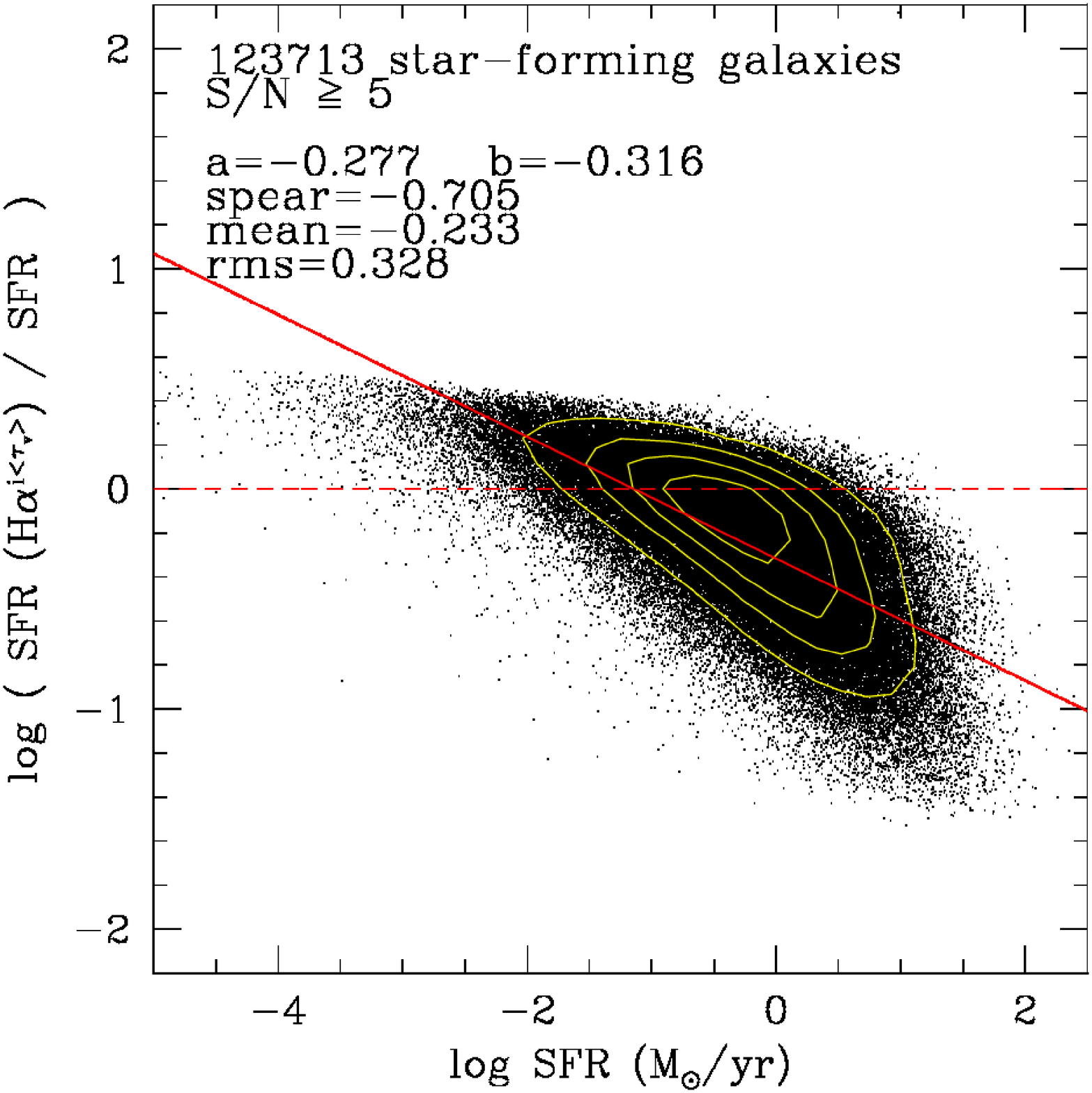}
\includegraphics[width=0.49\columnwidth]{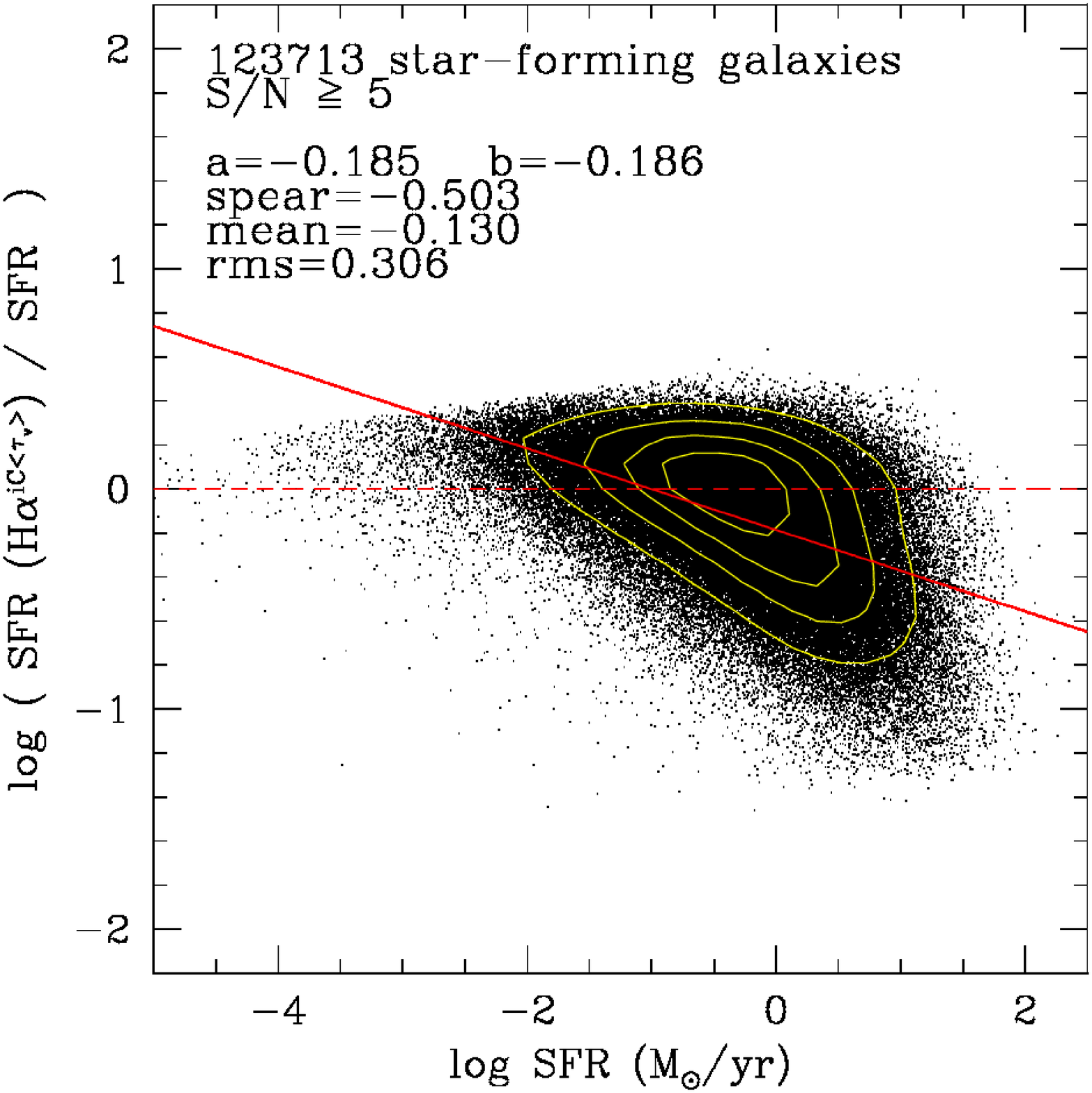}
\includegraphics[width=0.49\columnwidth]{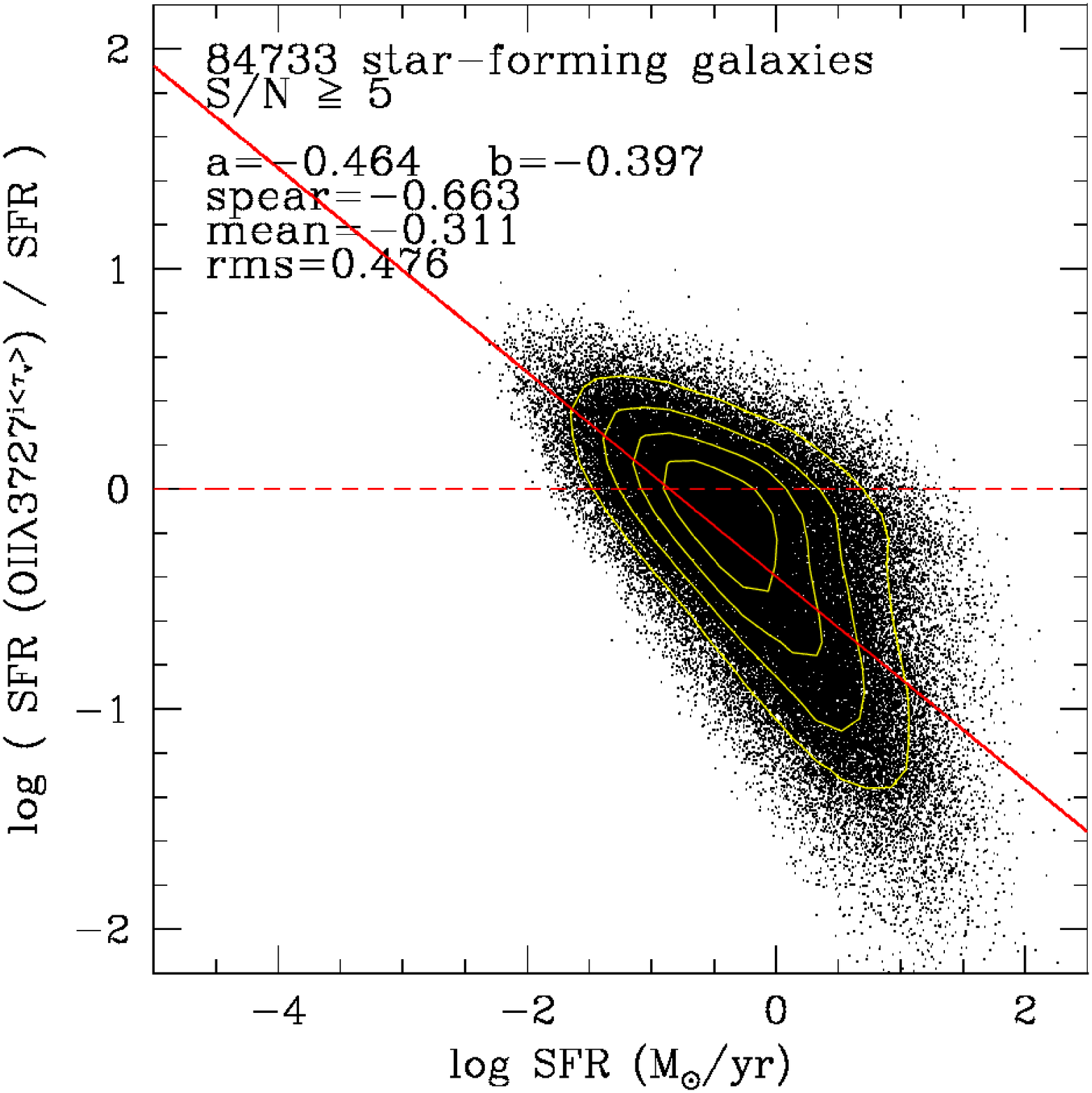}
\includegraphics[width=0.49\columnwidth]{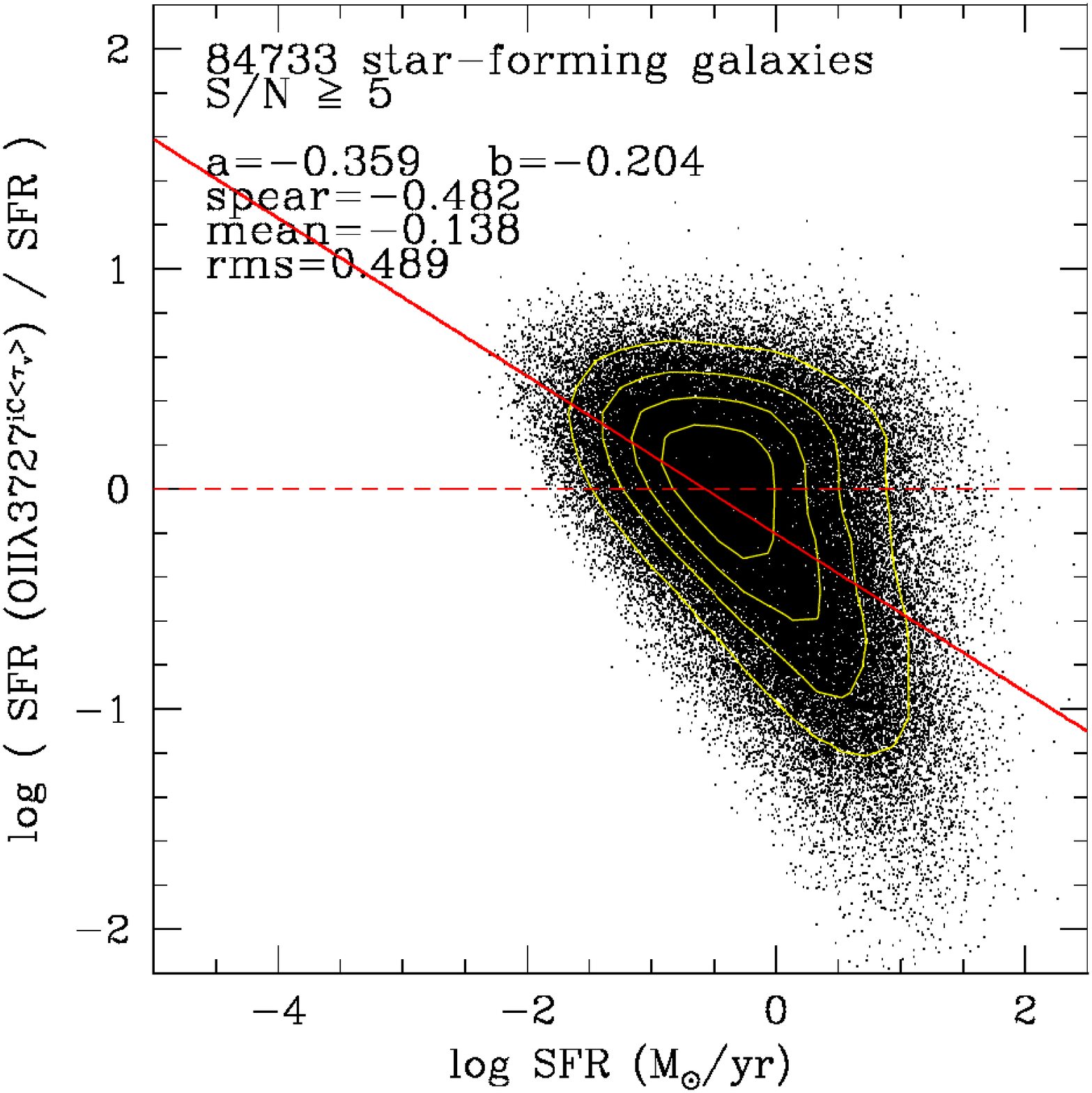}
\includegraphics[width=0.49\columnwidth]{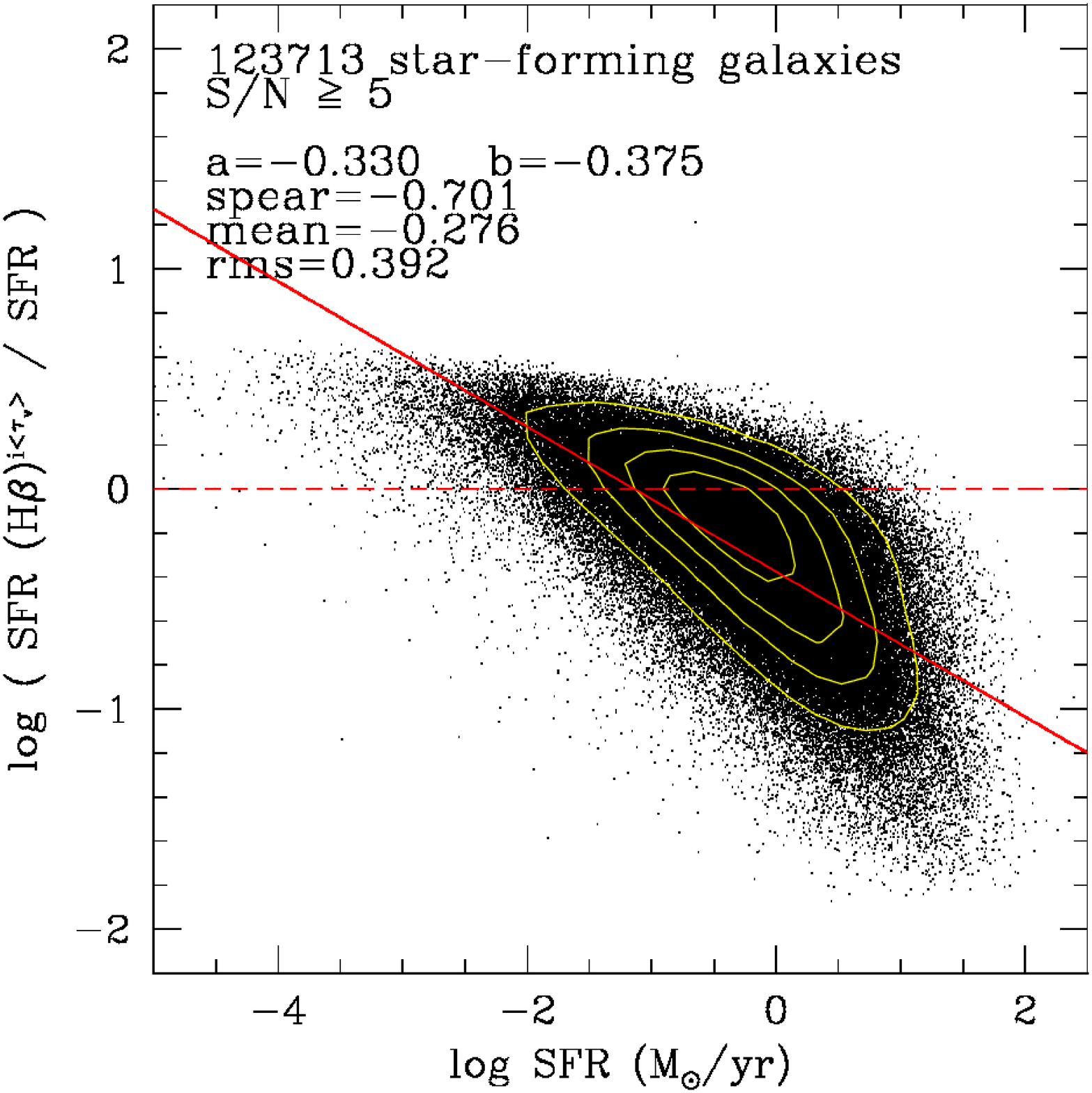}
\includegraphics[width=0.49\columnwidth]{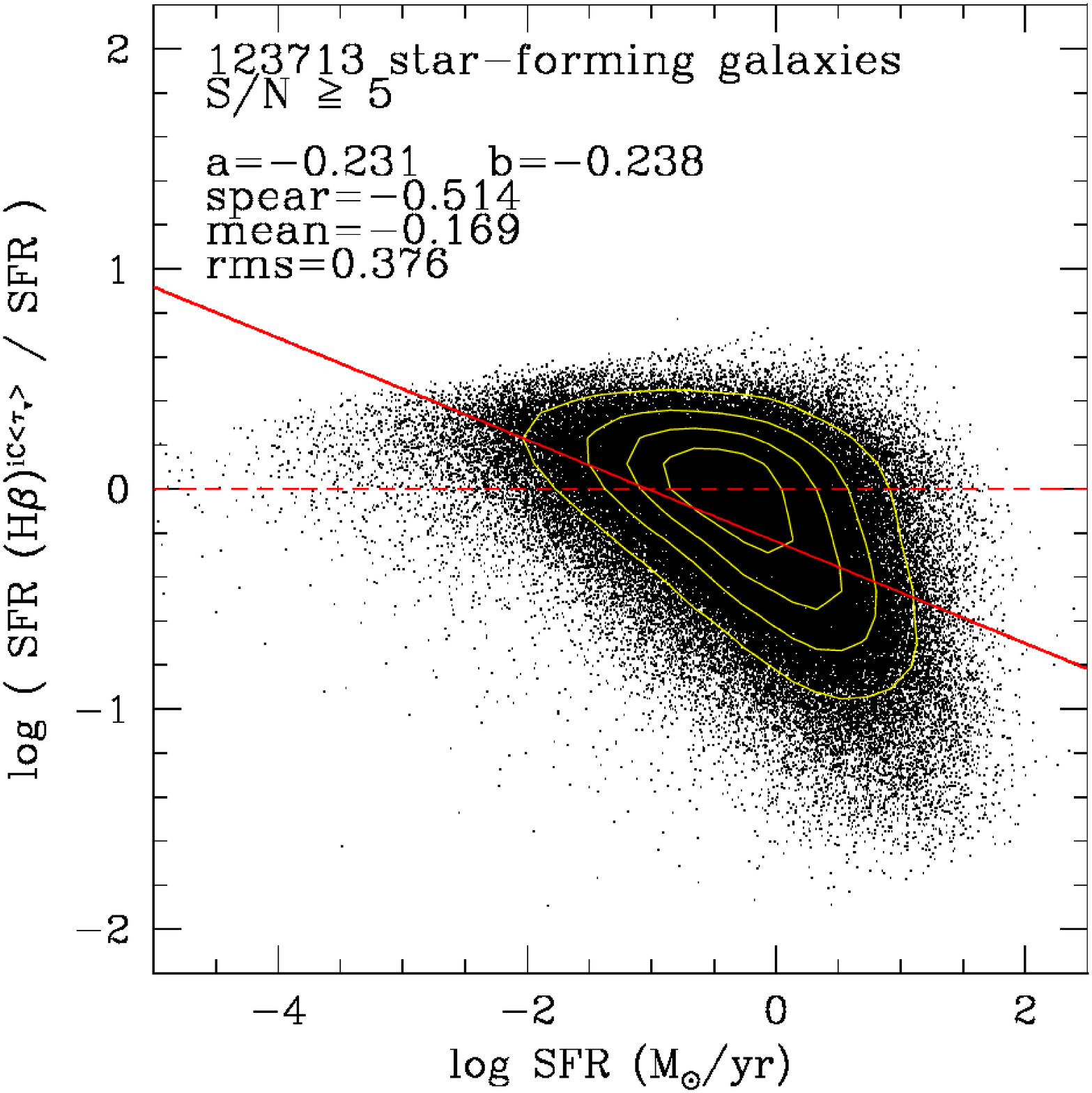}
\par\end{centering}

\caption{Same legend as in Fig.~\ref{fig:compstandard}. The studied calibrations
are the standard H$\alpha^{\mathrm{i}}$ (top-left), {[}O\noun{ii}]$^{\mathrm{i}}$
(middle-left), and H$\beta^{\mathrm{i}}$ (bottom-left) calibrations,
or the metallicity-unbiased H$\alpha^{\mathrm{iC}}$ (top-right),
{[}O\noun{ii}]$^{\mathrm{iC}}$ (middle-right), and H$\beta^{\mathrm{iC}}$
(bottom-right) calibrations, all applied on data corrected using an
assumed mean dust attenuation $A_{V}=1$.}

\label{fig:compmean}
\end{figure}

In this subsection, we discuss the quality of the standard calibrations
used on SDSS DR4 with an assumed mean correction for dust of $A_{V}=1$.

One important question we need to answer is: what kind of calibration
should we use on data corrected with a mean dust attenuation ? Should
we use the standard calibrations derived with the wrong assumption
of a constant intrinsic Balmer ratio (Sect.~\ref{sub:dustbalmer}),
or our new calibrations derived with the better CL01 estimates of
the dust (Sect.~\ref{sub:otherdust}) ? The answer basically depends
on the way the mean dust attenuation have been estimated (is it biased
towards the metallicity dependence of the intrinsic Balmer ratio or
not ?).

We thus plot the comparison of both calibrations in Fig.~\ref{fig:compmean}.
The main conclusion we draw from this figure is that the residuals
are not only \emph{highly dispersed} (rms in the range $0.3$-$0.5$
dex depending on the calibration) but also show a clear residual slope
in all cases ($-0.2$ to $-0.5$ depending on the calibration). This
residual slope is the signature of the correlation between the dust
attenuation and the SFR in galaxies. This correlation is obviously
not taken into account by the assumption of a constant dust attenuation
for the whole sample.

We note that the residual slopes are significant as shown by the Spearman
rank correlation coefficients ($-0.5$ to $-0.7$). It is clear also
that the residual slopes obtained with the H$\alpha^{\mathrm{iC}}$,
{[}O\noun{ii}]$^{\mathrm{iC}}$ , and H$\beta^{\mathrm{iC}}$ calibrations,
based on the CL01 estimates of the dust, are not only smaller but
also less significant (smaller Spearman rank correlation coefficient).

The method of the assumed mean correction thus leads to poor results,
either on galaxy-per-galaxy studies (high dispersion) or on statistical
studies (non-negligible residual slope). \emph{This method applied
with the {[}O}\emph{\noun{ii}}\emph{]}\emph{\noun{ }}\emph{line is
particularly inadvisable} (worst dispersion and steeper residual slope).

\subsubsection{Additional sources of biases\label{sub:meanplus}}

\begin{figure}
\begin{centering}
\includegraphics[width=0.8\columnwidth]{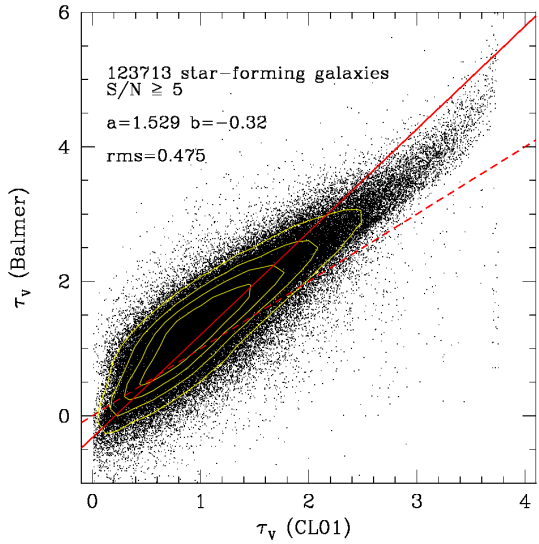}
\par\end{centering}

\caption{Relation between the dust attenuation estimated with the CL01 method
or with the Balmer decrement method, for star-forming galaxies in
the SDSS DR4 data. The dashed line is the $y=x$ curve. The solid
line is the $y=a\cdot x+b$ curve where $a$ and $b$ are the parameters
of the least-square fit (errors in $x$ and $y$), as given in the
plot. Isodensity contours are overplotted in white.}

\label{fig:relationtauv}
\end{figure}

We now discuss the additional effect of the use of a wrong mean correction.
Following Eq.~\ref{eq:intrflux} and~\ref{eq:defpower}, the bias
introduced on the SFR can be computed like that:\begin{equation}
\log\left(\frac{\mathrm{SFR^{w}}}{\mathrm{SFR}}\right)=\epsilon\times\frac{\tau_{V}^{\mathrm{w}}-\tau_{V}}{\ln(10)}\label{eq:wrongtau}\end{equation}
where $\tau_{V}^{\mathrm{w}}$ is the wrong assumed dust attenuation,
and SFR$^{\mathrm{w}}$ is the wrong derived SFR. Below, we investigate
three possible sources of bias.

The first bias comes from the assumption of $A_{V}=1$. The sample
used in this study actually shows a mean dust attenuation $\left\langle \tau_{V}^{\mathrm{CL01}}\right\rangle =1.21$
(see Sect.~\ref{sec:Description-of-the} and Fig.~\ref{fig:histotau}).
Converting from opacities to magnitudes, we obtain $\left\langle A_{V}^{\mathrm{CL01}}\right\rangle =\left\langle \tau_{V}^{\mathrm{CL01}}\right\rangle \times1.086=1.31$.
The difference between the assumed $A_{V}=1$ and the actual value
in the sample is partly responsible of the non-null means of the residuals
observed in Fig.~\ref{fig:compmean}: applying Eq.~\ref{eq:wrongtau},
we find a theoretical bias of the order of $-0.15$ dex, which is
similar to the observed values for the CL01 calibrations (Fig.~\ref{fig:compmean}
right).

The second bias comes from the dispersion in the distribution of dust
attenuation. In the SDSS DR4 data, the dispersion on the derived dust
attenuation is $0.64$, which means that the mean dust attenuation
may not be assumed with a better precision than $\pm0.64$. Thus it
implies a possible systematic shift of the order of $\pm0.30$ dex.
As stated in Sect.~\ref{sec:Description-of-the}, the mean dust attenuation
is $\left\langle \tau_{V}^{\mathrm{CL01}}\right\rangle =1.10$ instead
of $1.21$ when we add the selection on the {[}O\noun{ii}]\noun{ }line:
this implies a bias of the order of $\pm0.05$ dex.

Finally, we remind the reader that he has to think carefully of the
origin of the assumed mean correction before applying it, i.e. he
has to check whether this assumption comes from a dust attenuation
derived with a constant Balmer decrement, or from an unbiased estimation.
Having checked that, he would be able to apply the right calibrations.
Fig.~\ref{fig:relationtauv} gives for convenience the relation between
$\tau_{V}^{\mathrm{Balmer}}$ and the unbiased $\tau_{V}^{\mathrm{CL01}}$
in the SDSS DR4 data. We fit an empirical relation which follows this
formula:\begin{equation}
\tau_{V}^{\mathrm{Balmer}}=1.53\times\tau_{V}^{\mathrm{CL01}}-0.3\label{eq:tauvBalmerCL01}\end{equation}
This implies a theoretical bias of the order of $-0.25$ dex, which
is again similar to the observed values for the standard calibrations
(Fig.~\ref{fig:compmean} left).

Hence we conclude that the use of an assumed mean correction has to
be handled \emph{with great care}.

\subsection{Use of a self-consistent correction\label{sub:selfdust}}

As stated in previous subsection, the use of an assumed mean correction
has one main drawback: it does not take into account the variation
of the dust attenuation as a function of SFR. We thus discuss in this
subsection three possible ways to reduce this bias. The idea common
to these three possibilities is that the relation between the dust
attenuation and the SFR might be itself recovered from the data, i.e.
self-consistently.

We explore the self-consistent corrections obtained either from the
observed line luminosities, another parameter like the magnitude,
or a combination of two emission lines.

\subsubsection{Single-line calibrations}

\begin{figure}
\begin{centering}
\includegraphics[width=0.49\columnwidth]{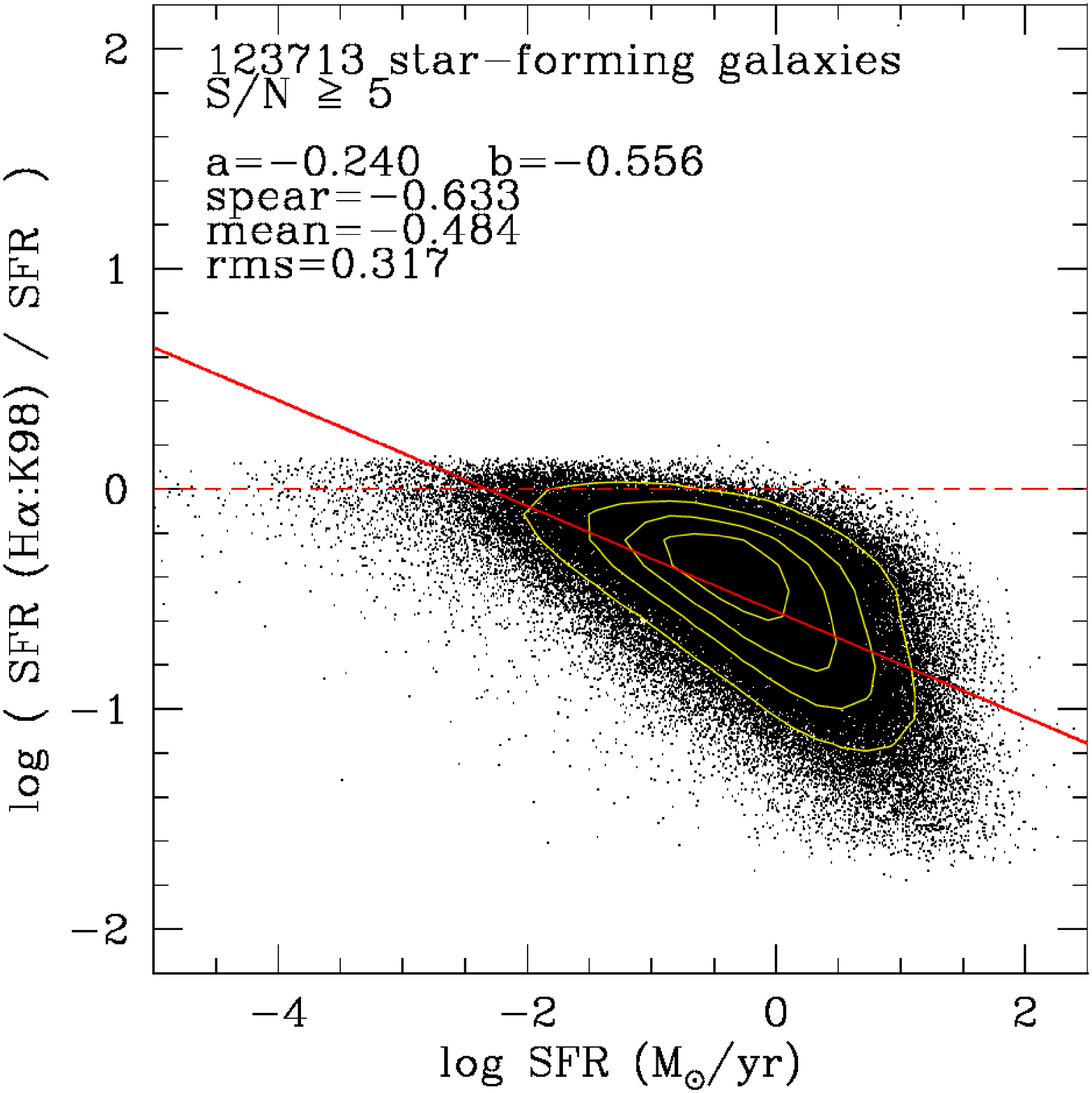}
\includegraphics[width=0.49\columnwidth]{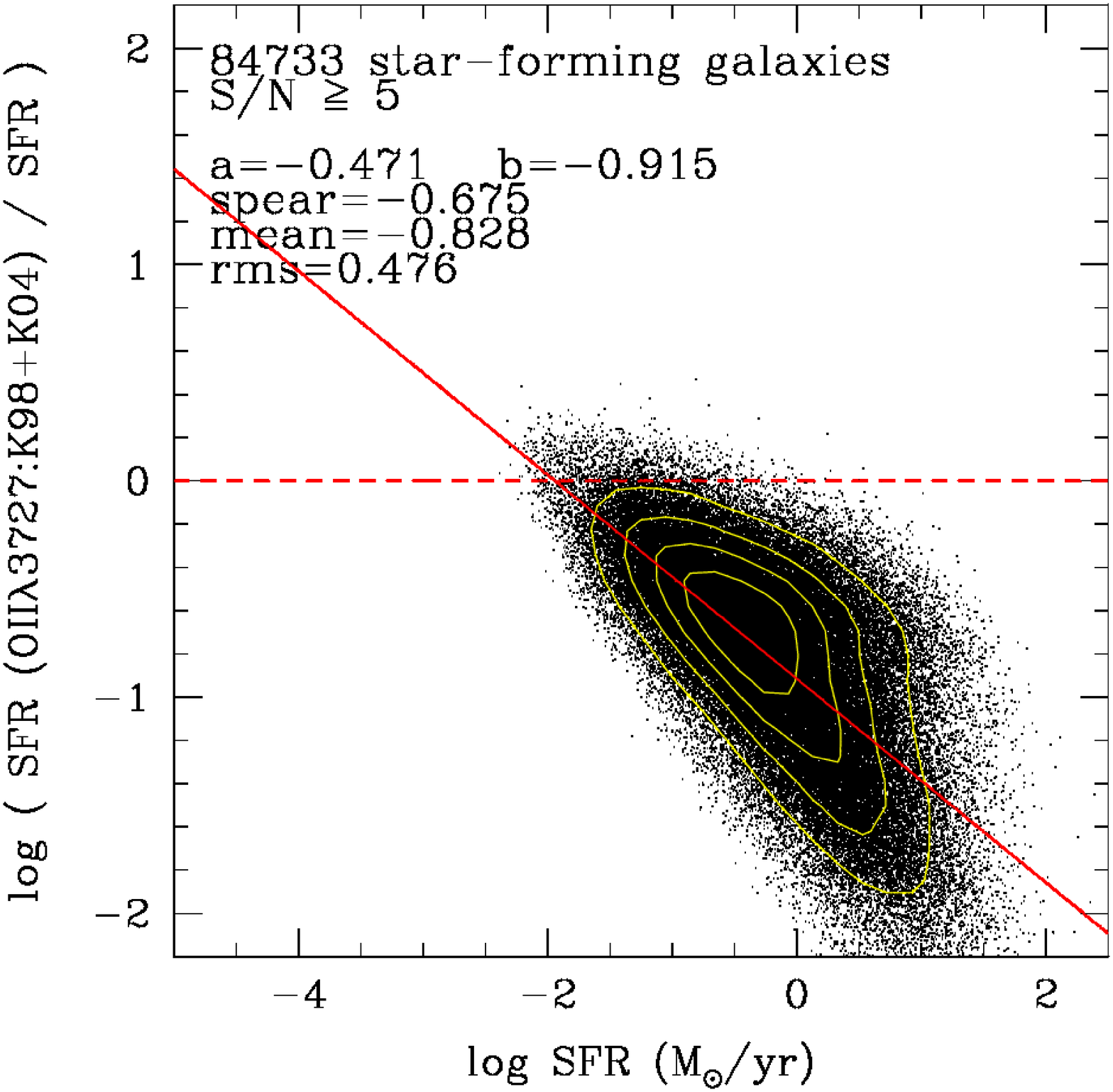}
\includegraphics[width=0.49\columnwidth]{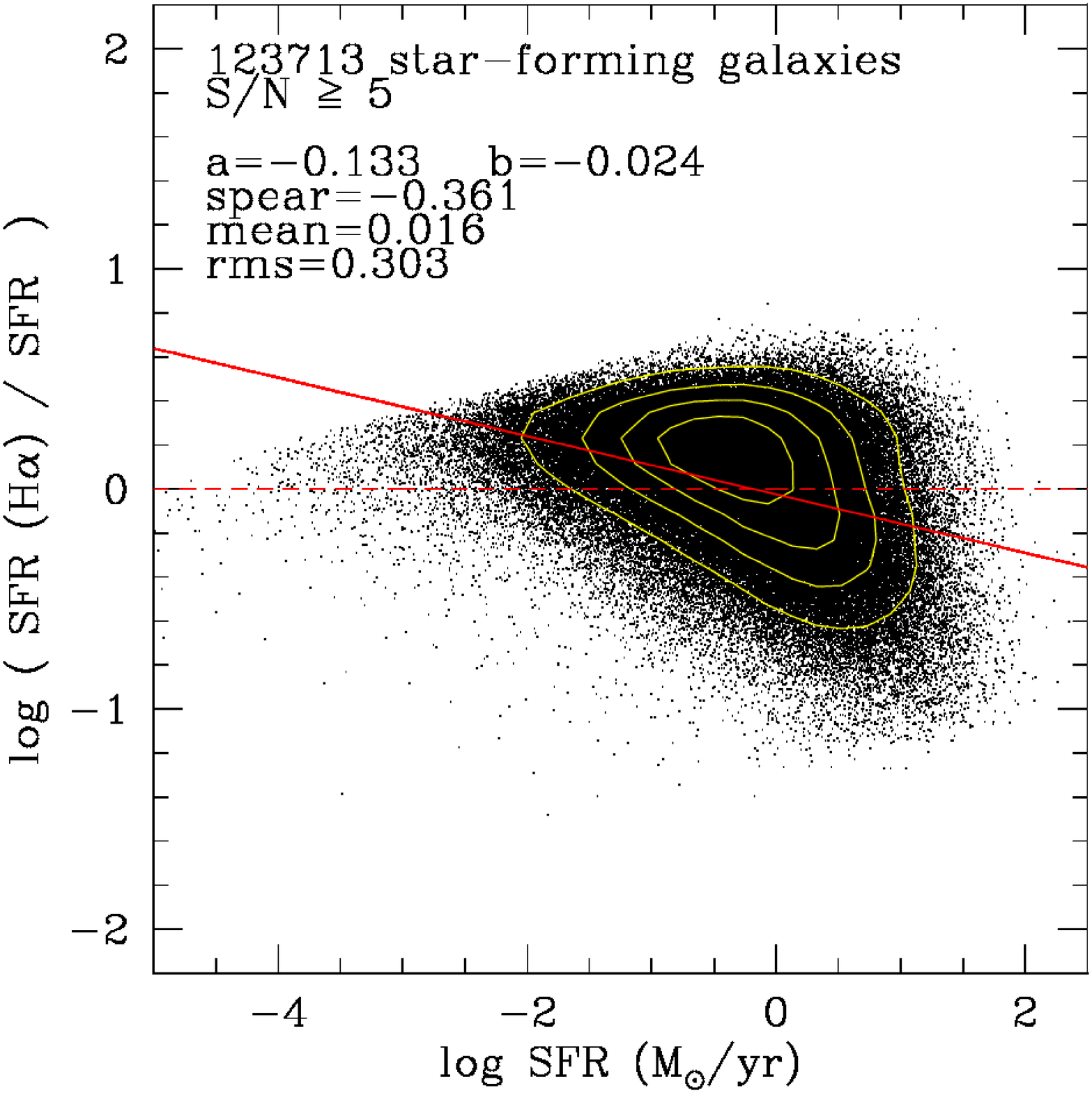}
\includegraphics[width=0.49\columnwidth]{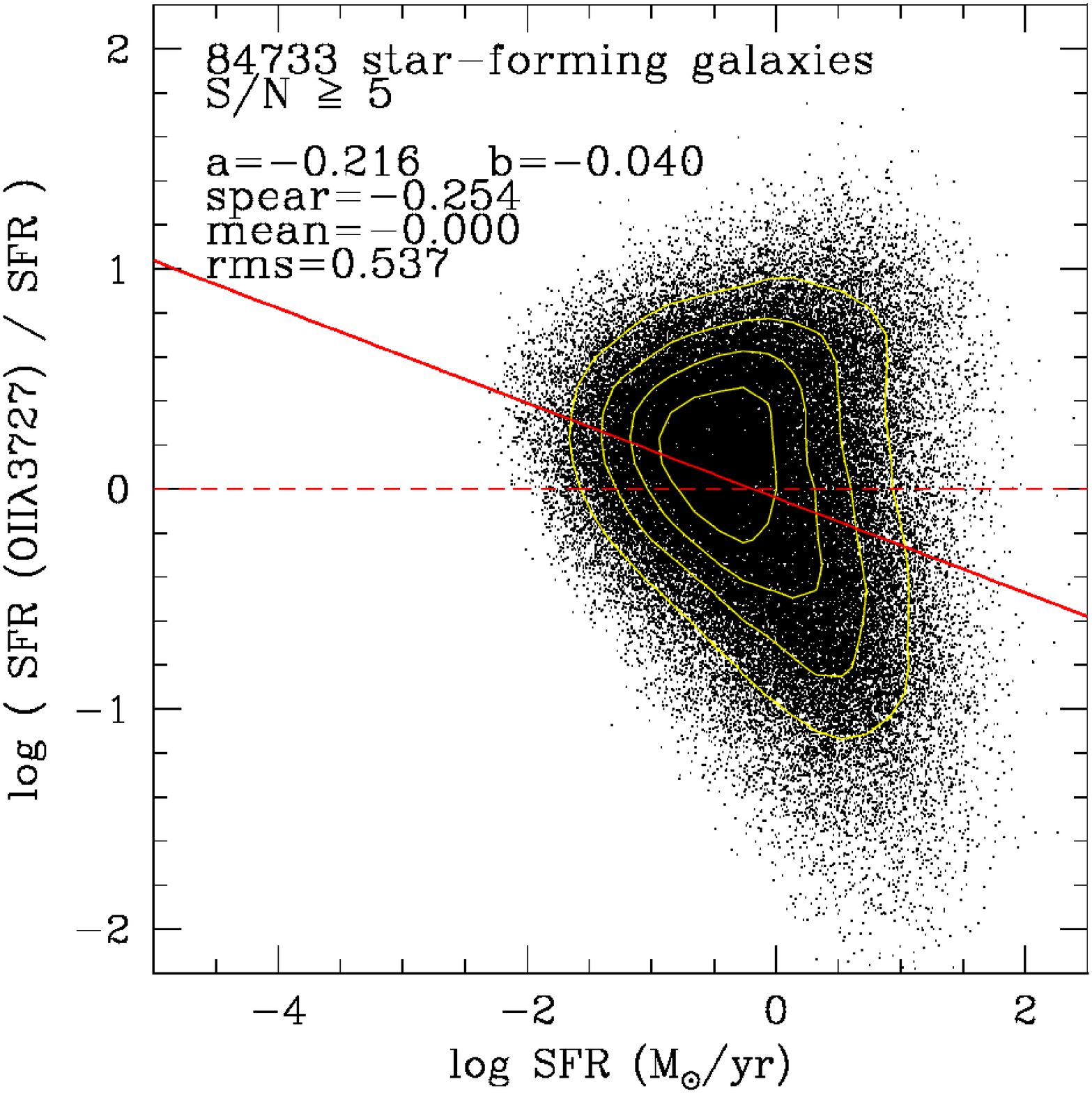}
\includegraphics[width=0.49\columnwidth]{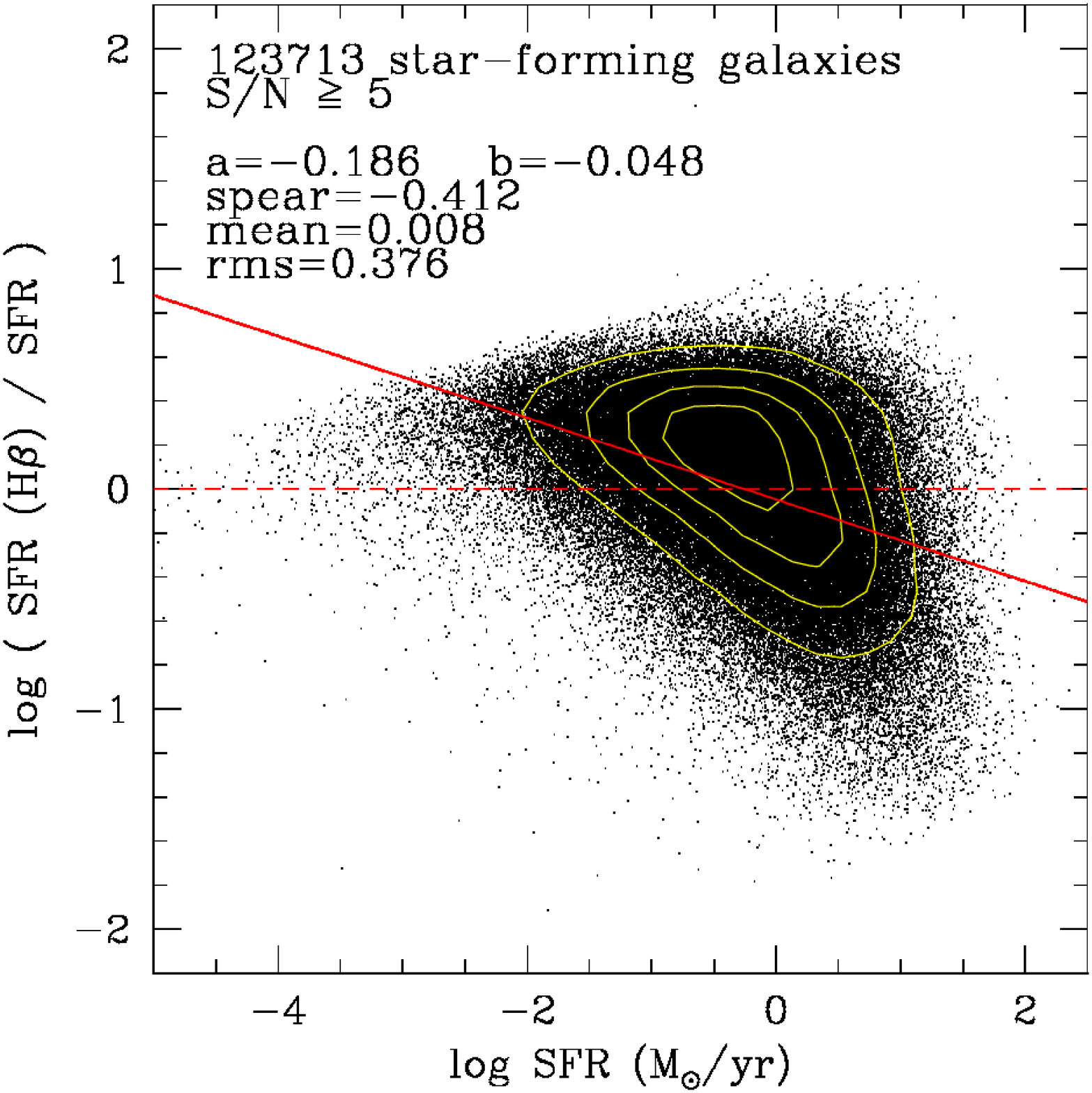}
\par\end{centering}

\caption{Same legend as in Fig.~\ref{fig:compstandard}. The studied calibrations
are: top-left: \citet{Kennicutt:1998ARA&A..36..189K} H$\alpha^{\mathrm{i}}$
applied on uncorrected data; top-right: \citet{Kennicutt:1998ARA&A..36..189K}
\citep[corrected by][]{Kewley:2004AJ....127.2002K} {[}O\noun{ii}]$^{\mathrm{i}}$
on uncorrected data; middle-left: our new H$\alpha$ calibration;
middle-right: our new {[}O\noun{ii}] calibration; bottom: our new
H$\beta$ calibration.}

\label{fig:compobs}
\end{figure}

\begin{figure*}
\begin{centering}
\includegraphics[width=0.27\linewidth]{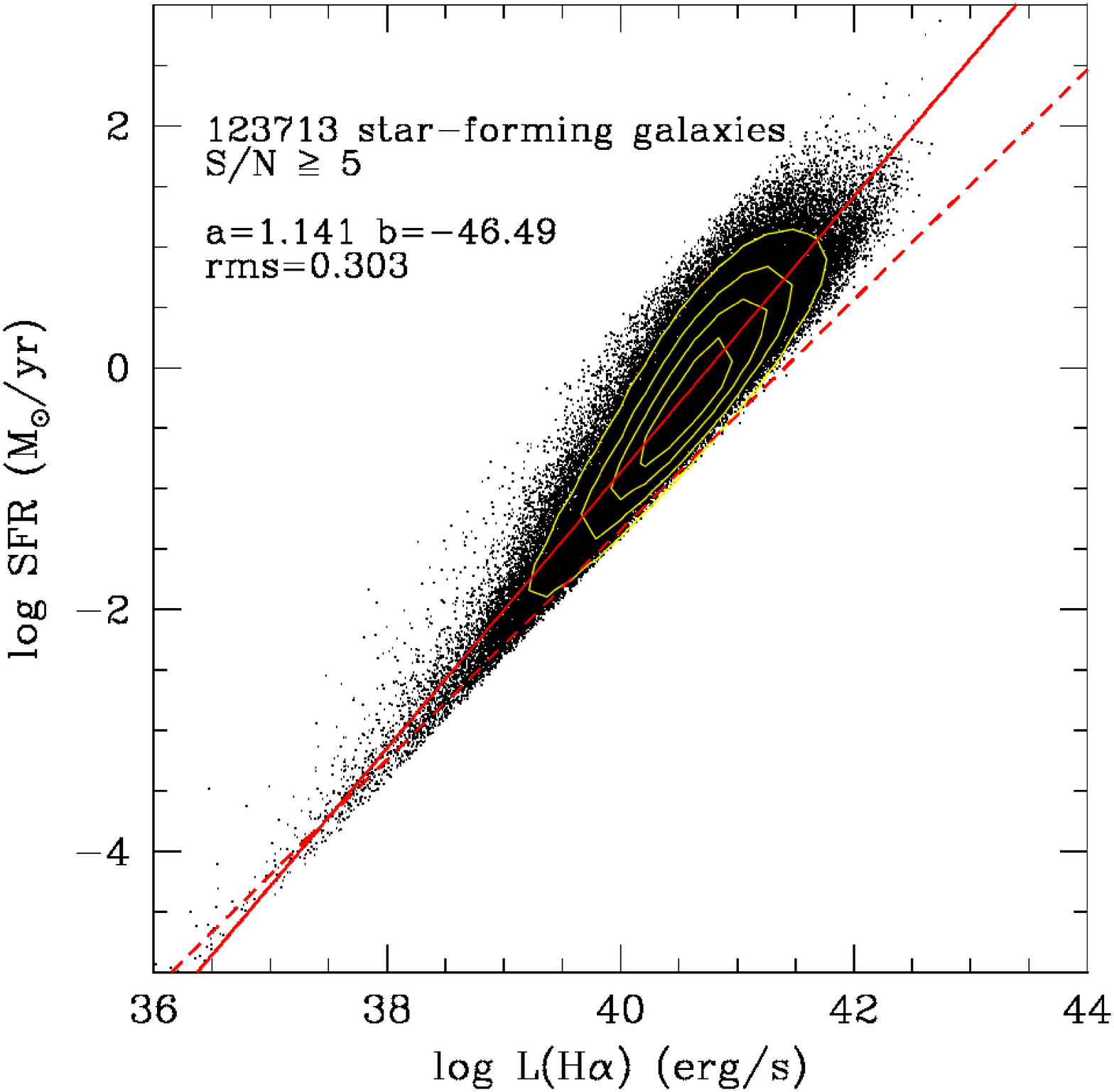}
\includegraphics[width=0.27\linewidth]{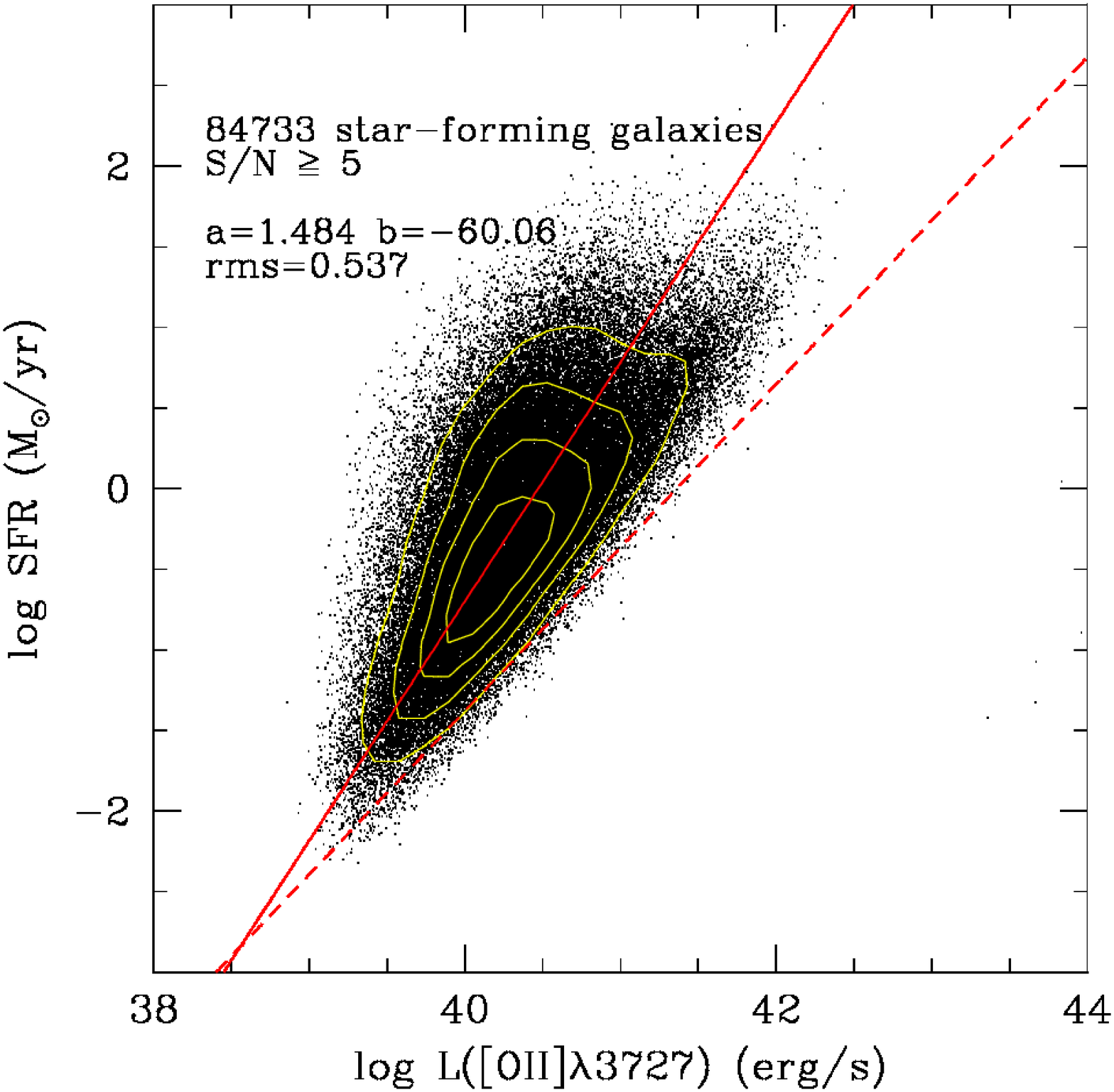}
\includegraphics[width=0.27\linewidth]{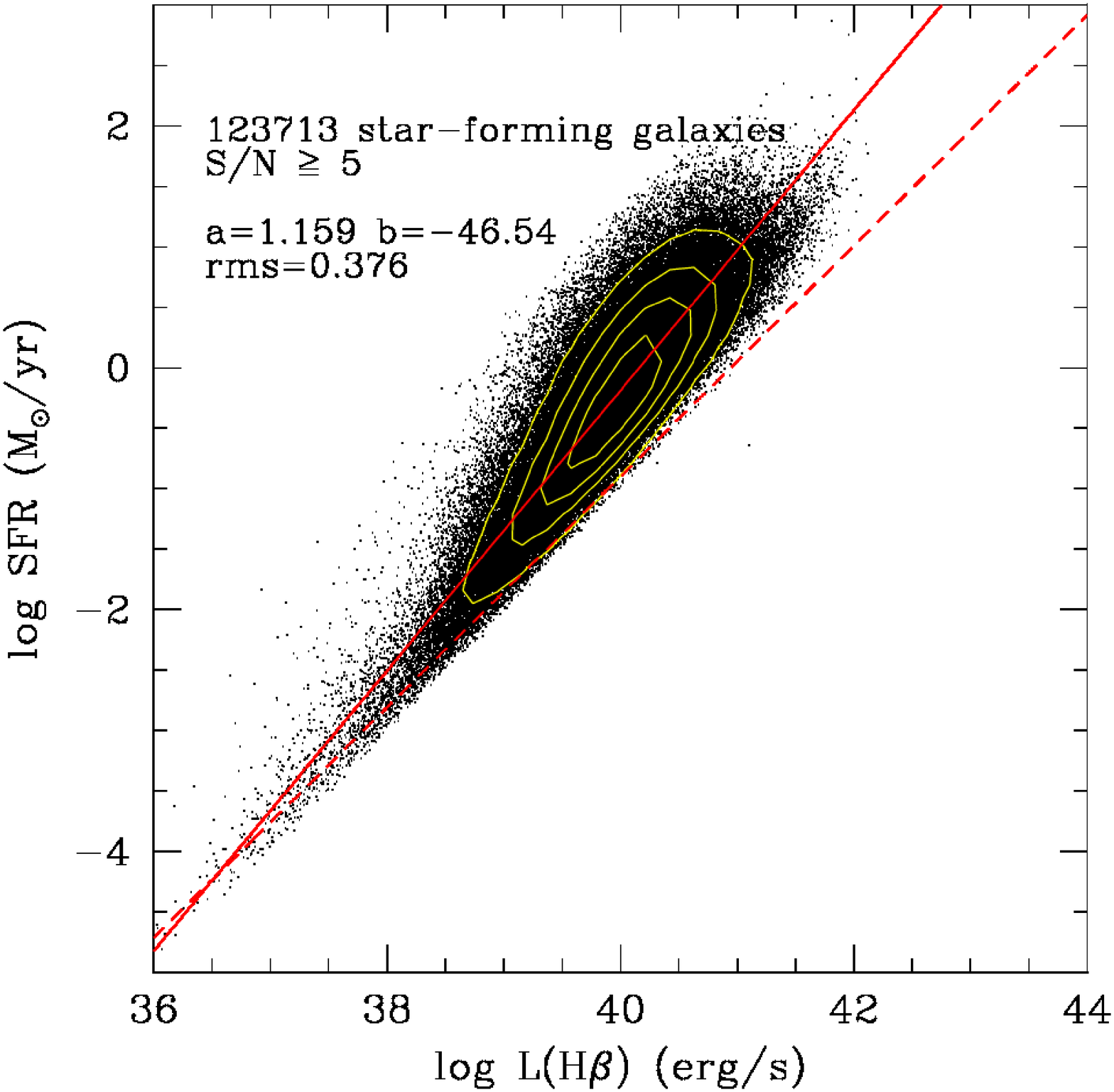}
\par\end{centering}

\caption{Relation between the SFR (logarithm of M$_{\odot}$/yr) and the H$\alpha$
(left), {[}O\noun{ii}] (center) or H$\beta$ (right) observed emission
lines luminosities (logarithm of erg/s) for star-forming galaxies
in the SDSS DR4 data. The solid line is a least-square fit to the
data (errors on $x$ and $y$). The dashed lines show the intrinsic
calibrations obtained with dust estimated with the Balmer decrement
method (see Fig.~\ref{fig:newstandard}). Isodensity contours are
overplotted in white.}

\label{fig:newobs}
\end{figure*}

The simplest way to apply a self-consistent correction for dust is
to relate the dust attenuation to the line luminosity itself. Hence,
it consists on calibrating the SFR directly with the observed, not
intrinsic, line luminosity with a two-parameter law as described in
Eq.~\ref{eq:defpower}.

Fig.~\ref{fig:compobs} (top-left) shows the results obtained when
using the standard \citet{Kennicutt:1992ApJ...388..310K} H$\alpha^{\mathrm{i}}$
calibration on data uncorrected for dust attenuation. We clearly see
that the dust attenuation causes an underestimate of the SFR (the
mean of the residuals is $-0.48$ dex), and that this effect is increasing
with SFR. Moreover the dispersion is $0.32$ dex, and there is a residual
slope of $-0.24$ which is significant (Spearman rank correlation
coefficient of $-0.63$).\textbf{ }Taking this into account, a two
parameter law is \emph{necessary}, differently from what we observed
on dust-corrected data in Sect.~\ref{sub:dustbalmer}.

Fig.\textbf{~}\ref{fig:newobs} (left) shows the correlation between
the observed H$\alpha$ emission-line luminosity and the SFR in SDSS
DR4 data. We obtain the following new calibration:

\begin{equation}
\left\{ \begin{array}{ccl}
\log\eta_{\mathrm{H}\alpha} & = & 46.49\pm0.04\\
\epsilon_{\mathrm{H}\alpha} & = & 1.141\pm0.001\end{array}\right.\label{eq:hanodust}\end{equation}
As shown in Fig.~\ref{fig:compobs} (middle-left), the dispersion
of this new calibration is $0.30$ dex, which tells us that the SFR
recovered without dust attenuation correction are almost twice more
dispersed than when an estimate of dust attenuation is available.
The mean of the residuals are now almost null ($0.02$ dex). This
calibration is better, in terms of the dispersion, than the results
obtained with an assumed mean correction (see Sect.~\ref{sub:assumedust},
dispersion of $0.33$ dex) but, it still \emph{quite poor}. 

Unfortunately the use of the two-parameters law, which allows in principle
to take into account the dependence between the dust attenuation and
the SFR, does not cancel completely the residual slope which is still
of $-0.13$. However it is clear, from the comparison of Fig.~\ref{fig:compmean}
and Fig.~\ref{fig:compobs} that this remaining residual slope is
much less significant (Spearman rank correlation coefficient of $-0.36$
instead of $-0.5$ to $-0.7$).

\medskip{}

No estimation of the dust attenuation would be available in most cases
where one has to use the {[}O\noun{ii}] or the H$\beta$ emission
line in order to compute a SFR. This is basically due to the fact
the H$\alpha$ line is commonly used to estimate the dust attenuation,
and that \emph{there is no reason to use another line if the H$\alpha$
line is measured}. 

Fig.~\ref{fig:compobs} (top-right) shows the results obtained when
using the standard \citet{Kennicutt:1992ApJ...388..310K} {[}O\noun{ii}]$^{\mathrm{i}}$
calibration on data uncorrected for dust attenuation. the dispersion
is $0.48$ dex which is high. There is also a significant systematic
shift of $-0.83$ dex, plus a significant (Spearman rank correlation
coefficient of $-0.68$) residual slope of $-0.47$. Fig.~\ref{fig:newobs}
(center) shows the relation between the observed {[}O\noun{ii}] emission
line luminosity and the SFR in SDSS DR4 data. We obtain the following
best-fit values:\begin{equation}
\left\{ \begin{array}{ccl}
\log\eta_{[\mathrm{O}\mathsc{ii}]} & = & 60.06\pm1.9\\
\epsilon_{[\mathrm{O}\mathsc{ii}]} & = & 1.484\pm0.05\end{array}\right.\label{eq:o2nodust}\end{equation}
 But, as we can see from Fig.~\ref{fig:compobs} (middle-right) the
dispersion ($0.54$ dex) is higher than the results obtained with
an assumed mean correction (see Sect.~\ref{sub:selfdust}, dispersion
of $0.48$ dex): the correlation between observed {[}O\noun{ii}] luminosity
and SFR is rather poor. Nevertheless, this new calibration has the
advantage of showing a smaller residual slope of $-0.22$, also much
less significant (Spearman rank correlation coefficient of $-0.25$).

Fig.~\ref{fig:newobs} (right) shows the relation between the observed
H$\beta$ emission line luminosity and the SFR in SDSS DR4 data, which
gives the following parameters:\begin{equation}
\left\{ \begin{array}{ccl}
\log\eta_{\mathrm{H}\beta} & = & 46.54\pm0.04\\
\epsilon_{\mathrm{H}\beta} & = & 1.159\pm0.001\end{array}\right.\label{eq:hbnodust}\end{equation}

As shown in Fig.~\ref{fig:compobs} (bottom), this new calibration
shows a small improvement as compared to the method of an assumed
mean correction: dispersion of $0.38$ dex instead of $0.39$ dex,
smaller residual slope of $-0.19$, smaller Spearman rank correlation
coefficient of $-0.41$ for this residual slope.

The two-parameters calibrations presented in this subsection do not
show significant residual slopes as compared to the method of the
assumed mean correction. Unfortunately, the gain of having no more
significant residual slopes is diminished by the still high dispersion.
We claim that these new self-consistent single-line calibrations are
however useful when one as no idea of the mean dust attenuation he
wants to assume. As stated before, the use of wrong mean correction
could lead to a non-negligible systematic bias.

\subsubsection{Discussion of previous works}

\begin{figure}
\begin{centering}
\includegraphics[width=0.8\columnwidth]{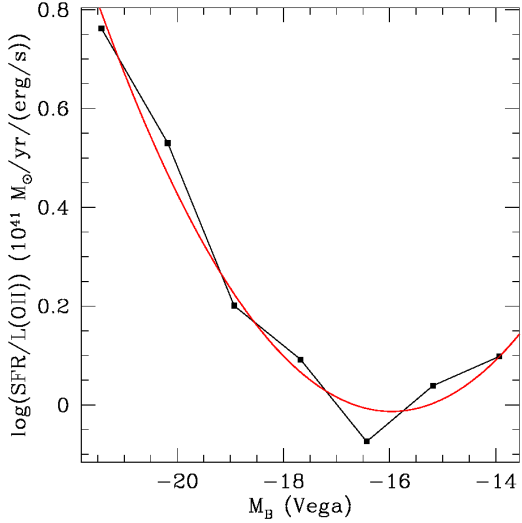}
\par\end{centering}

\caption{Relation between the SFR efficiency factor of the {[}O\noun{ii}] emission
line and the $B$-band $k$-corrected absolute magnitude (Vega system).
The data points are the values given in Table~2 of \citet{Moustakas:2006ApJ...642..775M}.
The solid curve is our 2nd degree polynomial least square fit with
the coefficients given in the text.}

\label{fig:calibM06}
\end{figure}

\begin{figure}
\begin{centering}
\includegraphics[width=0.49\columnwidth]{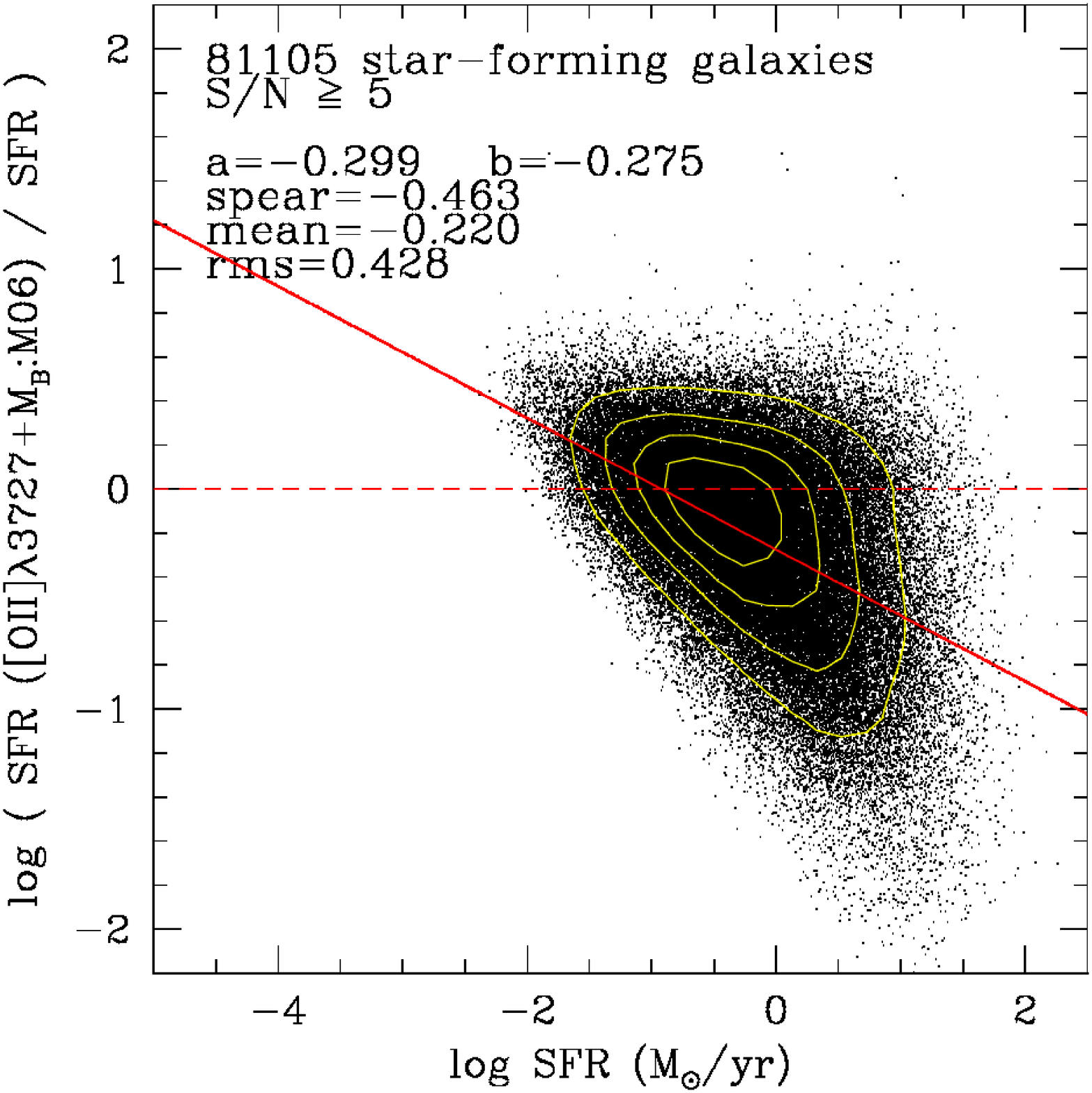}
\includegraphics[width=0.49\columnwidth]{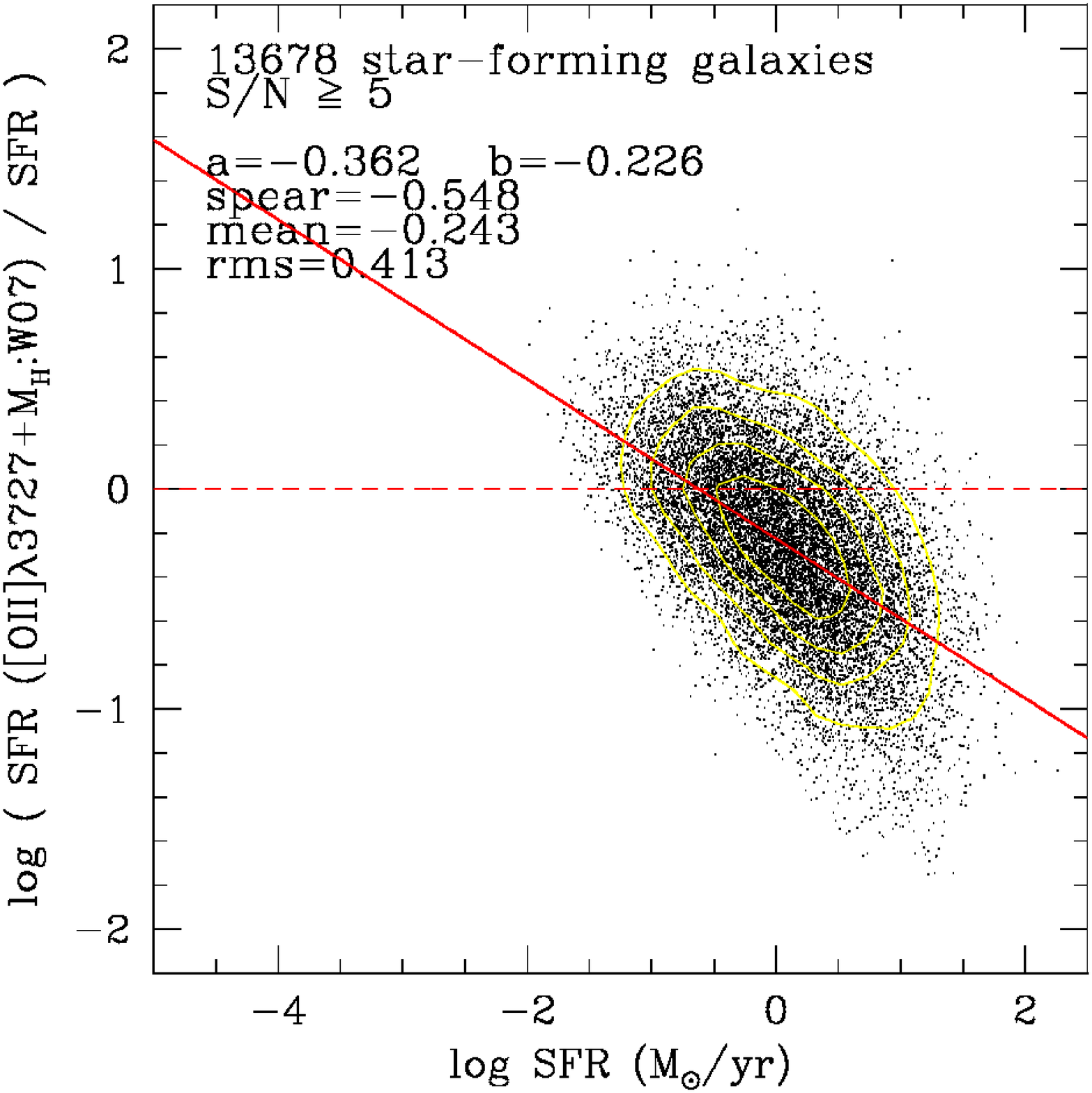}
\par\end{centering}

\caption{Same legend as in Fig.~\ref{fig:compstandard}. The studied calibrations
are: left: \citet{Moustakas:2006ApJ...642..775M} {[}O\noun{ii}]+$M_{B}$
calibration; right: \citet{Weiner:2006astro.ph.10842W} {[}O\noun{ii}]+$M_{H}$
calibration.}

\label{fig:compprevious}
\end{figure}

Another way to correct self-consistently for dust attenuation, as
long as for other parameters (e.g. metallicity, ionization degree,
...), is to use the absolute magnitude. Indeed, the absolute magnitude
provides an estimate of the dust attenuation through the general correlation
between mass and dust attenuation in galaxies. It also provides an
estimate of the metallicity through the luminosity-metallicity relation.
We note that these calibrations are only derived for the {[}O\noun{ii}]
line. Again, it is better to correct directly for dust attenuation
using the Balmer decrement when H$\alpha$ is observed.

The \citet{Moustakas:2006ApJ...642..775M} {[}O\noun{ii}] calibration
is based on a correction with $B$-band $k$-corrected absolute magnitude,
and starts now to be widely used in the literature. However, since
the correction is provided by the authors in a few number of discrete
points (see Table~2 of their paper), it is not clear how to interpolate
between them. The common approach now used in other works is to calculate
a global linear regression. Fig.~\ref{fig:calibM06} shows that this
approach is not valid in the whole domain. We thus provide a parabolic
fit which leads to the following final formula:\begin{equation}
\log\mathrm{SFR}=\log L(\mathrm{[O}\mathsc{ii}\mathrm{]})-41+6.86+0.862\times M_{B}+0.027\times M_{B}^{2}\label{eq:SFRM06}\end{equation}
where $M_{B}$ is the $B$-band $k$-corrected absolute magnitude
in the Vega system ($B_{\mathrm{Vega}}=B_{\mathrm{AB}}+0.09$). 

Fig.~\ref{fig:compprevious} (left) shows how the results obtained
with the \citet{Moustakas:2006ApJ...642..775M} calibration compares
with the reference CL01 SFR. This calibration shows a significant
systematic shift of $-0.22$ dex. However, the dispersion of $0.43$
dex is better than previous calibrations on the observed {[}O\noun{ii}]
luminosity. We observe also a non-negligible residual slope of $-0.30$
(Spearman rank correlation coefficient of $-0.46$), but not stronger
than the residual slope obtained with the method of the assumed mean
correction.

\medskip{}

The \citet{Weiner:2006astro.ph.10842W} {[}O\noun{ii}] calibration
is based on a correction with $H$-band $k$-corrected absolute magnitude.
The authors give the slope of the relation between the $\log(\mathrm{[O}\mathsc{ii}\mathrm{]}_{obs}/\mathrm{H}\alpha^{i})$
ratio and the $H$-band $k$-corrected absolute magnitude: $+0.23$
dex/mag. The zero-point is obtained by applying the $0.68$ dex $A(\mathrm{H}\alpha$)
extinction to the $\log(\mathrm{[O}\mathsc{ii}\mathrm{]}_{obs}/\mathrm{H}\alpha_{obs})$
ratio of $-0.32$ dex at $M_{H}=-21$. The final formula is:\begin{equation}
\log\mathrm{SFR}=\epsilon_{\mathrm{H}\alpha}^{\mathrm{i}}\left(\log L(\mathrm{[O}\mathsc{ii}\mathrm{]})-0.23\times M_{H}-3.83\right)-\log\eta_{\mathrm{H}\alpha}^{\mathrm{i}}\label{eq:SFRW07}\end{equation}
where $M_{H}$ is the $H$-band $k$-corrected absolute magnitude
in the AB system.

Fig.~\ref{fig:compprevious} (right) shows how the results obtained
with the \citet{Weiner:2006astro.ph.10842W} calibration compares
with the reference CL01 SFR. The mean of the residuals is $-0.24$
dex, the dispersion of $0.41$ dex is better than the calibration
on the observed {[}O\noun{ii}] luminosity, but is still significantly
high. Moreover, this smaller dispersion may be only due to the small
number of galaxies ($13\,678$) with an available measurement of $M_{H}$
in our sample. Finally, their is also a significant residual slope
in this calibration of $-0.36$ (Spearman rank correlation coefficient
of $-0.55$).

We conclude that the calibrations proposed by \citet{Moustakas:2006ApJ...642..775M}
and \citet{Weiner:2006astro.ph.10842W} lead to poor results on the
SDSS DR4 data. They do neither significantly better, nor significantly
worse, than the results obtained with the method of the assumed mean
correction, or with our direct calibrations between the SFR and observed
line luminosities.

\subsubsection{Two-lines calibrations}

\begin{figure*}
\begin{centering}
\includegraphics[width=0.27\linewidth]{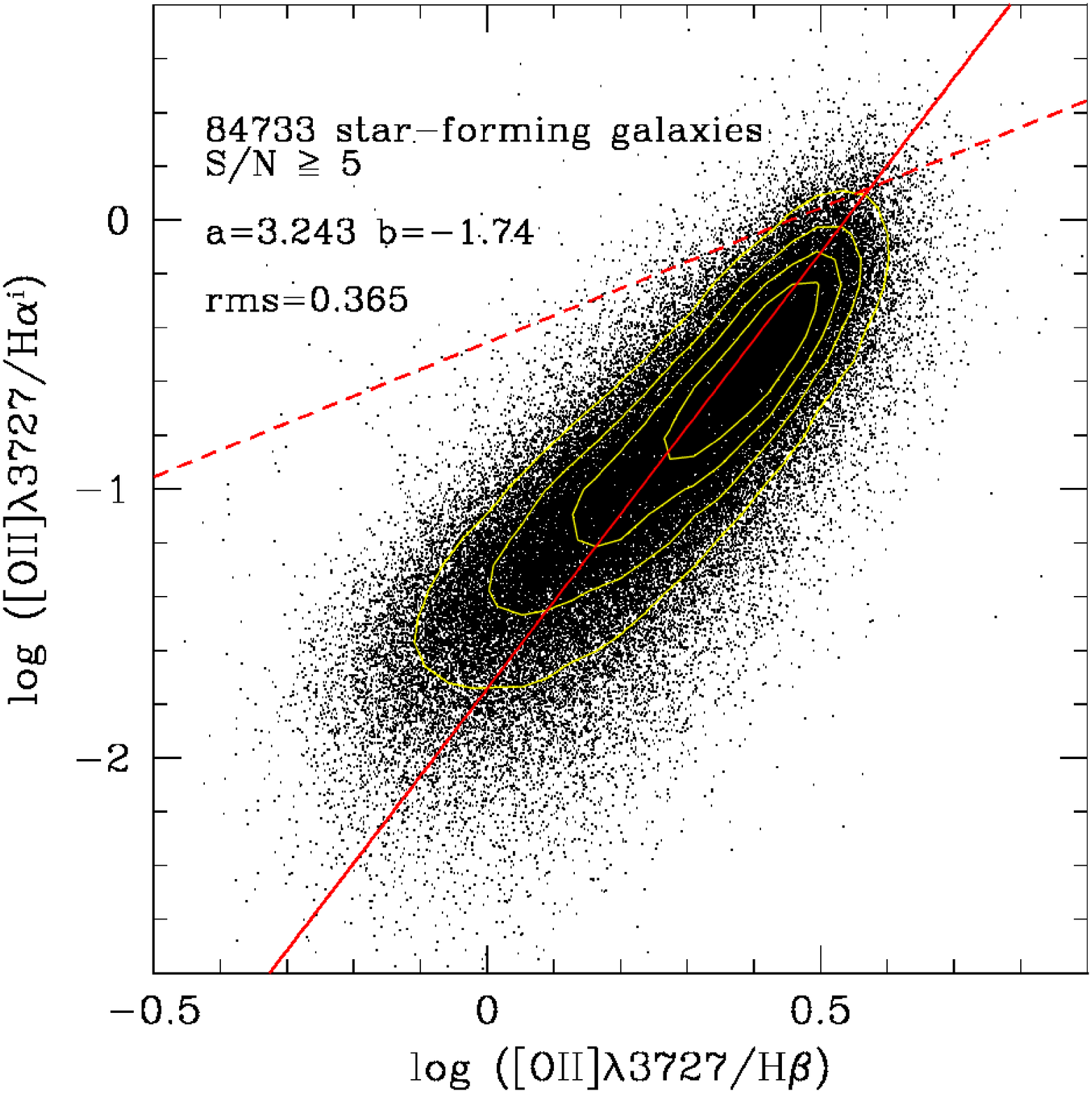} \includegraphics[width=0.27\linewidth]{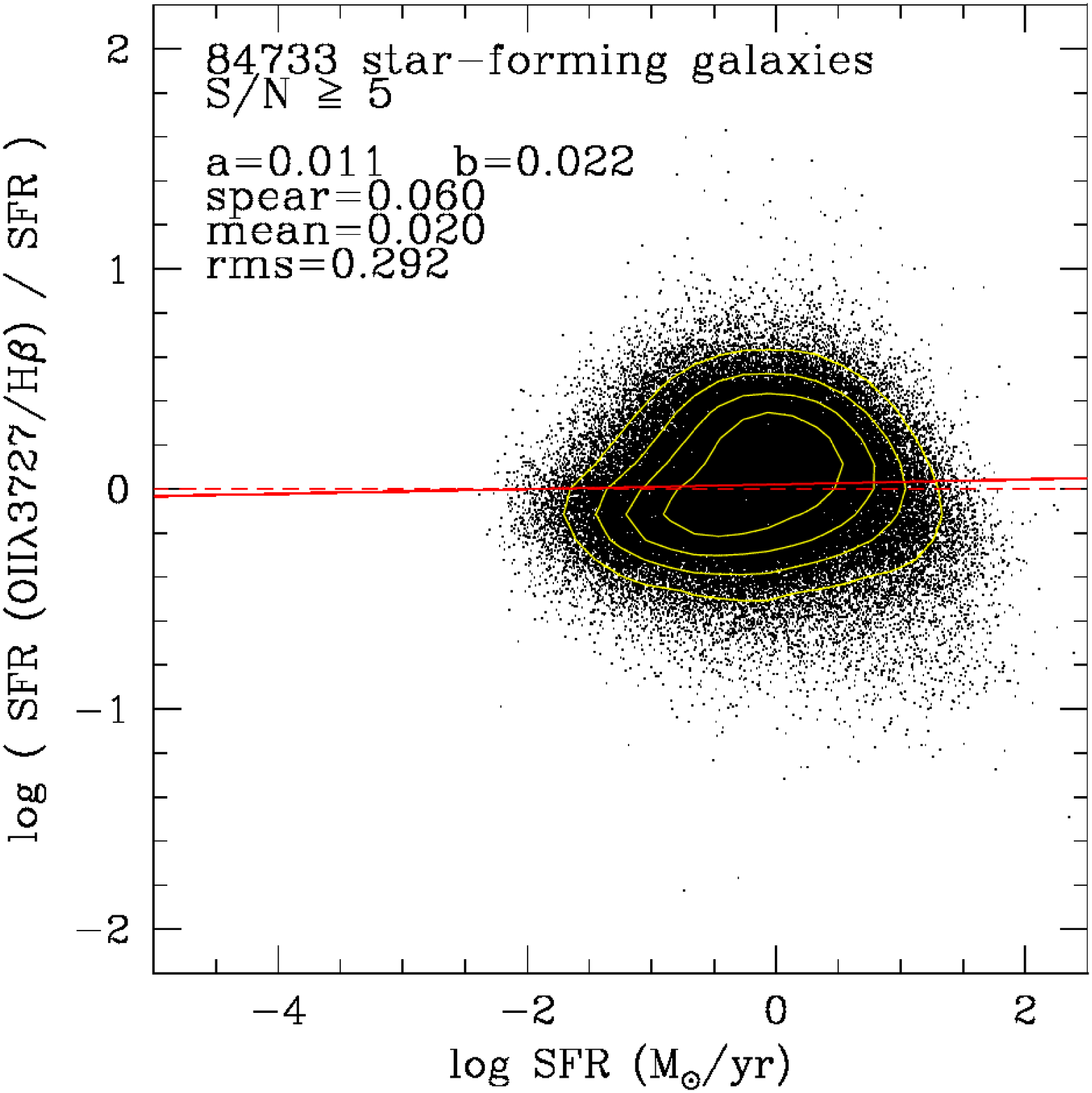}
\includegraphics[width=0.27\linewidth]{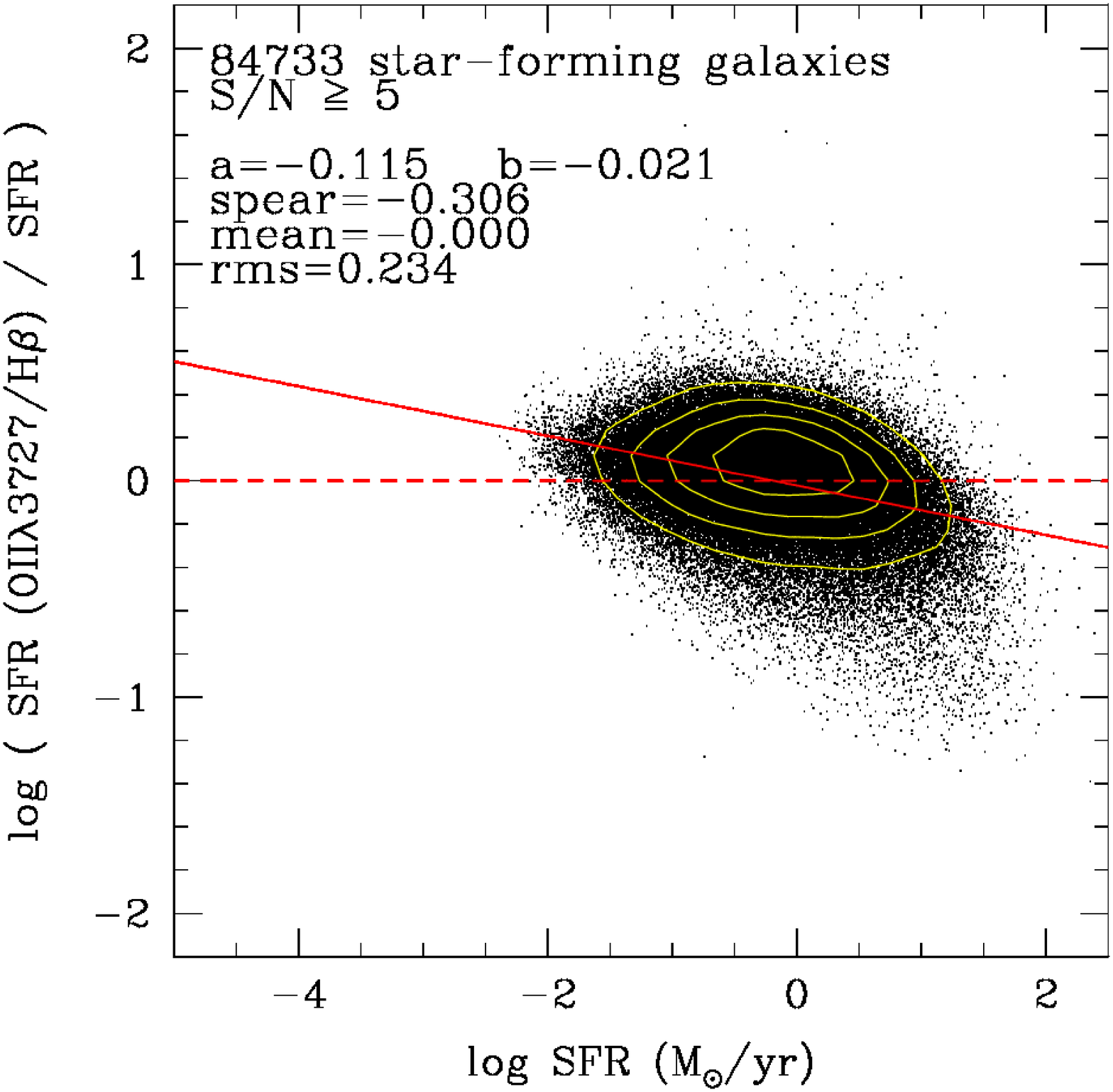}
\par\end{centering}

\caption{\emph{Left}: Relation between the {[}O\noun{ii}]/H$\alpha^{\mathrm{i}}$
(observed-to-intrinsic) and the {[}O\noun{ii}]/H$\beta$ (observed-to-observed)
line ratios for star-forming galaxies in the SDSS DR4 data. The solid
line is a least-square fit to the data (errors on $x$ and $y$).
The dashed line is the $\mathrm{H}\alpha=2.85\cdot\mathrm{H}\beta$
line. Isodensity contours are overplotted in white. \emph{Center and
right}: Same legend as in Fig.~\ref{fig:compstandard}. The studied
calibration are the new {[}O\noun{ii}]/H$\beta$ calibrations (center:
Eq.~\ref{eq:SFROIIHb}, right: Eq.~).}

\label{fig:newtwolines}
\end{figure*}

Finally we consider the general case where no estimation of the dust
attenuation is available. \citet{Moustakas:2006ApJ...642..775M} and
\citet{Weiner:2006astro.ph.10842W} have produced calibrations of
the {[}O\noun{ii}]/H$\alpha^{\mathrm{i}}$ line ratio against respectively
$B$-band or $H$-band absolute magnitudes. Following our idea of
using only emission-line measurements, we explore a possible correlation
between {[}O\noun{ii}]/H$\alpha^{\mathrm{i}}$ (observed-to-intrinsic)
and {[}O\noun{ii}]/H$\beta$ (observed-to-observed) line ratios. Fig.~\ref{fig:newtwolines}
(left) shows the best-fit relation, as defined by the following equation:{\small \begin{equation}
\begin{array}{cl}
\log\left([\mathrm{O}\mathsc{ii}]/\mathrm{H}\alpha^{\mathrm{i}}\right)= & 3.243(\pm0.3)\cdot\log\left([\mathrm{O}\mathsc{ii}]/\mathrm{H}\beta\right)-1.74(\pm0.09)\end{array}\label{eq:o2hanodust}\end{equation}
}The mean and the rms of the residuals around the fitted line are
respectively $-6\cdot10^{-10}$ dex and $0.37$ dex. We note that
the objects plotted upward the dashed line in Fig.~\ref{fig:newtwolines}
(left) have negative dust attenuation when we derive them from H$\alpha$/H$\beta$
ratio with the assumption of a constant intrinsic Balmer ratio.

Putting Eqs.~\ref{eq:defo2ha} and~\ref{eq:o2hanodust} together,
we obtain the following calibration formula:{\small \begin{equation}
\log\mathrm{SFR}=\epsilon_{\mathrm{H}\alpha}^{\mathrm{i}}\left(\log L(\mathrm{[O}\mathsc{ii}\mathrm{]})-3.243\cdot\log\left([\mathrm{O}\mathsc{ii}]/\mathrm{H}\beta\right)+1.74\right)-\log\eta_{\mathrm{H}\alpha}^{\mathrm{i}}\label{eq:SFROIIHb}\end{equation}
}{\small \par}

Fig.\textbf{~}\ref{fig:newtwolines} (center) shows the quality of
the recovered SFR as compared to the reference CL01 SFR. We see that
our new calibration show no significant systematic shift ($0.02$
dex), and no significant residual slope ($0.01$ with a Spearman rank
correlation coefficient of $0.06$). Moreover our new calibration
shows a dispersion of $0.29$ dex which, while being still higher
than the results obtained with a correction for dust attenuation,
is the best result obtained in Sect.~\ref{sec:SFR-calibration-without}
between all other possibilities: assumed mean dust correction, direct
calibration with observed line luminosities, or self-consistent correction
with the absolute magnitude.

We also explore the results obtained by doing a direct least-square
fit of the SFR as a function of the {[}O\noun{ii}] line luminosity
and the {[}O\noun{ii}]/H$\beta$ line ratio, without assuming the
underlying H$\alpha^{\mathrm{i}}$ calibration. We find the following
second-degree best-fit relation:{\small \begin{equation}
\begin{array}{r}
\log\mathrm{SFR}=-39.21+0.982\cdot\log L(\mathrm{[O}\mathsc{ii}\mathrm{]})-2.396\cdot\log\left([\mathrm{O}\mathsc{ii}]/\mathrm{H}\beta\right)\\
-0.3354\cdot\log\left([\mathrm{O}\mathsc{ii}]/\mathrm{H}\beta\right)^{2}\end{array}\label{eq:SFROIIHb2}\end{equation}
}{\small \par}

Fig.\textbf{~}\ref{fig:newtwolines} (right) shows the quality of
this other calibration as compared to the reference CL01 SFR. As compared
to Eq.~\ref{eq:SFROIIHb}, the SFR derived with Eq.~\ref{eq:SFROIIHb2}
show a null systematic shift and a smaller dispersion of $0.23$ dex.
Nevertheless this smaller dispersion is diminished by a residual slope
of $-0.12$ (Spearman rank correlation coefficient of $-0.31$).

Because it is much easier to use and gives better results, our calibrations
based on only two observed line fluxes ({[}O\noun{ii}] and H$\beta$)
should be preferred in most cases with no available correction for
dust attenuation. \emph{The minimum uncertainty when dealing with
data not corrected for dust attenuation is then only $0.23$ dex.
}We discuss in the following section the different cases where it
is better to use Eq.~\ref{eq:SFROIIHb} or Eq.~\ref{eq:SFROIIHb2}.

\subsection{The self-consistent correction on samples with different properties\label{sub:selfbaddust}}

It is clear at this stage of our study that all the calibrations derived
or studied in Sect.~\ref{sec:SFR-calibration-without} (including
those by \citealp{Moustakas:2006ApJ...642..775M} and \citealp{Weiner:2006astro.ph.10842W}),
which are based on observed quantities, depend on the properties of
the reference sample. Nevertheless, we emphasize that such statistically
significant sample such as the SDSS DR4 data, with a large selection
function, is not expected to be strongly biased in favor of one particular
population of galaxies.

It might be useful however to know how these calibrations are sensible
to the variations of the properties of the studied sample, especially
in terms of dust or metallicity.

\subsubsection{Dependence on dust}

\begin{figure*}
\begin{centering}
\includegraphics[width=0.19\linewidth]{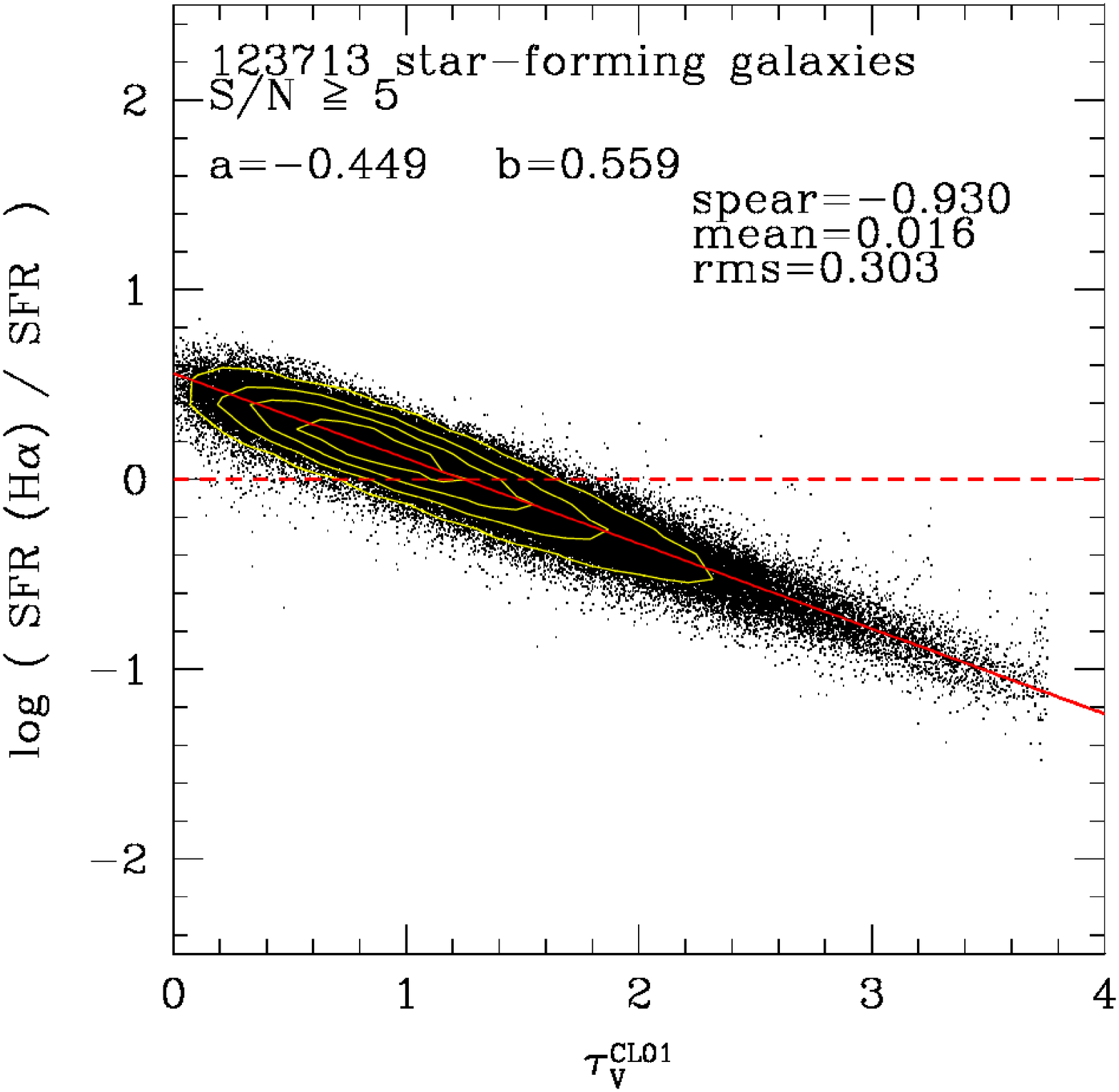}
\includegraphics[width=0.19\linewidth]{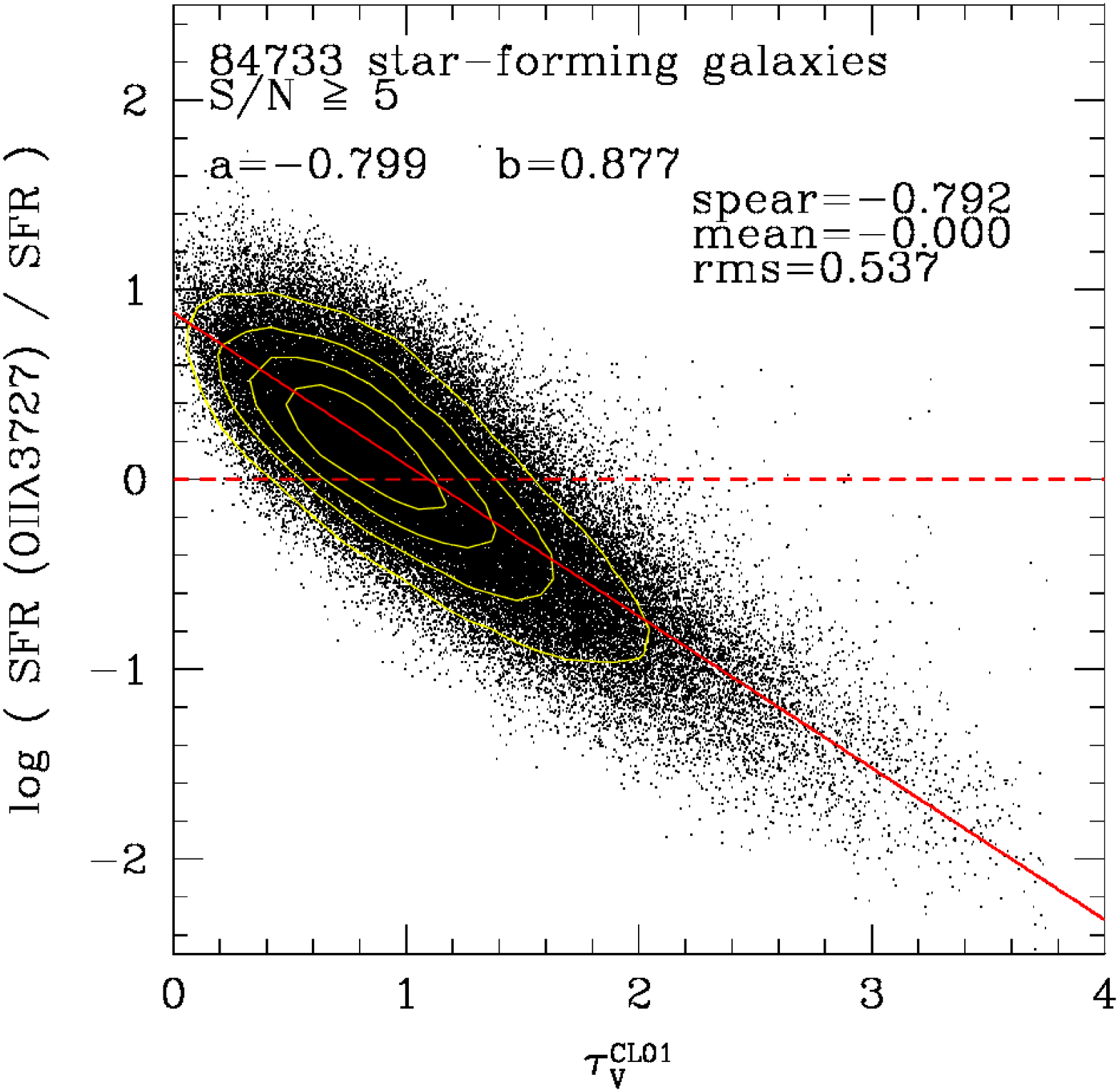}
\includegraphics[width=0.19\linewidth]{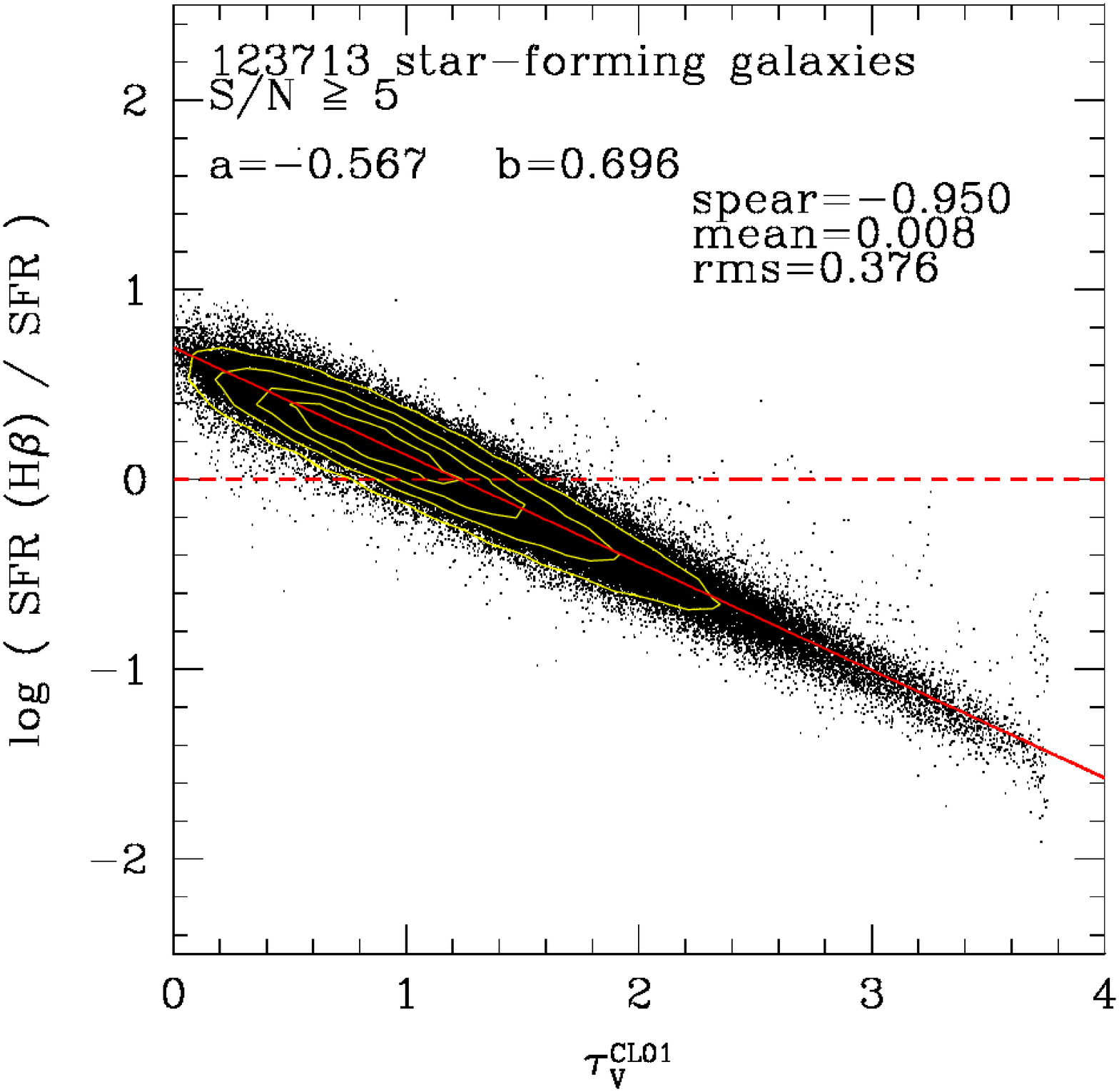}
\includegraphics[width=0.19\linewidth]{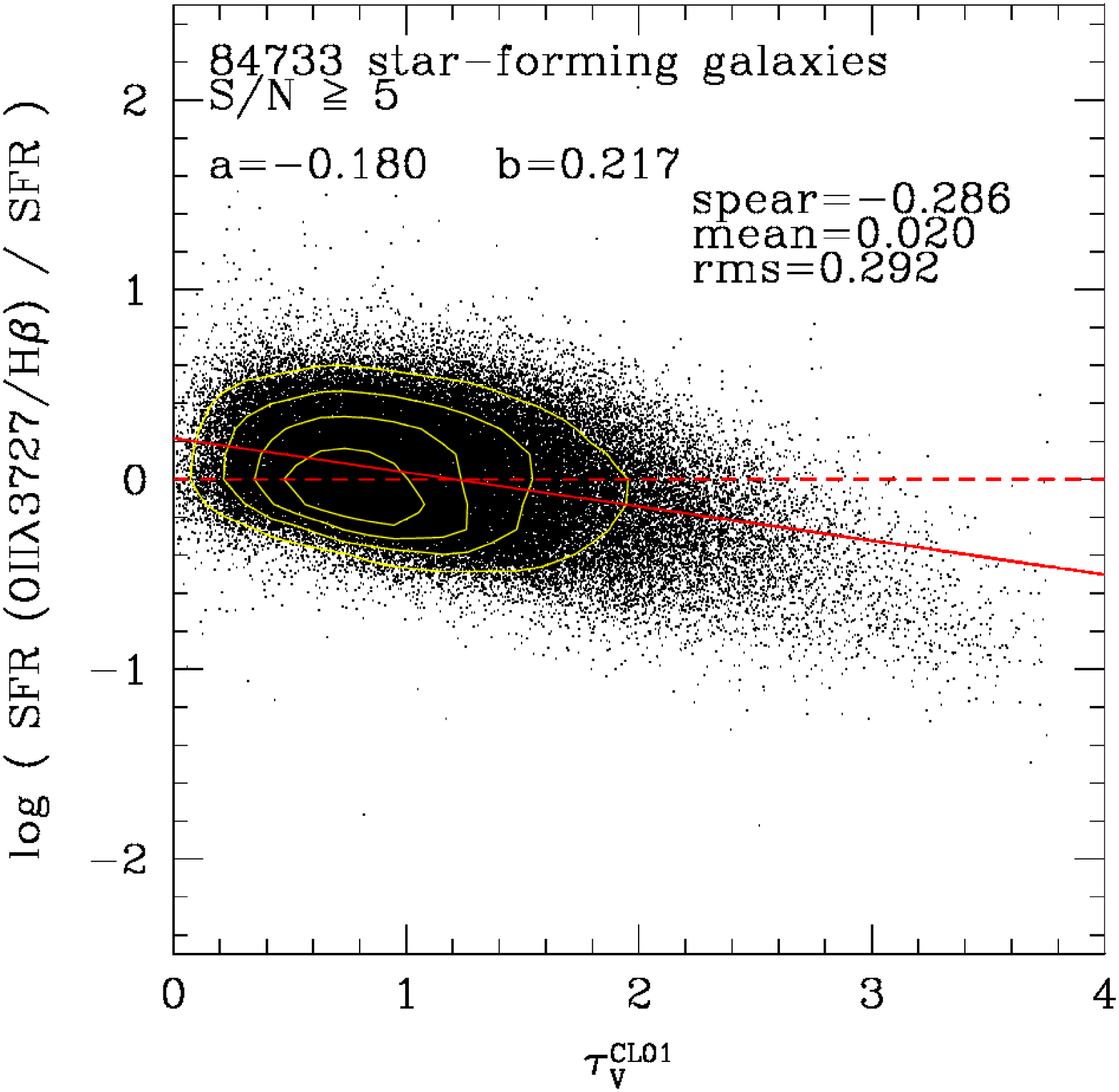}
\includegraphics[width=0.19\linewidth]{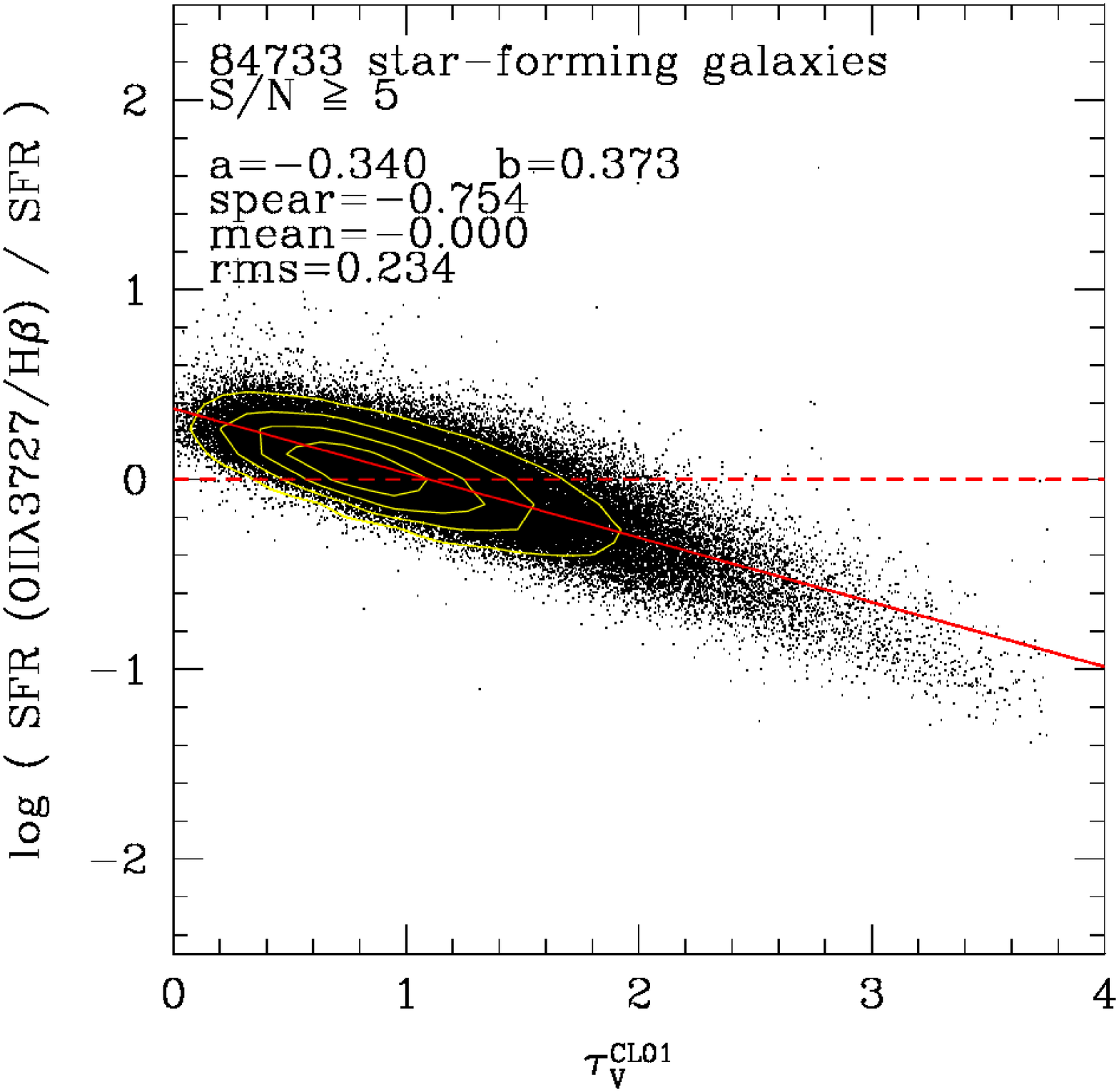}
\includegraphics[width=0.19\linewidth]{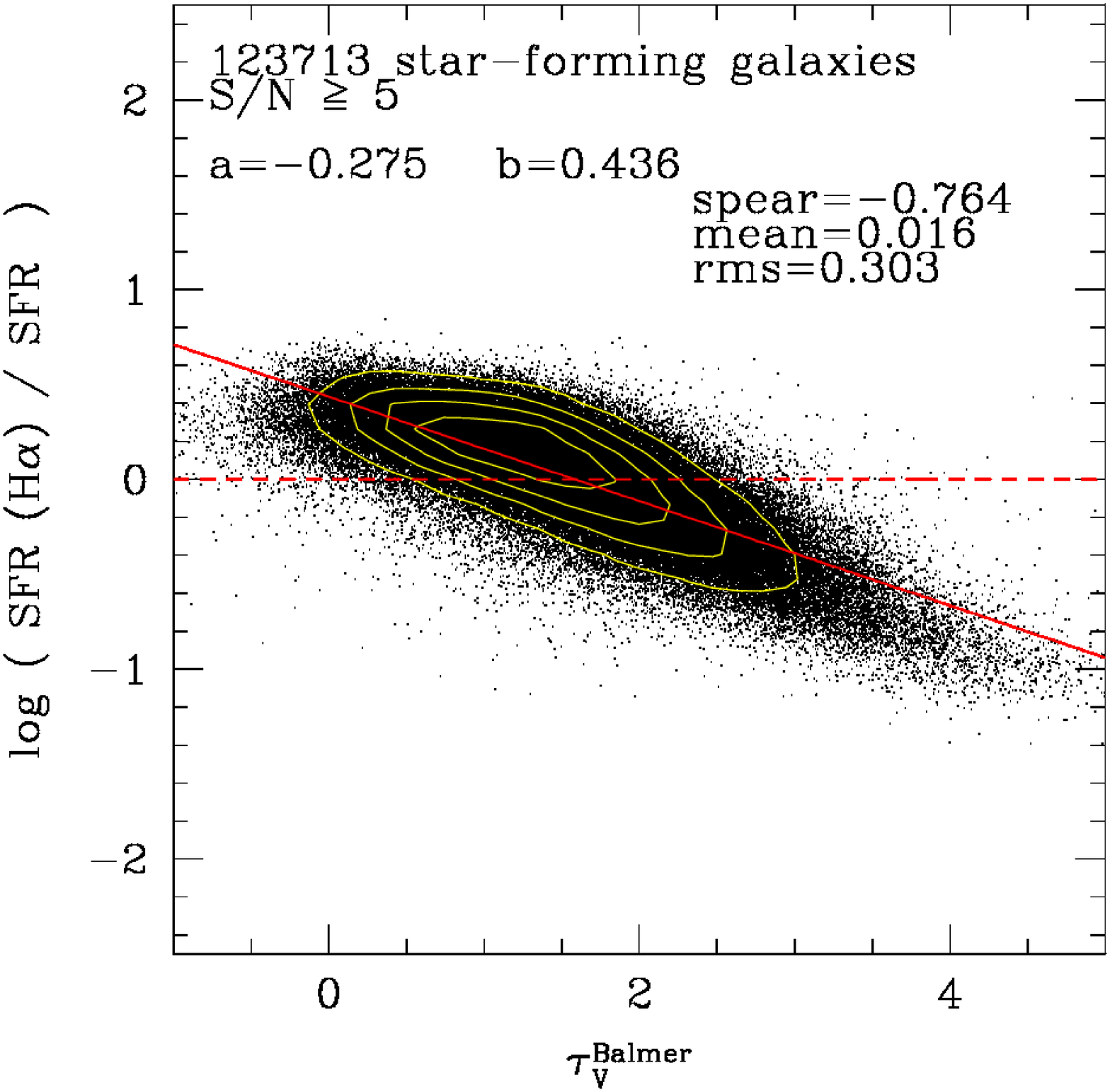}
\includegraphics[width=0.19\linewidth]{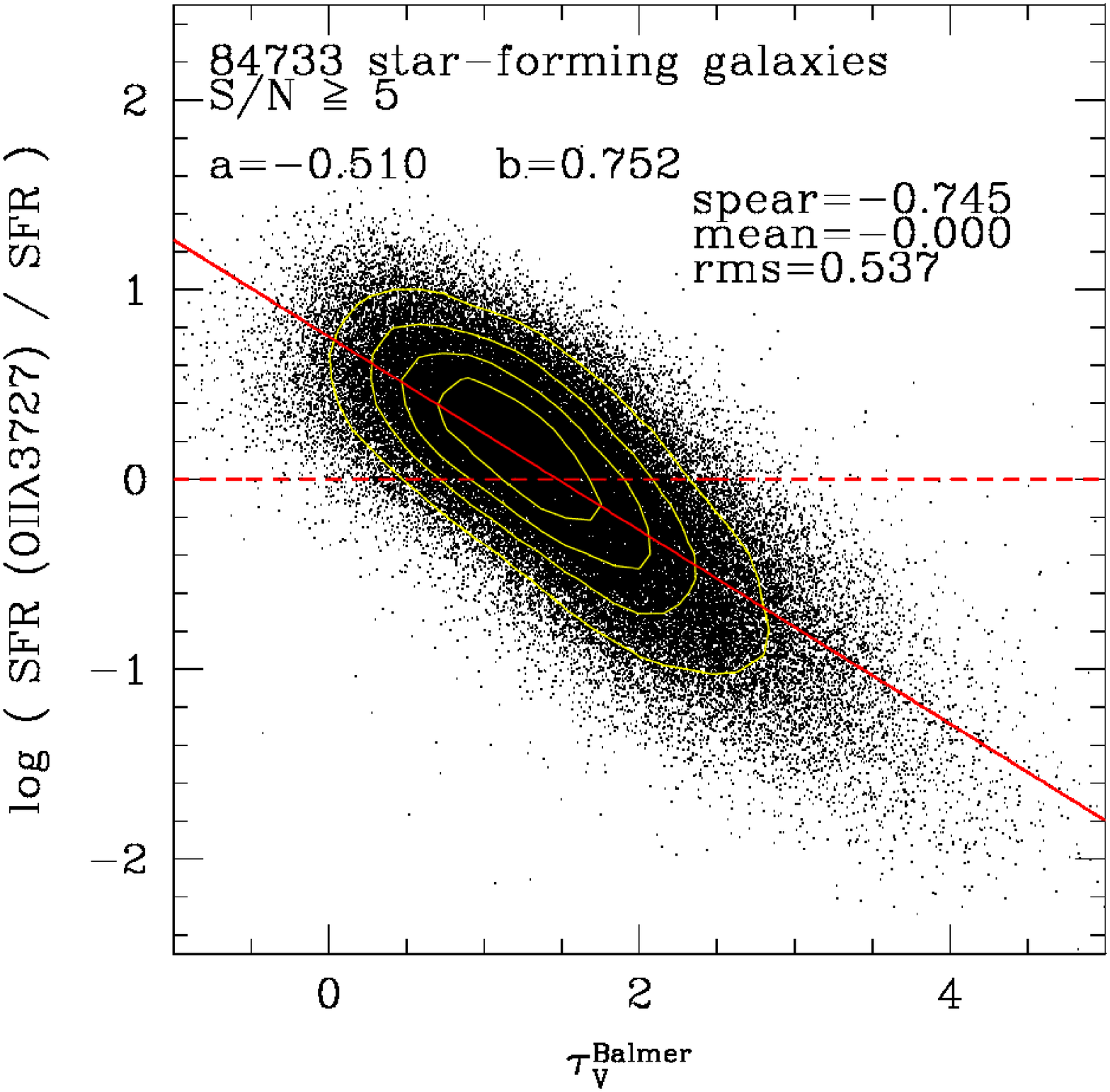}
\includegraphics[width=0.19\linewidth]{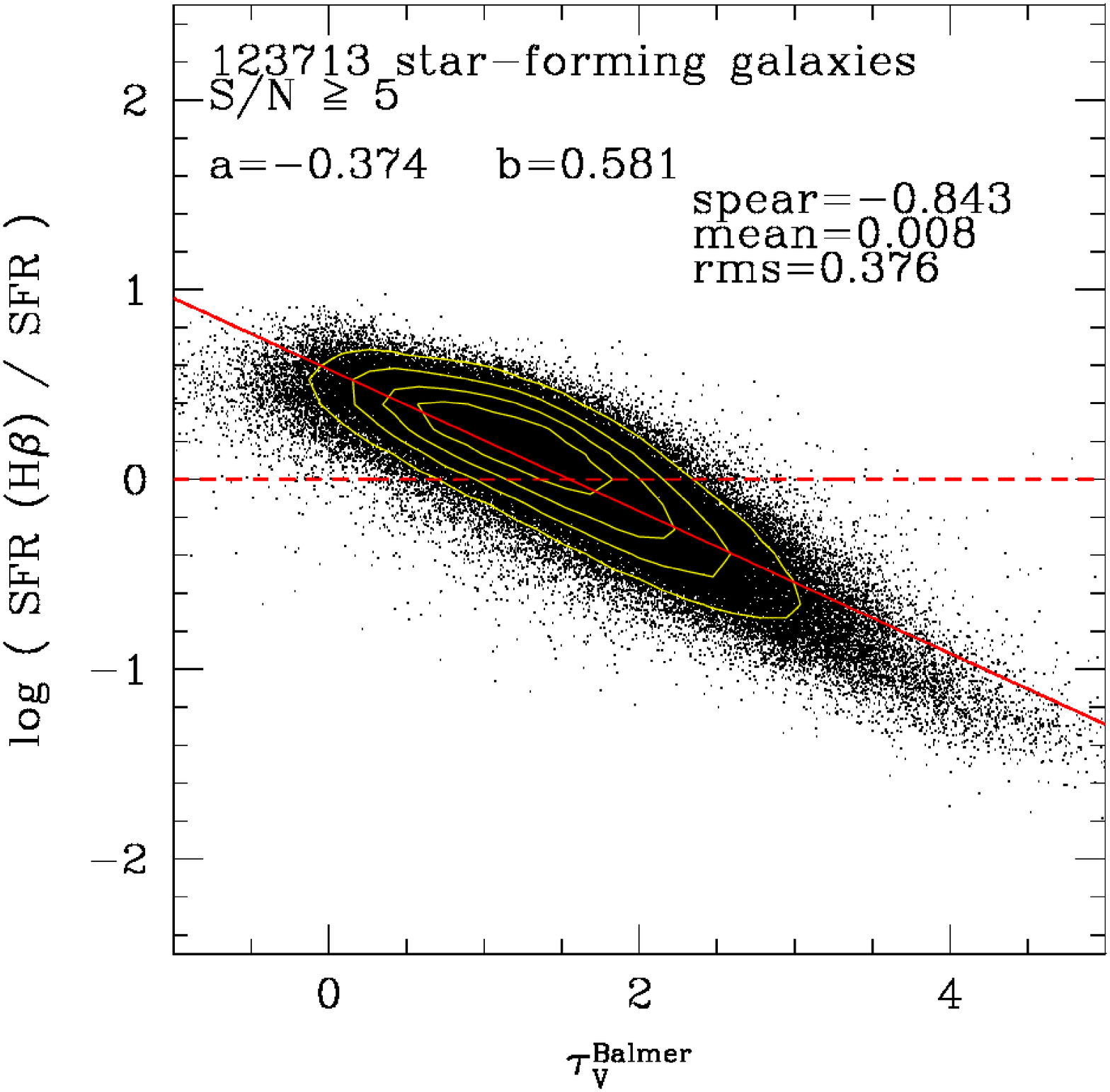}
\includegraphics[width=0.19\linewidth]{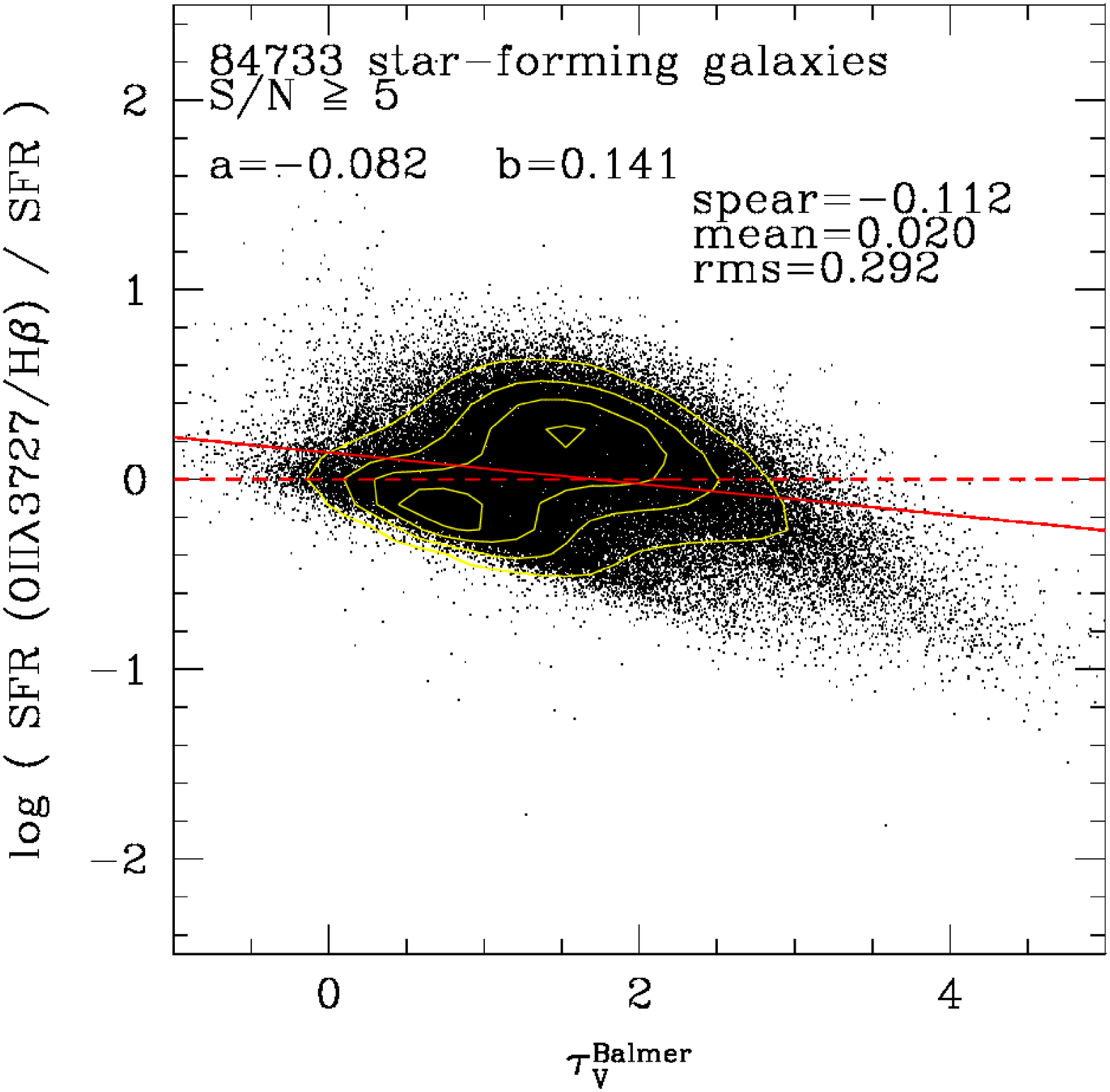}
\includegraphics[width=0.19\linewidth]{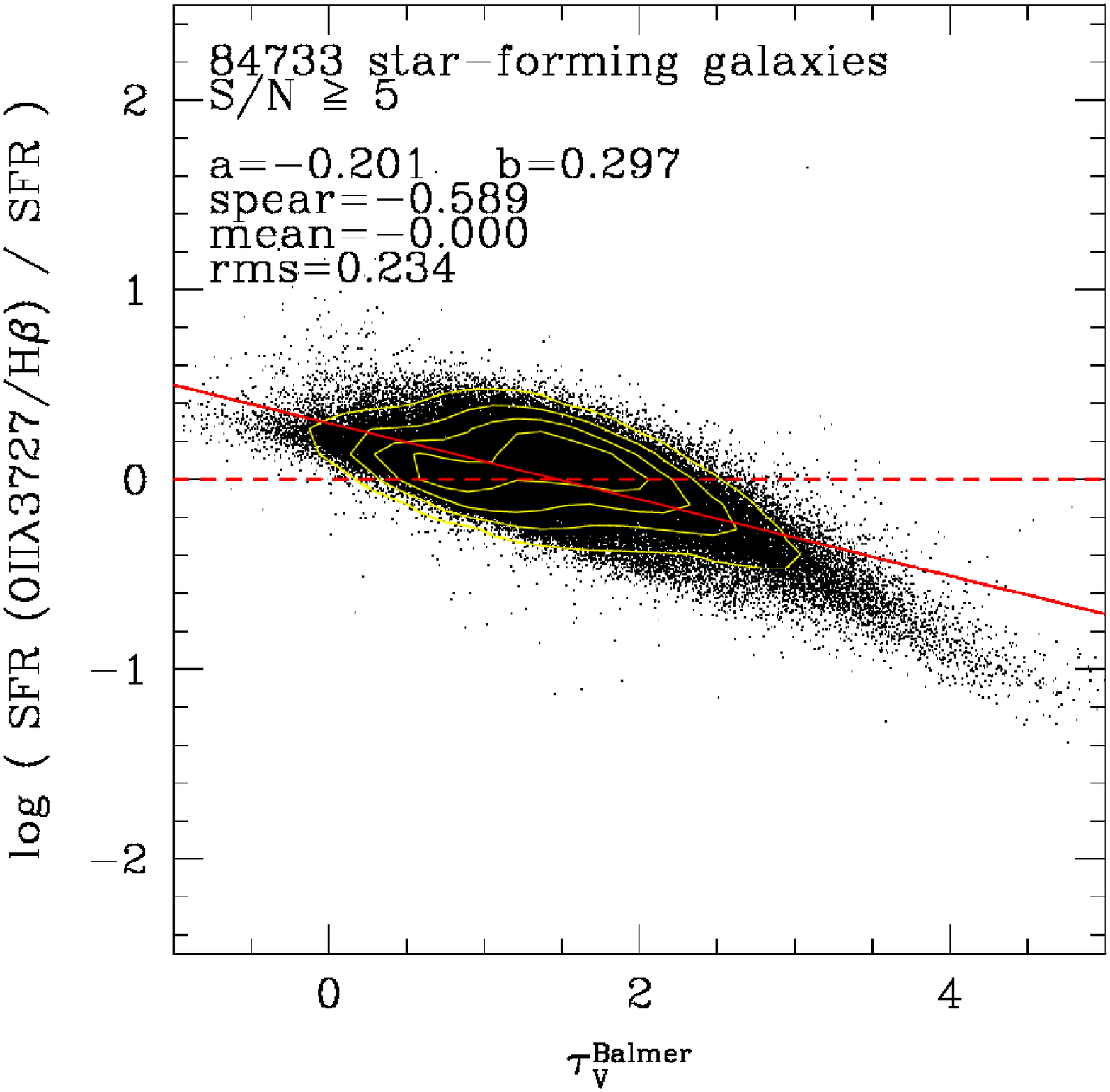} 
\par\end{centering}

\caption{Same legend as in Fig.~\ref{fig:compstandard}, except that the residuals
of the SFR calibration are now plotted as a function of the dust attenuation
estimated with the CL01 method (top) or with the Balmer decrement
method (bottom). The studied calibrations are, from left to right:
H$\alpha$, {[}O\noun{ii}], H$\beta$, {[}O\noun{ii}]/H$\beta$ (Eq.~\ref{eq:SFROIIHb}),
and {[}O\noun{ii}]/H$\beta$ (Eq.~\ref{eq:SFROIIHb2}). All calibrations
are related to observed quantities.}

\label{fig:compobstauC}
\end{figure*}

\begin{table}

\caption{Coefficients to be used to correct the SFR derived from an observed
calibration, as a function of the dust attenuation following Eq.~\ref{eq:corrSFRTau}.
The coefficients are given for a dust attenuation estimated with the
CL01 or the Balmer decrement method.}

\label{tab:corrTau}

\begin{centering}
\begin{tabular}{lrrrr}
\hline 
\hline
\hfill{}method & \multicolumn{2}{c}{CL01} & \multicolumn{2}{c}{Balmer}\tabularnewline
calibration & $a$ & $b$ & $a$ & $b$\tabularnewline
\hline
H$\alpha$ & $-0.45$ & $0.56$ & $-0.28$ & $0.44$\tabularnewline
{[}O\noun{ii}] & $-0.80$ & $0.88$ & $-0.51$ & $0.75$\tabularnewline
H$\beta$ & $-0.57$ & $0.70$ & $-0.37$ & $0.58$\tabularnewline
{[}O\noun{ii}]/H$\beta$ (Eq.~\ref{eq:SFROIIHb}) & $-0.18$ & $0.22$ & $-0.082$ & $0.14$\tabularnewline
{[}O\noun{ii}]/H$\beta$ (Eq.~\ref{eq:SFROIIHb2}) & $-0.34$ & $0.37$ & $-0.20$ & $0.30$\tabularnewline
\hline
\end{tabular}
\par\end{centering}

\end{table}

\begin{figure}
\begin{centering}
\includegraphics[width=0.49\columnwidth]{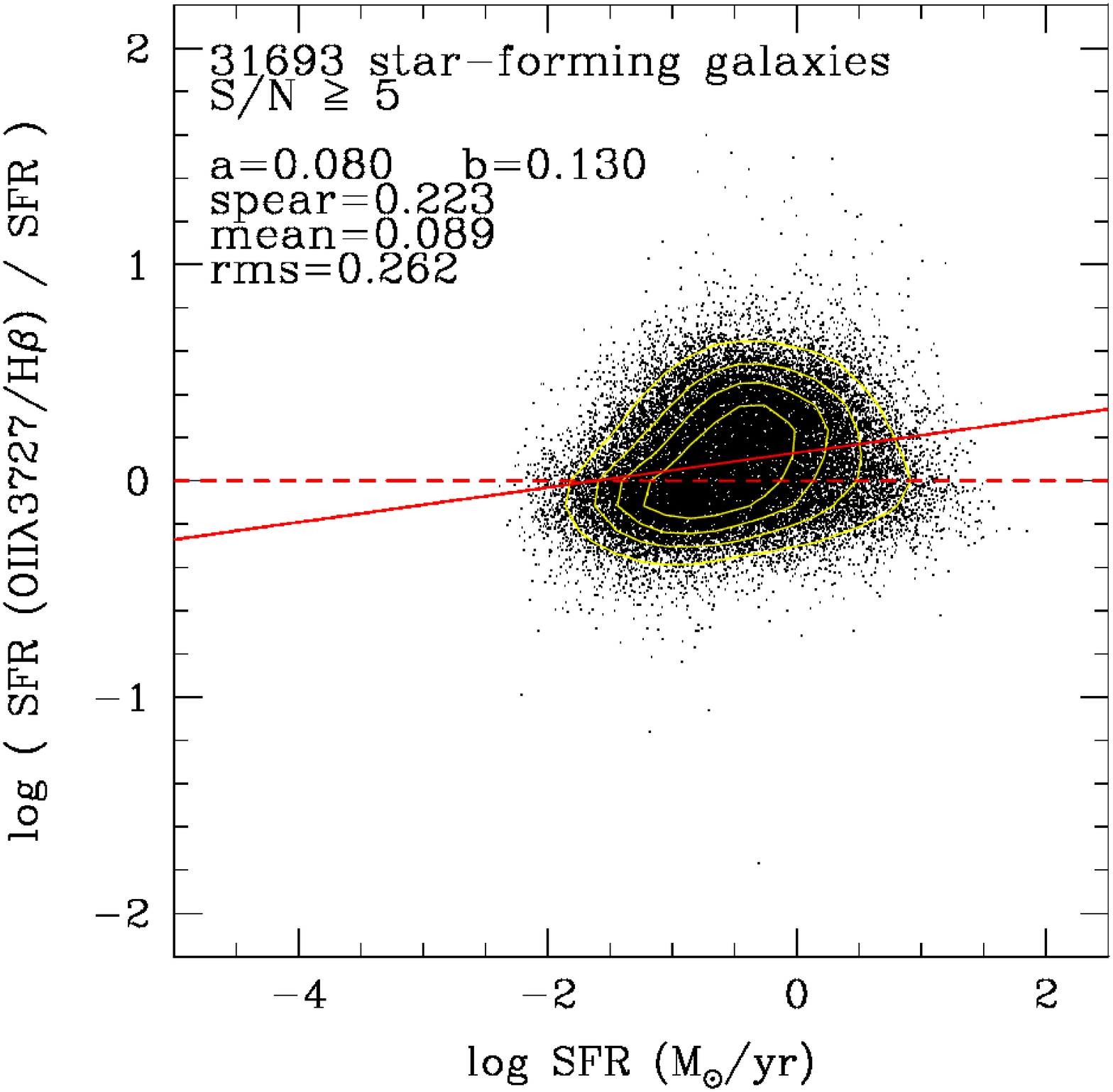}
\includegraphics[width=0.49\columnwidth]{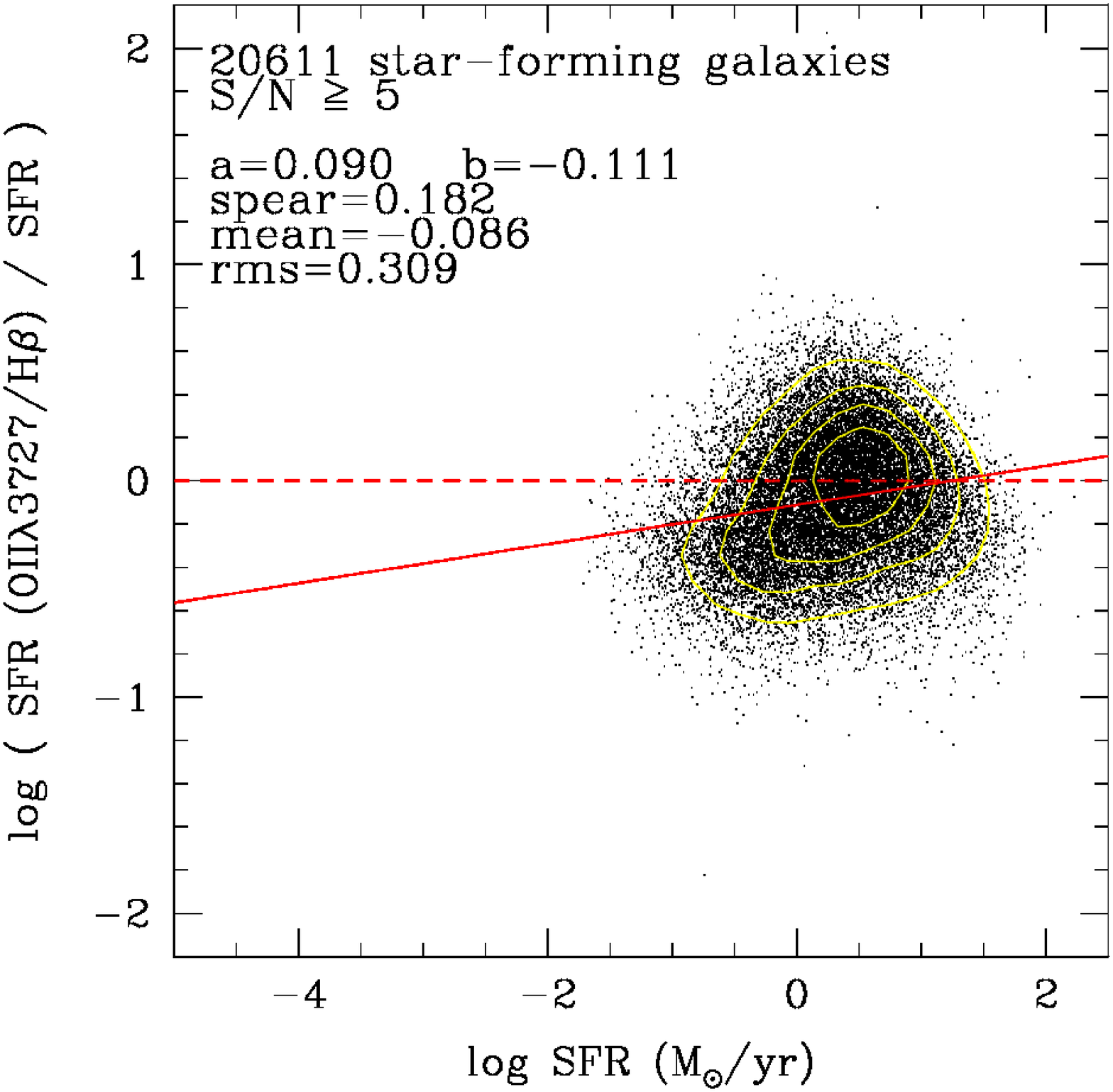}
\includegraphics[width=0.49\columnwidth]{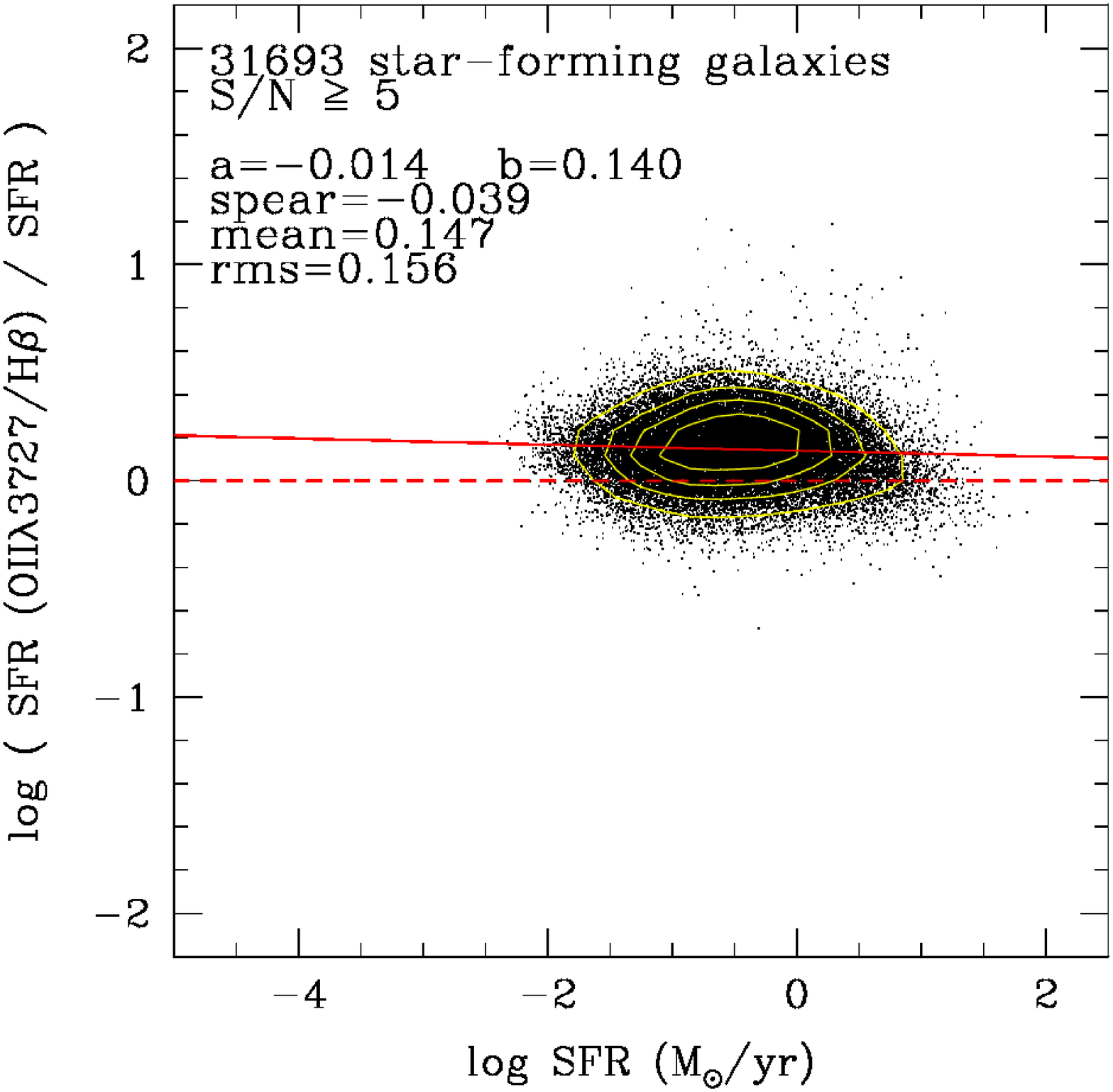}
\includegraphics[width=0.49\columnwidth]{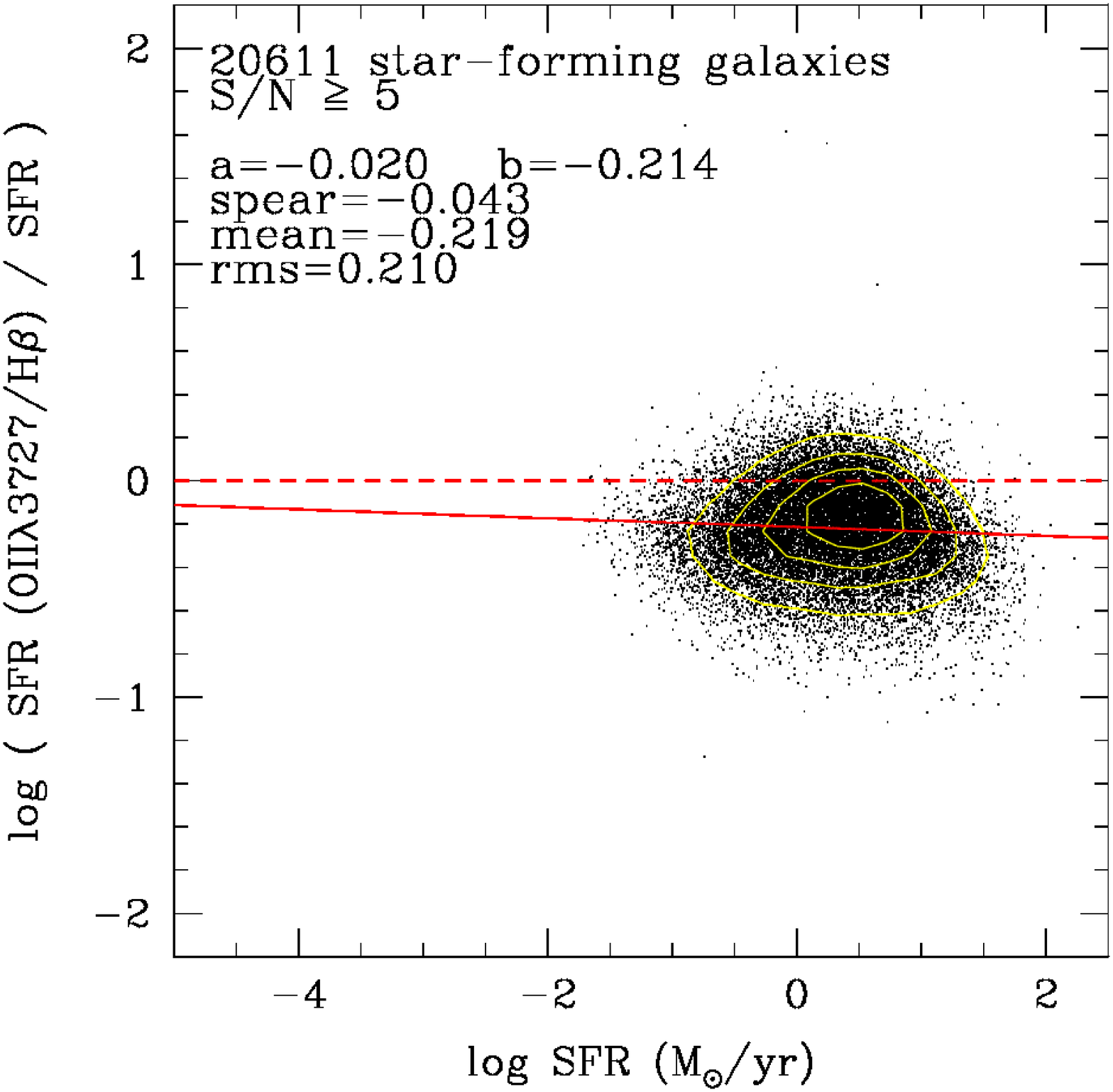}
\par\end{centering}

\caption{Same legend as in Fig.~\ref{fig:compstandard}. The studied calibration
is the new {[}O\noun{ii}]/H$\beta$ calibration given by Eq.~\ref{eq:SFROIIHb}
(top) or Eq.~\ref{eq:SFROIIHb2} (bottom), applied on the two dust
sub-samples defined in Sect.~\ref{sec:Description-of-the}. Left:
$\left\langle \tau_{V}^{\mathrm{CL01}}\right\rangle =0.63$, right:
$\left\langle \tau_{V}^{\mathrm{CL01}}\right\rangle =1.84$.}

\label{fig:compsubsample}
\end{figure}

We now study in Fig.~\ref{fig:compobstauC} the residuals of our
new calibrations, as compared to the reference CL01 SFR, as a function
of the dust attenuation estimated either with the CL01 method, or
with the assumption of a constant intrinsic Balmer ratio. This has
been done for the poor quality single-line calibrations (H$\alpha$,
{[}O\noun{ii}], and H$\beta$, see Fig.~\ref{fig:compobs}), and
for the better quality {[}O\noun{ii}]/ H$\beta$ calibration (see
Fig.~\ref{fig:newtwolines} center and right).

In all cases, the means of the residuals are very low, which is expected
as we always derived the best-fit calibration.

For the single-line calibrations, we obtain very significant correlations
between the residuals and the dust attenuation, with Spearman rank
correlation coefficients of the order of $-0.75$ to $-0.95$. Thus,
we can derive linear relations which may be used to correct these
calibrations for a desired studied sample, for which one has a rough
estimate of the mean dust attenuation. The corrective formula as derived
from Fig.~\ref{fig:compobstauC} is:\begin{equation}
\log\left(\mathrm{SFR}^{\mathrm{corr}}\right)=\log\left(\mathrm{SFR}\right)+a\times\tau_{V}+b\label{eq:corrSFRTau}\end{equation}

The $a$ and $b$ coefficients are summarized in Table~\ref{tab:corrTau}.
We note that this correction formula give a null correction for a
dust attenuation equal to the mean value in the SDSS DR4 sample: $\tau_{V}^{\mathrm{CL01}}=1.21$
in the H$\alpha$ and H$\beta$ cases, and for $\tau_{V}^{\mathrm{CL01}}=1.1$
in the {[}O\noun{ii}] case.

For the {[}O\noun{ii}]/H$\beta$ two-lines calibration given in Eq.~\ref{eq:SFROIIHb},
we see that the residuals do not show any more any strong significant
correlation with the dust attenuation: smaller slopes, and smaller
Spearman rank correlation coefficients of the order of $-0.1$ to
$-0.3$. This result shows that the {[}O\noun{ii}]/H$\beta$ calibration
given in Eq.~\ref{eq:SFROIIHb} is not significantly sensitive to
variations in the dust attenuation. This calibration can thus reliably
been used in any sample without applying a correction. This conclusion
is not true for the {[}O\noun{ii}]/H$\beta$ calibration given in
Eq.~\ref{eq:SFROIIHb2} which shows similar, but lower, residual
slopes as compared to single-line calibrations.

Fig.~\ref{fig:compsubsample} shows the {[}O\noun{ii}]/H$\beta$
calibrations applied on the two sub-samples defined in Sect.~\ref{sec:Description-of-the},
with two different dust properties: $\left\langle \tau_{V}^{\mathrm{CL01}}\right\rangle =0.63$
and $\left\langle \tau_{V}^{\mathrm{CL01}}\right\rangle =1.84$. It
confirms that the calibration defined in Eq.~\ref{eq:SFROIIHb} is
reliable when applied on samples with different dust properties: the
dispersion does not significantly changes, and the observed residual
slopes are not significant (Spearman rank correlation coefficients
of the order of $0.2$). In both cases, the observed systematic shifts
are in agreement with the correction formula given in Eq.~\ref{eq:corrSFRTau}
and the coefficients given in Table~\ref{tab:corrTau}.

\subsubsection{Dependence on metallicity}

\begin{figure*}
\begin{centering}
\includegraphics[width=0.19\linewidth]{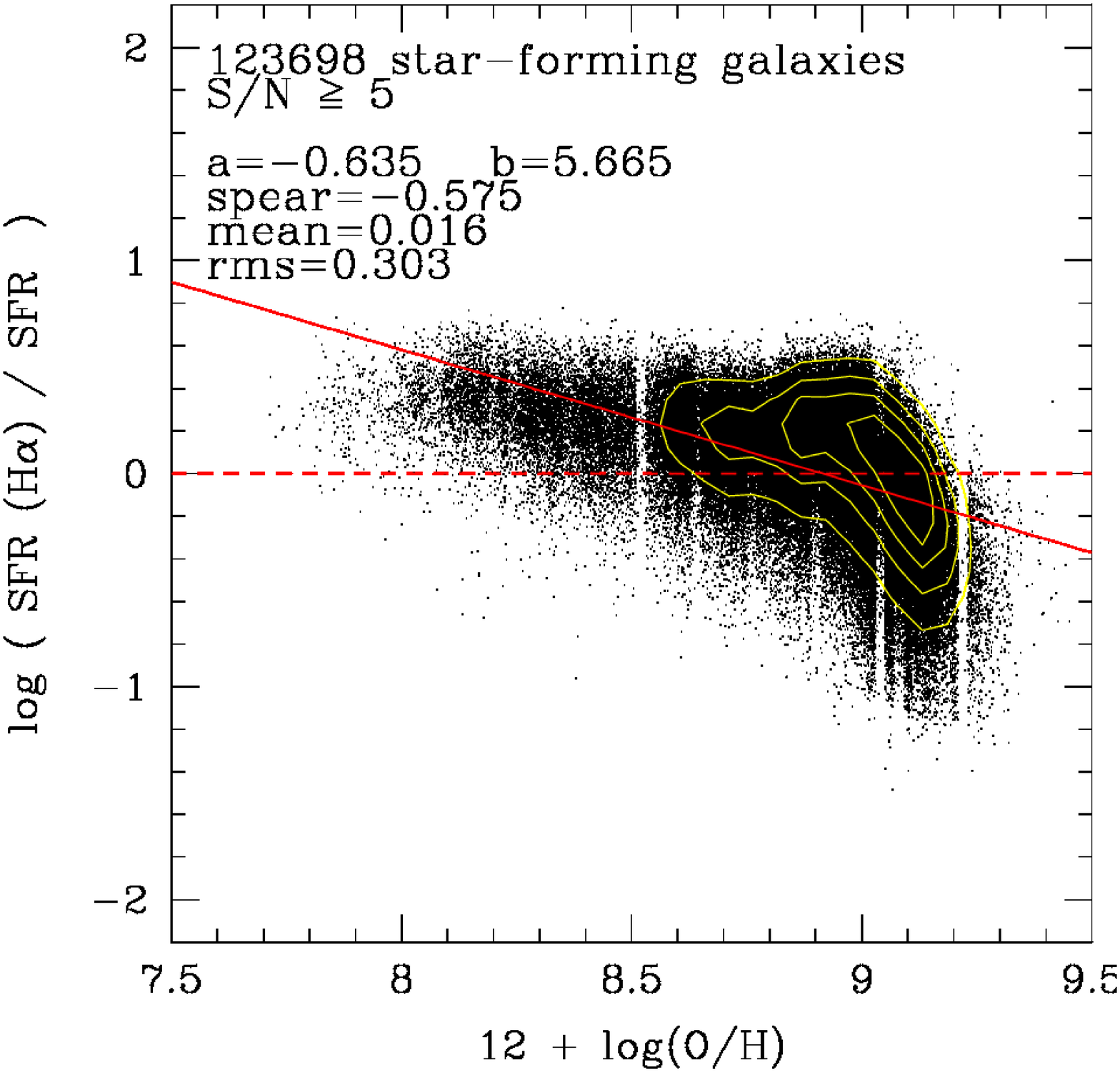}
\includegraphics[width=0.19\linewidth]{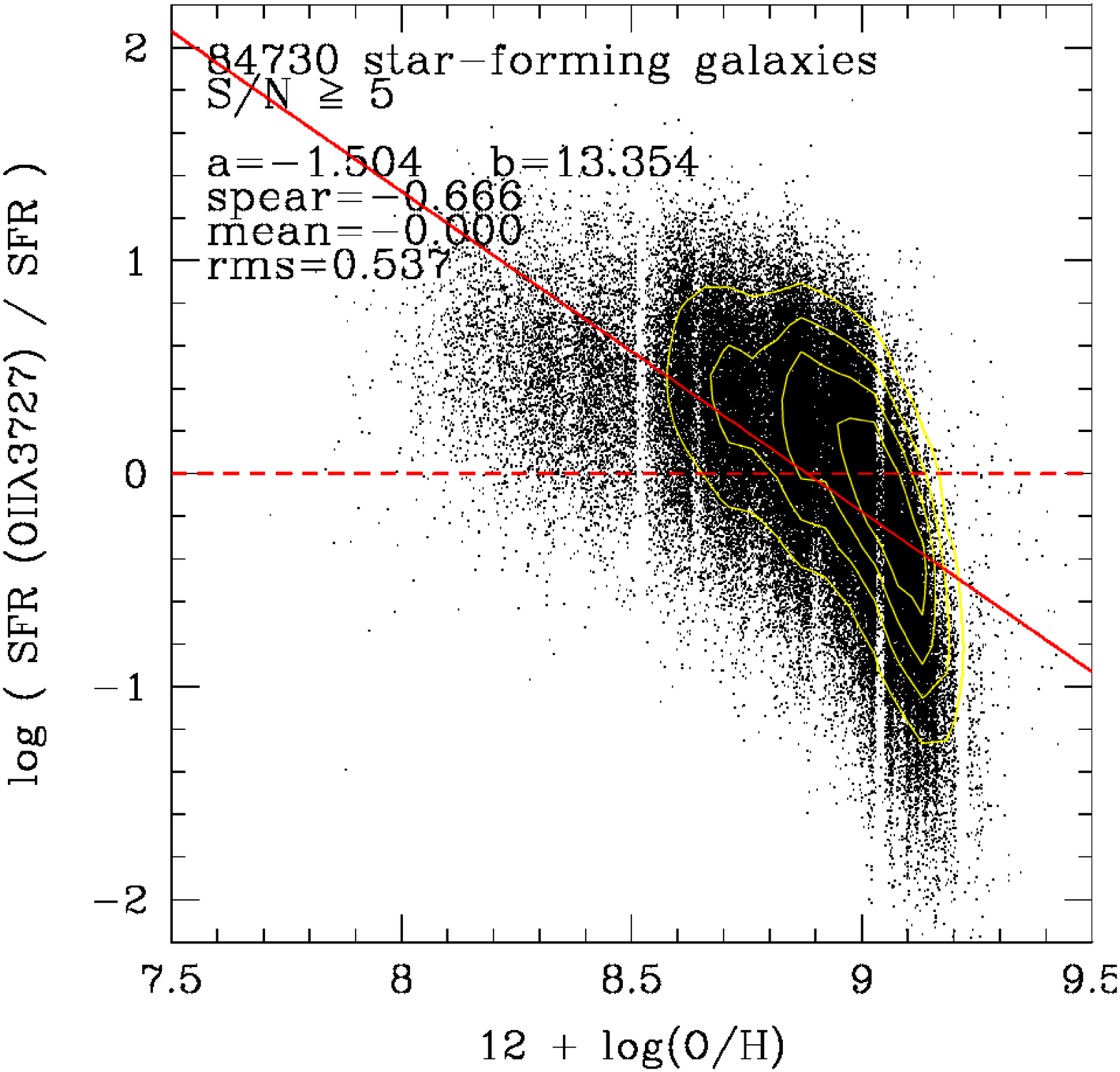}
\includegraphics[width=0.19\linewidth]{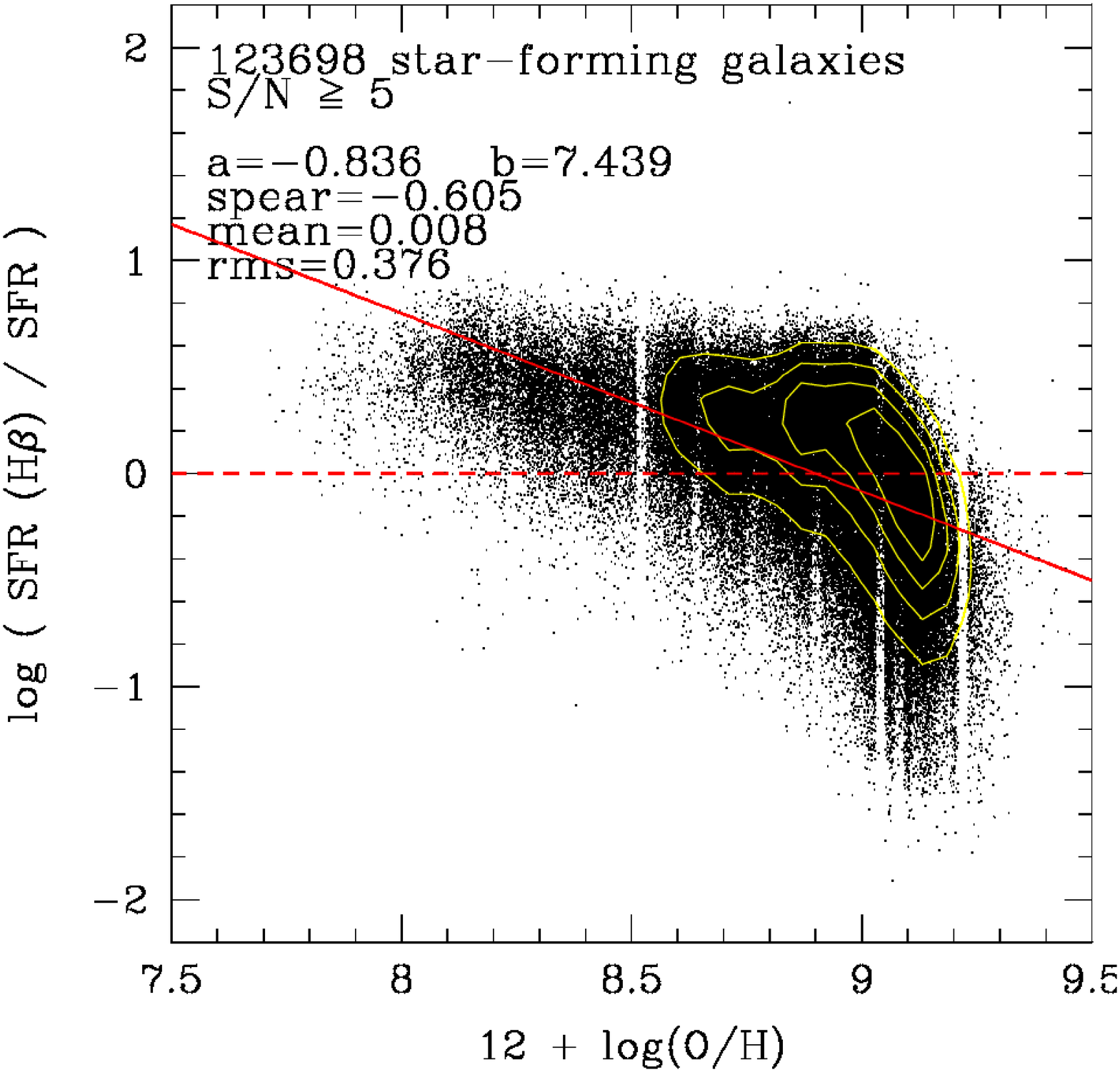}
\includegraphics[width=0.19\linewidth]{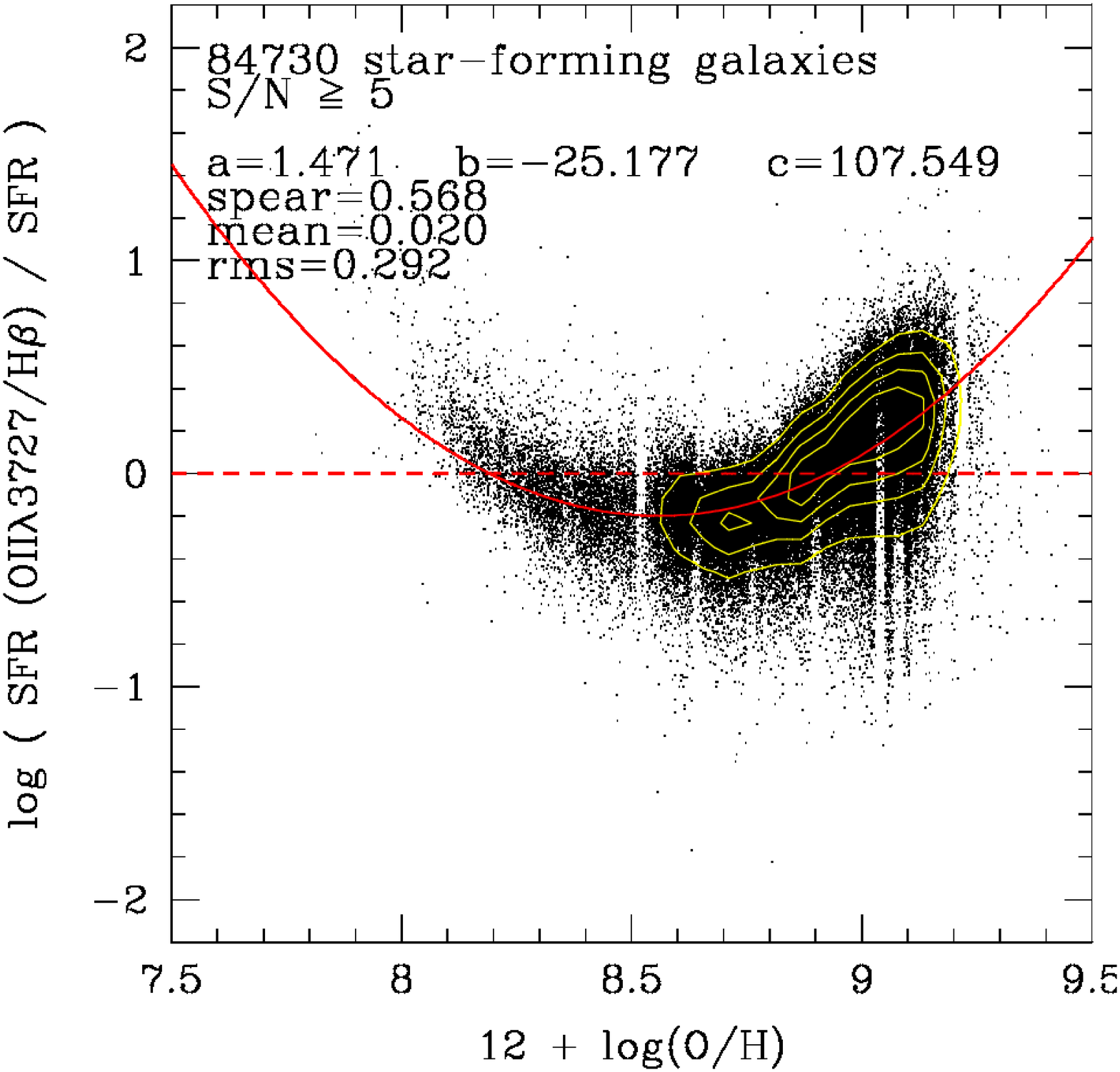}
\includegraphics[width=0.19\linewidth]{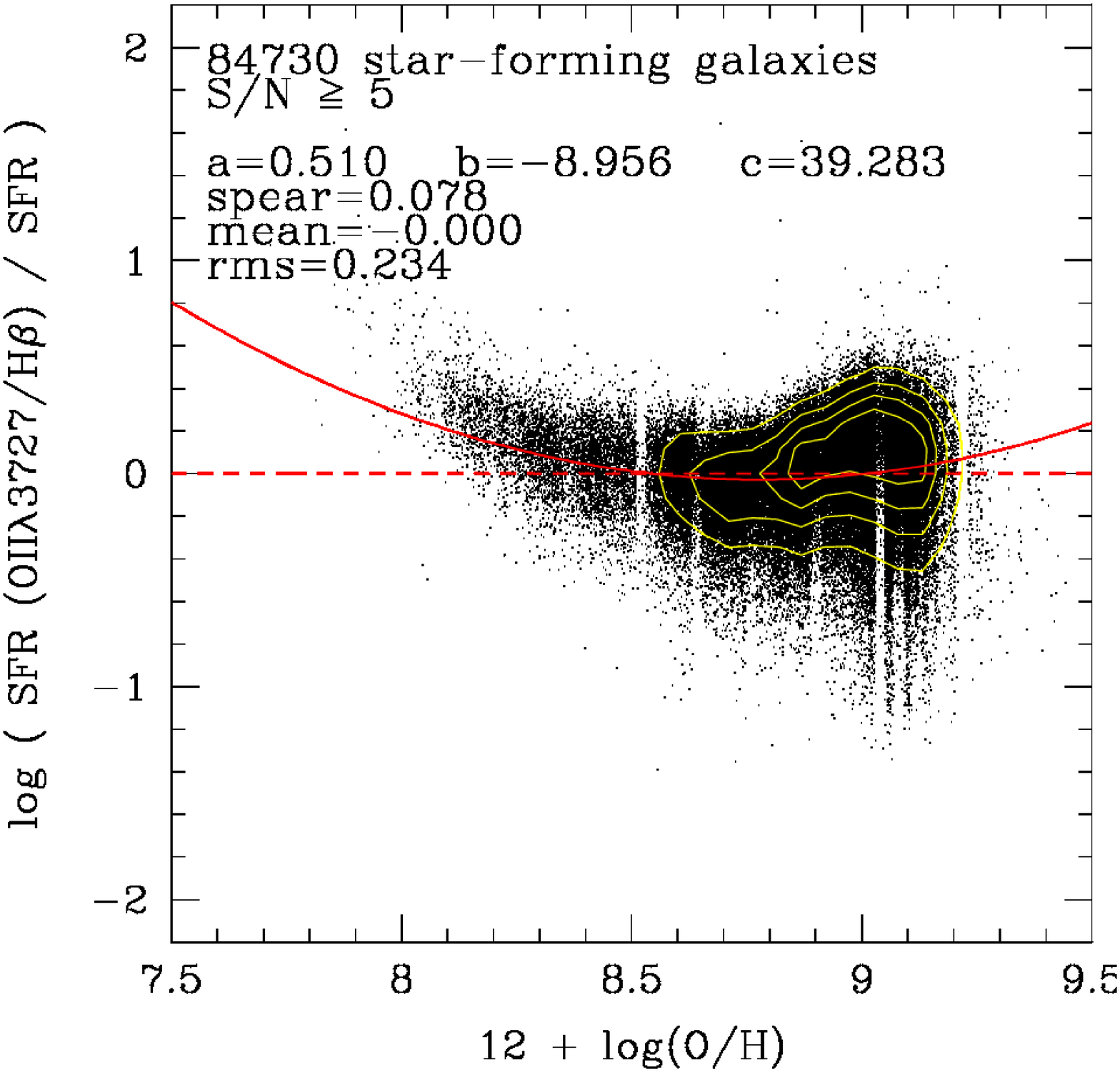}
\par\end{centering}

\caption{Same legend as in Fig.~\ref{fig:compstandard}, except that the residuals
of the SFR calibration are now plotted as a function of the gas-phase
oxygen abundance estimated with the CL01 method. The studied calibrations
are, from left to right: H$\alpha$, {[}O\noun{ii}], H$\beta$, {[}O\noun{ii}]/H$\beta$
(Eq.~\ref{eq:SFROIIHb}), and {[}O\noun{ii}]/H$\beta$ (Eq.~\ref{eq:SFROIIHb2}).
All calibrations are related to observed quantities.}

\label{fig:compobsmeta}
\end{figure*}

\begin{table}
\caption{Coefficients to be used to correct the SFR derived from an observed
calibration, as a function of the gas-phase oxygen abundance following
Eq.~. (first three calibrations) or Eq.~ (two last calibrations).}

\label{tab:corrMeta}

\begin{centering}
\begin{tabular}{lrrr}
\hline 
\hline
calibration & $a$ & $b$ & $c$\tabularnewline
\hline
H$\alpha$ & $-0.53$ & $4.80$ & \tabularnewline
{[}O\noun{ii}] & $-1.50$ & $13.35$ & \tabularnewline
H$\beta$ & $-0.74$ & $6.67$ & \tabularnewline
{[}O\noun{ii}]/H$\beta$ (Eq.~\ref{eq:SFROIIHb}) & $1.47$ & $-25.18$ & $107.5$\tabularnewline
{[}O\noun{ii}]/H$\beta$ (Eq.~\ref{eq:SFROIIHb2}) & $0.51$ & $-8.96$ & $39.3$\tabularnewline
\hline
\end{tabular}
\par\end{centering}
\end{table}

\begin{figure}
\begin{centering}
\includegraphics[width=0.49\columnwidth]{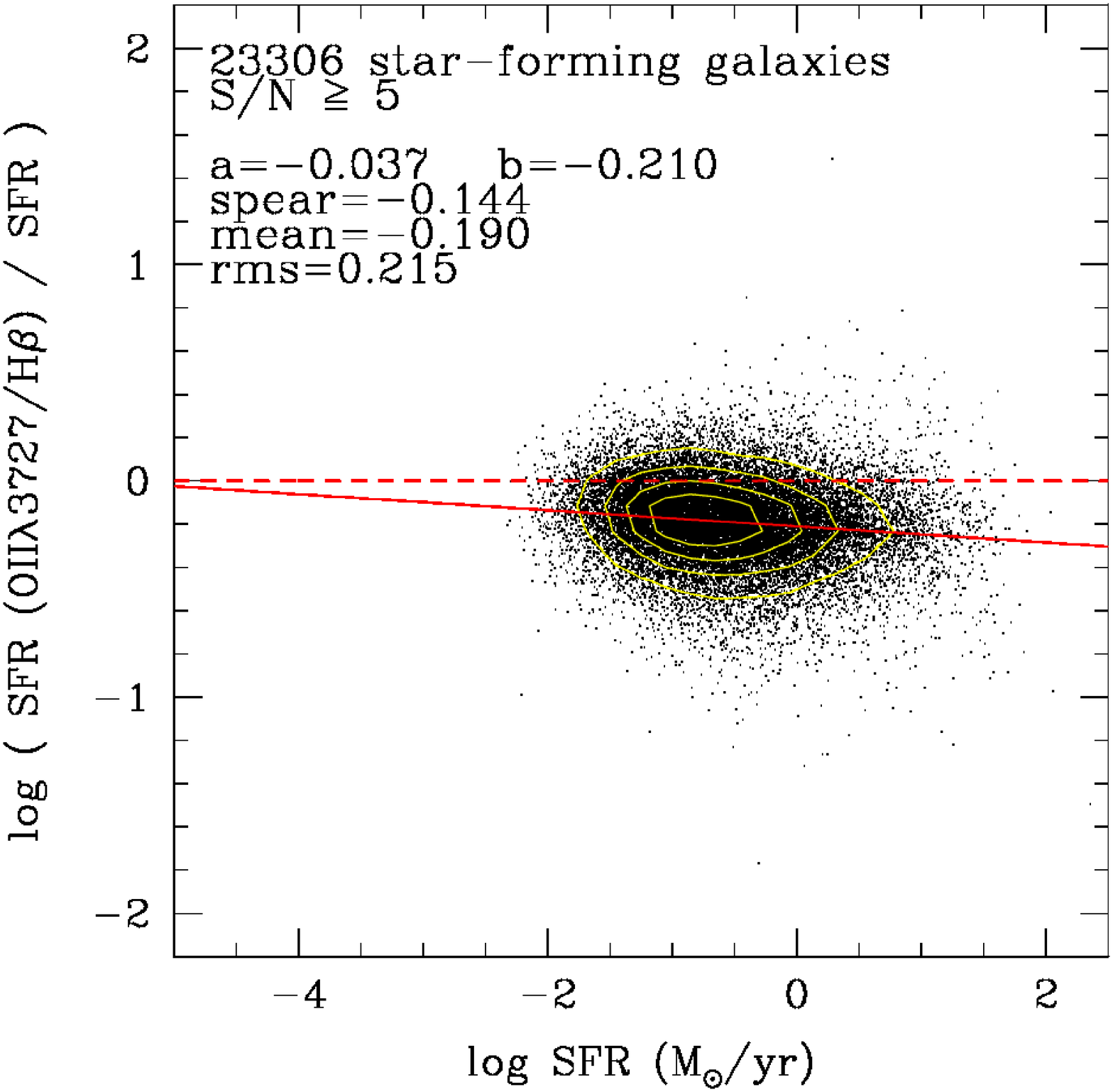}
\includegraphics[width=0.49\columnwidth]{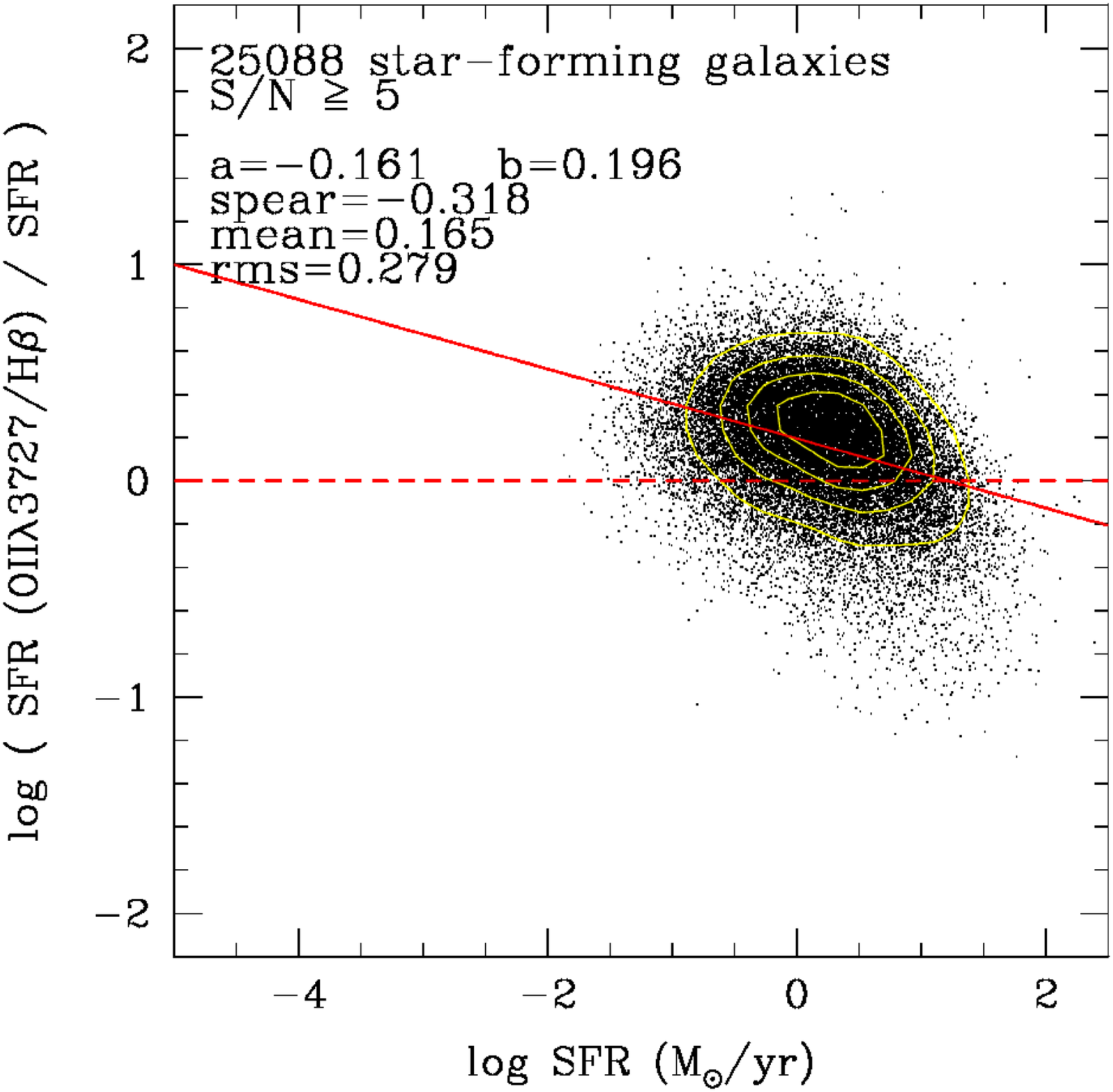}
\includegraphics[width=0.49\columnwidth]{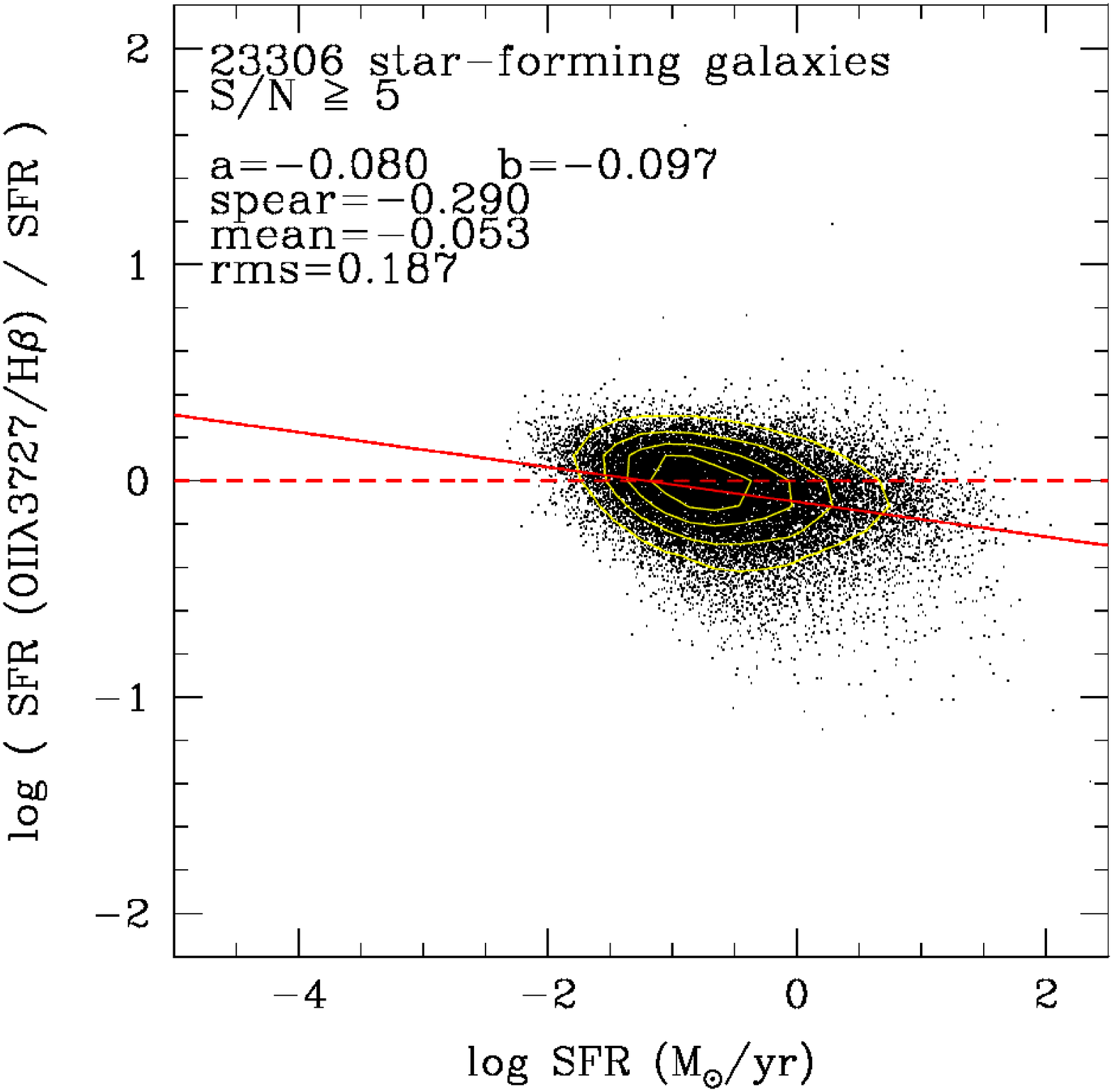}
\includegraphics[width=0.49\columnwidth]{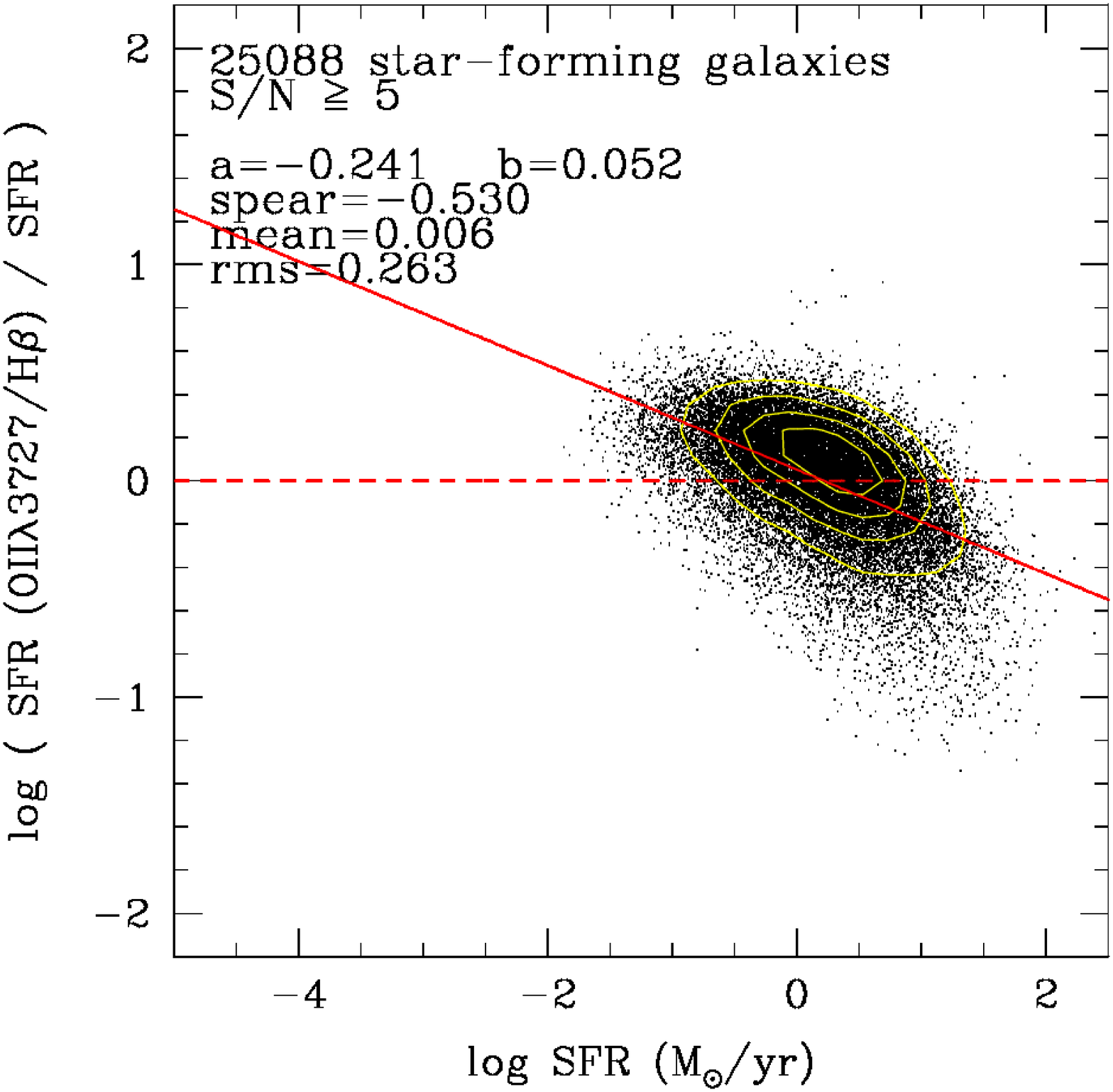}
\par\end{centering}

\caption{Same legend as in Fig.~\ref{fig:compstandard}. The studied calibration
is the new {[}O\noun{ii}]/H$\beta$ calibration given by Eq.~\ref{eq:SFROIIHb}
(top) or Eq.~\ref{eq:SFROIIHb2} (bottom), applied on the two metallicity
sub-samples defined in Sect.~\ref{sec:Description-of-the}. Left:
$\left\langle 12+\log(\mathrm{O/H})\right\rangle =8.68$, right: $\left\langle 12+\log(\mathrm{O/H})\right\rangle =9.07$.}

\label{fig:compsubsamplemeta}
\end{figure}

We also study the residuals of our new calibrations, as compared to
the reference CL01 SFR, as a function of the gas-phase oxygen abundance
estimated with the CL01 method. For the single-line calibrations,
Fig.~\ref{fig:compobsmeta} shows a significant correlation between
these two quantities with Spearman rank correlation coefficients of
the order of $-0.5$ to $-0.6$. Thus, we can derive linear relations
which may be used to correct these calibrations for a desired studied
sample, for which one has a rough estimate of the mean metallicity.
The corrective formula as derived from Fig.~\ref{fig:compobsmeta}
is:\begin{equation}
\log\left(\mathrm{SFR}^{\mathrm{corr}}\right)=\log\left(\mathrm{SFR}\right)+a\times x+b\label{eq:corrSFRMeta}\end{equation}
with $x=12+\log(\mathrm{O/H})$. The $a$ and $b$ coefficients are
summarized in Table~\ref{tab:corrMeta}. We note that this correction
formula give an approximately null correction for a metallicity equal
to the mean value in the SDSS DR4 sample: $12+\log(\mathrm{O/H})=8.89$.

For both {[}O\noun{ii}]/H$\beta$ calibrations, we have fitted a second
degree curve to the residual, which leads to the following corrective
formula:

\begin{equation}
\log\left(\mathrm{SFR}^{\mathrm{corr}}\right)=\log\left(\mathrm{SFR}\right)+a\times x^{2}+b\times x+c\label{eq:corrSFRMeta2}\end{equation}
with $x=12+\log(\mathrm{O/H})$. The $a$,$b$ and $c$ coefficients
are summarized in Table~\ref{tab:corrMeta}.

For the {[}O\noun{ii}]/H$\beta$ two-lines calibration given in Eq.~\ref{eq:SFROIIHb2},
we see that the residuals do not show any more any strong significant
correlation with the metallicity, with a Spearman rank correlation
coefficients of $0.08$. This result shows that the {[}O\noun{ii}]/H$\beta$
calibration given in Eq.~\ref{eq:SFROIIHb2} is not significantly
sensitive to variations in the metallicity. This conclusion is not
true for the {[}O\noun{ii}]/H$\beta$ calibration given in Eq.~\ref{eq:SFROIIHb}
which shows a Spearman correlation coefficient of $0.57$. 

Fig.~\ref{fig:compsubsamplemeta} shows the {[}O\noun{ii}]/H$\beta$
calibrations applied on the two sub-samples defined in Sect.~\ref{sec:Description-of-the},
with two different metallicity properties: $\left\langle 12+\log(\mathrm{O/H})\right\rangle =8.68$
and $\left\langle 12+\log(\mathrm{O/H})\right\rangle =9.07$. It confirms
that the calibration defined in Eq.~\ref{eq:SFROIIHb2} is reliable
when applied on samples with different dust properties: the dispersion
does not significantly changes. In both cases, the observed systematic
shifts are in agreement with the correction formula given in Eq.~\ref{eq:corrSFRMeta2}
and the coefficients given in Table~\ref{tab:corrMeta}.

We conclude that:

\begin{itemize}
\item The {[}O\noun{ii}]/H$\beta$ calibration given in Eq.~\ref{eq:SFROIIHb}
is weakly sensitive to variations in dust attenuation. It can be applied
on samples with an assumed metallicity, using the corrective formula
given in Eq.~\ref{eq:corrSFRMeta2}.
\item The {[}O\noun{ii}]/H$\beta$ calibration given in Eq.~\ref{eq:SFROIIHb2}
is not sensitive to variations in metallicity. It can be applied on
samples with an assumed dust attenuation, using the corrective formula
given in Eq.~\ref{eq:corrSFRTau}.
\end{itemize}

\section{Conclusions\label{sec:Conclusions}}

We draw the following conclusions from our study:

\begin{itemize}
\item As already shown in many previous studies, we confirm from SDSS DR4
data the the best emission-line calibration of the SFR is based on
H$\alpha^{\mathrm{i}}$, this one being corrected for dust attenuation
(see Sect.~\ref{sub:improved}). This calibration has an uncertainty
of $0.17$ dex.
\item If the dust has been estimated from the \emph{wrong} assumption of
a constant intrinsic Balmer ratio, the standard one-parameter \citet{Kennicutt:1998ARA&A..36..189K}
scaling law gives good enough results. Nevertheless, the SDSS DR4
data has shown a $-0.11$ dex systematic shift which can be corrected
either by applying a mean shift in the opposite direction, or by using
a two-parameter power law (see Fig.~\ref{fig:compstandard}, Fig.~\ref{fig:newstandard}
and Eq.~\ref{eq:ha}).
\item When H$\alpha^{\mathrm{i}}$ is not observed but a correction for
dust attenuation is still available, the \citet{Kennicutt:1998ARA&A..36..189K}
law based on {[}O\noun{ii}]$^{\mathrm{i}}$ (and corrected for dust
by \citealp{Kewley:2004AJ....127.2002K}) gives good results but shows
a higher dispersion (see Fig.~\ref{fig:compstandard}).
\item This dispersion may be reduced by taking into account the dependence
of {[}O\noun{ii}]$^{\mathrm{i}}$ on the metallicity. However, the
metallicity correction proposed by \citet{Kewley:2004AJ....127.2002K}
does not end to a significant improvement, while it relies on a very
uncertain calibration of the metallicity, as shown by \citet{Kewley:2008arXiv0801.1849K}
(see Fig.~\ref{fig:o2haoh} and Fig.~\ref{fig:compo2hbi}).
\item A very good correction of this effect is rather obtained by using
the {[}O\noun{ii}]$^{\mathrm{i}}$ and H$\beta^{\mathrm{i}}$ lines
together or, even better, H$\beta^{\mathrm{i}}$ alone (see Eq.~\ref{eq:defo2ha},
Eq.~\ref{eq:o2hadust} and Eq.~\ref{eq:hahbdust}).
\item We caution the reader against the use of the inadequate SFR calibration
when data is corrected for dust estimated from another method (e.g.
CL01 method). If the method to derive dust attenuation is not biased
towards the wrong assumption of a constant intrinsic Balmer ratio,
then a different SFR calibration should be used (see Sect.~\ref{sub:metalunbiased},
Fig.~\ref{fig:newCL01}, Eq.~\ref{eq:haiC}, Eq.~\ref{eq:o2iC},
and Eq.~\ref{eq:hbiC}).
\item We advise the reader not to use dust estimated from SED fitting in
order to calculate SFR from emission lines. This method leads to highly
uncertain results because of the high dispersion in the stellar-to-gas
attenuation ratio (see Fig.~\ref{fig:seddust}). We emphasize that
dust attenuation estimated from SED fitting would be reliable only
if it is applied to the light emitted by stars, not to emission lines.
\item When no estimation of the dust attenuation is available, it is common
in the literature to assume a mean correction, typically $A_{V}=1$.
We advise the reader to use such method with great care for two reasons:
\emph{(i)} the SFRs recovered with the assumption of a constant dust
attenuation are biased by a non-negligible residual slope, coming
from the correlation between dust attenuation and SFR (see Sect.~\ref{sub:meanav1}
and Fig.~\ref{fig:compmean}); \emph{(ii)} additional biases come
from the choice of the assumed dust attenuation which is likely not
to be the right one. The choice of a wrong assumed dust attenuation
leads to non-negligible systematic shifts (see Sect.~\ref{sub:meanplus}).
\item We have derived new direct calibrations between the SFR and the observed
line luminosities which are still quite poor in terms of dispersion,
even if they do not show any more a significant residual slope (see
Fig.~\ref{fig:compobs}, Fig.~\ref{fig:newobs}, Eq.~\ref{eq:hanodust},
Eq.~\ref{eq:o2nodust} and Eq.~\ref{eq:hbnodust}).
\item The calibrations proposed by \citet{Moustakas:2006ApJ...642..775M}
or \citet{Weiner:2006astro.ph.10842W} which include a correction
based on the $B$-band or $H$-band $k$-corrected absolute magnitudes
do neither significantly better nor significantly worse (see Fig.~\ref{fig:compprevious}).
\item We derive two new two-lines calibrations based only on the {[}O\noun{ii}]
and H$\beta$ observed line flux. These calibrations give the best
results among all calibrations based on observed quantities (i.e.
not reliably corrected for dust attenuation). It shows no systematic
shift, no significant residual slope, and a reduced dispersion (see
Fig.~\ref{fig:newtwolines}, Eq.~\ref{eq:SFROIIHb} and Eq.~\ref{eq:SFROIIHb2}).
The minimum uncertainty with data not corrected for dust attenuation
is $0.23$ dex.
\item We have studied the relation between the residuals of our calibrations
based on observed quantities, and the dust attenuation (see Fig.~\ref{fig:compobstauC}).
There is a clear correlation for single-line calibrations, which can
be corrected (see Eq.~\ref{eq:corrSFRTau} and Table~\ref{tab:corrTau}).
We have also found a correlation between the residuals of our calibrations
based on observed quantities (see Fig.~\ref{fig:compobsmeta}), which
can be corrected for single-line calibrations using Eq.~\ref{eq:corrSFRMeta}
and Table~\ref{tab:corrMeta}.
\item Among our two new {[}O\noun{ii}]/H$\beta$ calibrations, the one defined
by Eq.~\ref{eq:SFROIIHb} is designed to be used on a sample with
an assumed metallicity and an unknown dust attenuation, while the
one defined by Eq.~\ref{eq:SFROIIHb2} give better results on samples
with unknown metallicity and an assumed dust attenuation.
\end{itemize}
\begin{acknowledgements}
We thank warmly J. Brinchmann, G. Zamorani, S. Charlot and T. Contini
for valuable discussions concerning this work. We thank the anonymous
referee for his very useful comments which have significantly improved
the paper. F. Lamareille thanks the Osservatorio Astronomico di Bologna
and the COSMOS consortium for the receipt of a post-doctoral fellowship.
The physical properties of SDSS galaxies were produced by a collaboration
of researchers (currently or formerly) from the MPA and the JHU. The
team is made up of Stéphane Charlot, Guinevere Kauffmann and Simon
White (MPA), Tim Heckman (JHU), Christy Tremonti (University of Arizona
- formerly JHU) and Jarle Brinchmann (Centro de Astrofísica da Universidade
do Porto - formerly MPA).
\end{acknowledgements}
\bibliographystyle{aa}
\bibliography{my}

\end{document}